\documentclass[a4paper,11pt]{article}

\usepackage{amsthm,amsmath,amssymb,amsfonts}
\usepackage{amscd}
\usepackage[bookmarksnumbered=true]{hyperref}

\renewcommand{\arraystretch}{1.5}

\usepackage{color}

\usepackage{tikz}

\usetikzlibrary{arrows,positioning,cd,calc,math}
\usetikzlibrary{matrix}
\usetikzlibrary{decorations.pathreplacing,decorations.markings}

\usepackage{titlesec}
\titleformat*{\section}{\Large \normalfont \bfseries}
\titleformat*{\subsection}{\large \normalfont \bfseries}
\usepackage{setspace}

\numberwithin{equation}{section}

\theoremstyle{definition}
\newtheorem{Remark}{Remark}[section]

\def\be{\begin{equation}\begin{gathered}}
\def\ee{\end{gathered}\end{equation}}
\def\T2{\mathbb{T}^2}
\newcommand{\rf}[1]{(\ref{#1})}

\newcommand{\nc}{\newcommand}
\nc{\ra}{\rangle}
\nc{\la}{\langle}
\nc{\seteq}{\mathbin{:=}}
\nc{\simto}{\xrightarrow{\,\sim\,}}

\nc{\Vir}{\textsf{Vir}}
\nc{\Asl}{\widehat{\mathfrak{sl}}}
\nc{\rmF} {\mathrm{F}}
\nc{\calM}{\mathcal{M}}
\nc{\calA}{\mathcal{A}}
\nc{\calF}{\mathcal{F}}
\nc{\calL}{\mathcal{L}}
\nc{\calU}{\mathcal{U}}
\nc{\calE}{\mathcal{E}}
\nc{\calV}{\mathcal{V}}
\nc{\calH}{\mathcal{H}}
\nc{\sfV}{\textsf{V}}
\nc{\tr}{\mathrm{Tr}}
\nc{\NS}{\scriptscriptstyle{\textsf{NS}}}

\def\colA{cyan}
\def\colB{magenta}

\def\Tr{\mathrm{Tr}\,}

\def\diag{\mathrm{diag}\,}
\def\M1{\mathbf{1}}
\def\XXZ{\mathrm{XXZ}}
\def\RLL{\mathrm{RLL}}
\def\T2{\mathbb{T}^2}
\def\Td2{\hat{S}}
\def\UqM{U_q({\mathfrak{gl}_M})}
\def\UqhM{U_q({\widehat{\mathfrak{gl}}_M})}
\def\k{\kappa}

\def\E{\mathrm{{E}}}

\def\la{\times}
\def\lb{+}
\def\ya{x^{\la}}
\def\yb{x^{\lb}}
\def\yabold{\mathbf{x}^{\la}}
\def\ybbold{\mathbf{x}^{\lb}}
\def\mua{\mu^{\la}}
\def\mub{\mu^{\lb}}
\def\lbr{[}
\def\rbr{]}
\def\cof{\mathbf{y}}
\def\c{\mathbf{u}}
\def\cofa{\cof^{\la}}
\def\cofb{\cof^{\lb}}
\def\ta{\tau^{\la}}
\def\tba{\bar{\tau}^{\la}}
\def\tb{\tau^{\lb}}
\def\tbb{\bar{\tau}^{\lb}}
\def\z{z}

\def\avars{\tau}
\def\Z{Z}

\def\Kast{\mathfrak{D}}
\def\GLb{\widehat{\mathrm{PGL}}^\sharp}
\def\Wext{\widetilde{W}}

\def\XG{\mathcal{X}_\Gamma}
\def\Gd{\mathcal{G}_{\Delta}}
\def\Gdp{\mathcal{G}'_{\Delta}}
\def\Gq{\mathcal{G}_\mathcal{Q}}

\def\sc{\mathrm{s}}
\def\H1{\mathrm{H}_1}

\usepackage[colorinlistoftodos]{todonotes}

\title{Cluster integrable systems and spin chains}

\author{A. Marshakov$^{1,2,3,4}$\footnote{andrei.marshakov@gmail.com}, M. Semenyakin$^{1,2}$ \footnote{semenyakinms@gmail.com}}

\date{$^1$ Center for Advanced Studies, Skoltech, Moscow, Russia\\$^2$ Department of Mathematics, NRU HSE, Moscow, Russia \\$^3$ Institute for Theoretical and Experimental Physics, Moscow, Russia \\$^4$ Theory Department of Lebedev Physics Institute, Moscow, Russia}

\begin{document}
\tikzset{
	styleDiagonals/.style={color=violet!50,very thin},
	styleFill/.style={fill,gray,opacity=0.2},
	styleArrow/.style={
		postaction={decorate},
		decoration={markings,mark=at position 0.75 with {\arrow{Stealth[scale=1]}}},
	},
	styleArrowShort/.style={
		postaction={decorate},
		decoration={markings,
		mark=at position 0.5 with {\arrow{Stealth[scale=1]}}
		}
	},
	styleArrowShifted/.style={
		postaction={decorate},
		decoration={markings,
		mark=at position 0.35 with {\arrow{Stealth[scale=1]}}
		}
	},
	styleTextEdges/.style={
		scale = 0.8
	},
	styleQuiverEdge/.style={
		line width=0.7pt, 
		postaction={decorate},
		decoration={markings,
		mark=at position 0.6 with {\arrow{Stealth[scale=1]}}
		}
	},
	quiverVertex/.style={fill=black,thick,radius=0.07},
	quiverVertexBlue/.style={fill=blue,thick,radius=0.07},
	blackCircle/.style={fill=black,thick,radius=0.1,inner sep=0},
	whiteCircle/.style={fill=white,thick,radius=0.1,inner sep=0}
}

\tikzset{
toda_one_text/.pic=
{
\tikzset{
	scale=1
}

\draw[black,styleArrow, thick] (1,1)--(0,0);
\draw[black,styleArrow, thick] (1,1)--(2,2);
\draw[black,styleArrowShifted, thick] (1,1)--(1,3);
\draw[black,styleArrow, thick] (1,1)--(1,0);
\draw[thick] (1,2) -- (1,4);
\draw[thick] (0,2)--(2,4);

\draw[whiteCircle] (1,3) circle;
\draw[blackCircle] (1,1) circle;

}
}

\tikzset{
toda_one_text_noarrow/.pic=
{
\tikzset{
	scale=1
}

\draw[black, thick] (1,1)--(0,0);
\draw[black, thick] (1,1)--(2,2);
\draw[black, thick] (1,1)--(1,3);
\draw[black, thick] (1,1)--(1,0);
\draw[thick] (1,2) -- (1,4);
\draw[thick] (0,2)--(2,4);

\draw[whiteCircle] (1,3) circle;
\draw[blackCircle] (1,1) circle;

}
}

\tikzset{
toda_two_text/.pic=
{
\tikzset{
	scale=1
}

\draw[black,styleArrow, thick] (3,1)--(1,3);
\draw[black,styleArrow, thick] (3,1)--(1,1);
\draw[black,styleArrow, thick] (3,1)--(4,2);
\draw[black,styleArrow, thick] (3,1)--(2,0);
\draw[black,styleArrow, thick] (3,3)--(4,4);
\draw[black,styleArrow, thick] (3,3)--(1,3);
\draw[thick] (1,1) -- (0,0);
\draw[thick] (1,3)--(0,2);
\draw[thick] (2,4)--(1,3);

\draw[whiteCircle] (1,1) circle;
\draw[whiteCircle] (1,3) circle;

\draw[blackCircle] (3,1) circle;
\draw[blackCircle] (3,3) circle;
}
}

\tikzset{
toda_casi_hor/.pic=
{
\tikzset{
	scale=1
}

\draw[blue,styleArrow, thick] (0,0)--(1,1);
\draw[orange,styleArrow, thick] (2,2)--(1,1);
\draw[orange,styleArrowShifted, thick] (1,1)--(1,3);
\draw[blue,styleArrow, thick] (1,1)--(1,0);
\draw[orange, thick] (1,2) -- (1,3);
\draw[blue, thick] (1,3) -- (1,4);
\draw[orange, thick] (0,2)--(1,3);
\draw[blue, thick] (2,4)--(1,3);

\draw[whiteCircle] (1,3) circle;
\draw[blackCircle] (1,1) circle;

}
}

\tikzset{
toda_casi_hor_twist/.pic=
{
\tikzset{
	scale=1
}

\draw[green,styleArrow, thick] (1,1)--(0,0);
\draw[green,styleArrowShifted, thick] (1,3)--(1,1);
\draw[green, thick] (2,4)--(1,3);

\draw[red,styleArrow, thick] (1,1)--(2,2);
\draw[red,styleArrow, thick] (1,0)--(1,1);
\draw[red, thick] (1,3) -- (1,4);
\draw[red, thick] (0,2)--(1,3);

\draw[whiteCircle] (1,3) circle;
\draw[blackCircle] (1,1) circle;

}
}

\tikzset{
quiver_cross/.pic=
{
\tikzset{
	scale=1, font = \small 
}

\draw[styleQuiverEdge] (0,1) -- (1,1);
\draw[styleQuiverEdge] (1,1) -- (1,0);
\draw[styleQuiverEdge] (1,1) -- (1,2);
\draw[styleQuiverEdge] (2,1) -- (1,1);

\draw[quiverVertex] (0,1) circle;
\draw[quiverVertex] (1,0) circle;
\draw[quiverVertex] (1,2) circle;
\draw[quiverVertex] (2,1) circle;
\draw[quiverVertex] (1,1) circle;

}
}

\tikzset{
quiver_cross_skew/.pic=
{
\tikzset{
	scale=1, font = \small 
}

\draw[styleQuiverEdge] (2,2) -- (1,1);
\draw[styleQuiverEdge] (1,1) -- (2,0);
\draw[styleQuiverEdge] (1,1) -- (0,2);
\draw[styleQuiverEdge] (0,0) -- (1,1);

\draw[quiverVertex] (0,0) circle;
\draw[quiverVertex] (2,2) circle;
\draw[quiverVertex] (0,2) circle;
\draw[quiverVertex] (2,0) circle;
\draw[quiverVertex] (1,1) circle;

}
}

\tikzset{
bipart_text/.pic=
{
\tikzset{
	scale=0.75, font = \small 
}


\draw[black,styleArrow, thick] (0,2)--(2,4);
\draw[black,styleArrow, thick] (0,2)--(2,0);
\draw[black,styleArrow, thick] (4,2)--(2,4);
\draw[black,styleArrow, thick] (4,2)--(2,0);

\draw[whiteCircle] (2,4) circle;
\draw[whiteCircle] (2,0) circle;
\draw[blackCircle] (0,2) circle;
\draw[blackCircle] (4,2) circle;

}
}

\tikzset{
bipart_casi_hor/.pic=
{
\tikzset{
	scale=0.75, font = \small 
}

\draw[orange, styleArrow, thick] (2,4)--(0,2);
\draw[orange, styleArrow, thick] (4,2)--(2,4);
\draw[blue, styleArrow, thick] (0,2)--(2,0);
\draw[blue, styleArrow, thick] (2,0)--(4,2);

\draw[whiteCircle] (2,4) circle;
\draw[whiteCircle] (2,0) circle;
\draw[blackCircle] (0,2) circle;
\draw[blackCircle] (4,2) circle;

}
}

\tikzset{
bipart_casi_vert/.pic=
{
\tikzset{
	scale=0.75, font = \small 
}

\draw[red, styleArrow, thick] (0,2)--(2,4);
\draw[red, styleArrow, thick] (2,0)--(0,2);
\draw[green, styleArrow, thick] (4,2)--(2,0);
\draw[green, styleArrow, thick] (2,4)--(4,2);

\draw[whiteCircle] (2,4) circle;
\draw[whiteCircle] (2,0) circle;
\draw[blackCircle] (0,2) circle;
\draw[blackCircle] (4,2) circle;

}
}

\tikzset{
bipart/.pic=
{
\tikzset{
	font = \small 
}

\draw[thick] (1,2)--(0,1);
\draw[thick] (2,1)--(1,2);
\draw[thick] (0,1)--(1,0);
\draw[thick] (1,0)--(2,1);

\draw[fill=black,thick,radius=0.066, inner sep=0] (1,2) circle;
\draw[fill=black,thick,radius=0.066, inner sep=0] (1,0) circle;
\draw[fill=white,thick,radius=0.066, inner sep=0] (0,1) circle;
\draw[fill=white,thick,radius=0.066, inner sep=0] (2,1) circle;

}
}


\tikzset{
Wt_triv_gr/.pic=
{
	\fill[color=gray!10] (0,0)--(1,0)--(1,1)--(0,1)--cycle;
	\draw (0,0)--(1,0);
	\draw (0,1)--(1,1);
}
}
\tikzset{
Wt_triv_gr_no_top/.pic=
{
	\fill[color=gray!10] (0,0)--(1,0)--(1,1)--(0,1)--cycle;
	\draw (0,0)--(1,0);
}
}
\tikzset{
Wt_triv_gr_no_bot/.pic=
{
	\fill[color=gray!10] (0,0)--(1,0)--(1,1)--(0,1)--cycle;
	\draw (0,1)--(1,1);
}
}
\tikzset{
Wt_triv_gr_no_top_bot/.pic=
{
	\fill[color=gray!10] (0,0)--(1,0)--(1,1)--(0,1)--cycle;
}
}

\tikzset{
Wt_triv_wh/.pic=
{
	\fill[color=white!10] (0,0)--(1,0)--(1,1)--(0,1)--cycle;
	\draw (0,0)--(1,0);
	\draw (0,1)--(1,1);
}
}
\tikzset{
Wt_triv_wh_no_top/.pic=
{
	\fill[color=white!10] (0,0)--(1,0)--(1,1)--(0,1)--cycle;
	\draw (0,0)--(1,0);
}
}
\tikzset{
Wt_triv_wh_no_bot/.pic=
{
	\fill[color=white!10] (0,0)--(1,0)--(1,1)--(0,1)--cycle;
	\draw (0,1)--(1,1);
}
}
\tikzset{
Wt_triv_wh_no_top_bot/.pic=
{
	\fill[color=white!10] (0,0)--(1,0)--(1,1)--(0,1)--cycle;
}
}

\tikzset{
Wt_lower_left_gr/.pic=
{
	\fill[color=gray!10] (0,0)..controls(0.5,0)..(1,1)--(0,1)--cycle;
	\draw (0,0)..controls(0.5,0)..(1,1)--(0,1);
}
}

\tikzset{
Wt_lower_left_wh/.pic=
{
	\fill[color=gray!10] (0,0)..controls(0.5,0)..(1,1)--(1,0)--cycle;
	\draw (0,0)..controls(0.5,0)..(1,1)--(0,1);
}
}

\tikzset{
Wt_lower_right_gr/.pic=
{
	\fill[color=gray!10] (1,0)..controls(0.5,0)..(0,1)--(1,1)--cycle;
	\draw (1,0)..controls(0.5,0)..(0,1)--(1,1);
}
}

\tikzset{
Wt_lower_right_wh/.pic=
{
	\fill[color=gray!10] (1,0)..controls(0.5,0)..(0,1)--(0,0)--cycle;
	\draw (1,0)..controls(0.5,0)..(0,1)--(1,1);
}
}

\tikzset{
Wt_upper_left_gr/.pic=
{
	\fill[color=gray!10] (0,1)..controls(0.5,1)..(1,0)--(0,0)--cycle;
	\draw (0,1)..controls(0.5,1)..(1,0)--(0,0);
}
}

\tikzset{
Wt_upper_left_wh/.pic=
{
	\fill[color=gray!10] (0,1)..controls(0.5,1)..(1,0)--(1,1)--cycle;
	\draw (0,1)..controls(0.5,1)..(1,0)--(0,0);
}
}

\tikzset{
Wt_upper_right_gr/.pic=
{
	\fill[color=gray!10] (1,1)..controls(0.5,1)..(0,0)--(1,0)--cycle;
	\draw (1,1)..controls(0.5,1)..(0,0)--(1,0);
}
}

\tikzset{
Wt_upper_right_wh/.pic=
{
	\fill[color=gray!10] (1,1)..controls(0.5,1)..(0,0)--(0,1)--cycle;
	\draw (1,1)..controls(0.5,1)..(0,0)--(1,0);
}
}

\tikzset{
Wt_LambdaL_gr/.pic=
{
	\fill[color=gray!10] (0,0)..controls(0.5,0)..(1,1)--(0,1)--cycle;
	\draw (0,0)..controls(0.5,0)..(1,1);
}
}

\tikzset{
Wt_LambdaL_wh/.pic=
{
	\fill[color=gray!10] (0,0)..controls(0.5,0)..(1,1)--(1,0)--cycle;
	\draw (0,0)..controls(0.5,0)..(1,1);
}
}

\tikzset{
Wt_LambdaU_gr/.pic=
{
	\fill[color=gray!10] (0,0)..controls(0.5,1)..(1,1)--(0,1)--cycle;
	\draw (0,0)..controls(0.5,1)..(1,1);
}
}

\tikzset{
Wt_LambdaU_wh/.pic=
{
	\fill[color=gray!10] (0,0)..controls(0.5,1)..(1,1)--(1,0)--cycle;
	\draw (0,0)..controls(0.5,1)..(1,1);
}
}

\tikzset{
thurston_def/.pic=
{
\tikzset{
	scale=0.5, font = \small 
}

\draw[black,styleArrow] (0,3)--(3,6);
\draw[black,styleArrow] (0,3)--(3,0);
\draw[black,styleArrow] (6,3)--(3,6);
\draw[black,styleArrow] (6,3)--(3,0);

\draw[whiteCircle] (3,6) circle;
\draw[whiteCircle] (3,0) circle;
\draw[blackCircle] (0,3) circle;
\draw[blackCircle] (6,3) circle;

}
}

\tikzset{
thurston_left/.pic=
{
\tikzset{
	scale=0.5, font = \small 
}

\draw[black,styleArrowShort] (6,3)--(4.5,4.5);
\draw[black,styleArrowShort] (6,3)--(4.5,1.5);

}
}

\tikzset{
thurston_top/.pic=
{
\tikzset{
	scale=0.5, font = \small 
}


\draw[black,styleArrowShort] (1.5,1.5)--(3,0);
\draw[black,styleArrowShort] (4.5,1.5)--(3,0);

}
}

\tikzset{
thurston_right/.pic=
{
\tikzset{
	scale=0.5, font = \small 
}


\draw[black,styleArrow] (0,3)--(3,6);
\draw[black,styleArrow] (0,3)--(3,0);
\draw[black,styleArrowShort] (4.5,4.5)--(3,6);
\draw[black,styleArrowShort] (4.5,1.5)--(3,0);

\draw[whiteCircle] (3,6) circle;
\draw[whiteCircle] (3,0) circle;
\draw[blackCircle] (0,3) circle;

}
}

\tikzset{
thurston_bottom/.pic=
{
\tikzset{
	scale=0.5, font = \small 
}

\draw[black,styleArrow] (0,3)--(3,6);
\draw[black,styleArrowShort] (0,3)--(1.5,1.5);
\draw[black,styleArrow] (6,3)--(3,6);
\draw[black,styleArrowShort] (6,3)--(4.5,1.5);

\draw[whiteCircle] (3,6) circle;
\draw[blackCircle] (0,3) circle;
\draw[blackCircle] (6,3) circle;

}
}

\tikzset{
thurston_right_bottom/.pic=
{
\tikzset{
	scale=0.5, font = \small 
}

\draw[black,styleArrow] (0,3)--(3,6);
\draw[black,styleArrow] (0,3)--(1.5,1.5);
\draw[black,styleArrowShort] (4.5,4.5)--(3,6);

\draw[whiteCircle] (3,6) circle;
\draw[blackCircle] (0,3) circle;

}
}


\maketitle

\begin{abstract}\vspace*{2pt}
\noindent
We discuss relation between the cluster integrable systems and spin chains in the context of their correspondence with 5d supersymmetric gauge theories.
It is shown that $\mathfrak{gl}_N$ $\XXZ$-type  spin chain on $M$ sites is isomorphic to a cluster integrable system with $N \times M$ rectangular Newton polygon and $N \times M$ fundamental domain of a 'fence net' bipartite graph. The Casimir functions of the Poisson bracket, labeled by the zig-zag paths on the graph, correspond to the inhomogeneities, on-site Casimirs and twists of the chain, supplemented by total spin. The symmetricity of cluster formulation implies natural spectral duality, relating $\mathfrak{gl}_N$-chain on $M$ sites with the $\mathfrak{gl}_M$-chain on $N$ sites.
For these systems we construct explicitly a subgroup of the cluster mapping class group $\Gq$ and show that it acts by permutations of zig-zags and, as a consequence, by permutations of twists and inhomogeneities. Finally, we derive Hirota bilinear equations, describing dynamics of the tau-functions or A-cluster variables under the action of some generators of $\Gq$.

\end{abstract}

\tableofcontents

\section{Introduction}

In the seminal paper \cite{SW:1994} Seiberg and Witten found 'exact solution' to 4d $\mathcal{N}=2$ super-symmetric gauge theory in the strong coupling regime. More strictly, the IR effective couplings
were constructed geometrically, from the period integrals on a complex curve, whose moduli are determined by the condensates and bare couplings of the UV gauge theory.
Shortly after, it has been also realized \cite{GKMMM} that natural language for the Seiberg-Witten theory is given by classical integrable systems. In such context the pure supersymmetric gauge theories (with only $\mathcal{N}=2$ vector supermultiplets) correspond to the Toda chains, while integrable systems for the gauge theories with fundamental matter multiplets are usually identified with classical spin chains of $XXX$-type.

The next important step was proposed in \cite{S:1996}, where this picture has been lifted to $5d$. Then it has been shown that transition from $4d$ to $5d$ (actually -- four plus one compact dimensions) results in 'relativization' of the integrable systems \cite{N:1996} (in the sense of Ruijsenaars \cite{R:1990}). In the simplest case of $SU(2)$ pure Yang-Mills theory, or affine Toda chain with two particles, instead of the Hamiltonian
\begin{equation}
H_{4d}=p^2 + e^{q}+Z e^{-q},
\end{equation}
corresponding to $4d$ theory, one has to consider
\begin{equation}
H_{5d}=e^p + e^{-p} + e^{q} + Z e^{-q},
\end{equation}
or the Hamiltonian of relativistic Toda chain, which describes effective theory for $5d$ pure  $SU(2)$ Yang-Mills \footnote{The slightly misleading term 'relativistic' appears here due to formal similarity of momentum dependence to the rapidities of a massive relativistic particle in $1+1$ dimensions.}.
It has been also shown that $5d$ theories with fundamental matter correspond to $\XXZ$-type spin chains (see e.g. \cite{MM:1997} and references therein).

Relativistic Toda chains lead to natural relation of this story with the integrable systems on the Poisson submanifolds in Lie groups, or more generally to the \emph{cluster} integrable systems -- recently discovered class of integrable systems of relativistic type \cite{GK:2011,M:2012,FM:2014}. Direct relation between cluster integrable systems and $5d$ gauge theories has been proposed in \cite{BGM:2017}. It was shown there that for the case of Newton polygons with single internal point, dynamics of discrete flow is governed by q-Painlev\'e equations and their bilinear form is solved by Nekrasov $5d$ dual partition functions (for other examples of $5d$ gauge
 theories the same phenomenon was considered in \cite{Jimbo:2017, BGT:2017, BGM:2018})\footnote{Other relations between $5d$ supersymmetric gauge theories and cluster integrable systems (involving exact   spectrum of quantized cluster integrable systems, BPS counting and toric Calabi-Yau quantization) were discussed in \cite{FHM}, \cite{Gaiotto}, \cite{ORV} correspondingly. They seem to be related to our case and we are going to return to these issues elsewhere.
}.

\paragraph{Cluster integrable systems}

Any convex polygon $\Delta$ with vertices in  $\mathbb{Z}^2\subset\mathbb{R}^2$ can be considered as a Newton polygon of polynomial $f_\Delta(\lambda,\mu)$, and equation
\be
\label{SC}
f_\Delta(\lambda,\mu) = \sum_{(a,b)\in\Delta}\lambda^a\mu^b f_{a,b}=0.
\ee
defines a plane (noncompact) spectral curve in $\mathbb{C}^\times\times \mathbb{C}^\times$. The genus $g$
of this curve is equal to the number of integral points strictly inside the polygon $\Delta$.

According to \cite{GK:2011},\cite{FM:2014} a convex Newton polygon $\Delta$, modulo action of $SA(2,\mathbb{Z})=SL(2,\mathbb{Z})\ltimes \mathbb{Z}^2$, defines a cluster integrable system, i.e. an integrable system on X-cluster Poisson variety $\mathcal{X}$ of dimension $\dim_\mathcal{X} = 2S$, where $S$ is area of the polygon $\Delta$. The Poisson structure can be encoded by quiver $\mathcal{Q}$ with $2S$ vertices. Let  $\epsilon_{ij}$ be the number of arrows from $i$-th to $j$-th vertex ($\epsilon_{ji} = -\epsilon_{ij}$) of $\mathcal{Q}$, then logarithmically constant Poisson bracket has the form
\begin{equation}
\label{PB}
\{ x_i, x_j\} = \varepsilon_{ij}x_ix_j,\ \ \ \ \{x_i\}\in \left(\mathbb{C}^\times\right)^{2S}.
\end{equation}
The product of all cluster variables $\prod_i x_i$ is a Casimir for the Poisson bracket \eqref{PB}. Setting it to be
\be
\label{eq:q}
q=\prod_i x_i=1
\ee
and fixing values of other Casimirs, corresponding to the boundary points of Newton polygon $I\in\bar{\Delta}$ (their total number is $B-3$, since equation \rf{SC} is defined modulo multiplicative renormalization of spectral parameters $\lambda$, $\mu$ and $f_\Delta(\lambda,\mu)$ itself), one obtains symplectic leaf.

The properly normalized coefficients, corresponding to the internal points, are integrals of motion
in involution
\be
\label{Pinv}
\{ f_{a,b}(x),f_{c,d}(x) \} = 0,\ \ \ \ \ (a,b), (c,d)\in\Delta
\ee
w.r.t. the Poisson bracket \rf{PB}. By Pick theorem one has
\be
\label{Pick}
2S-1= (B-3)+2g
\ee
where $g$ is the number of internal points (or genus of the curve \rf{SC}), or the number of independent integrals of motion. So the number of independent integrals of motion is half of the dimension of symplectic leaf, and the system is integrable. One of distinguished features of the cluster integrable systems is that their integrals of motion are the Laurent polynomials of (generally -- fractional powers) in the cluster variables.\\

There are several different ways to get explicit form of the spectral curve equation \rf{SC}:
\begin{itemize}
  \item Compute the dimer partition function (with signs) for a bipartite graph on a torus. One possible form of it is a characteristic equation
  \be
  \det \mathfrak{D}(\lambda,\mu) = 0
  \ee
  for the Kasteleyn-Dirac operator on a bipartite graph $\Gamma\subset\T2$, depending on two 'quasimomenta' $\lambda,\mu\in \mathbb{C}^\times$;
  \item Alternatively, one can get the same equation \rf{SC} as a Lax-type equation of a spectral curve,
 with the Lax operator coming from affine Lie group construction, identifying cluster variety with a Poisson submanifold in the co-extended affine group.
\end{itemize}
Short exposition of the first construction of cluster integrable system, relevant for this paper, is contained Appendix~\ref{ss:cluster}.

\paragraph{Classical integrable chains}{
Integrability of classical $\mathfrak{gl}_{M}$ chains of $\XXZ$ type is based on the that their $M\times M$ Lax matrices satisfy the following  classical $\RLL$ relation
\begin{equation}
\label{rLLintro}
\{L(\lambda)\otimes L(\mu)\} = \k[r(\lambda/\mu),L(\lambda)\otimes L(\mu)]
\end{equation}
with the classical (trigonometric) $r$-matrix~\footnote{See details of derivation of Lax matrix from quantum algebra and notations in Appendix \ref{ss:limit}.}
\begin{equation}
\label{eq:rmat1}
r(\lambda) = -\dfrac{\lambda^{1/2}+\lambda^{-1/2}}{\lambda^{1/2}-\lambda^{-1/2}} \sum_{i\neq j} E_{ii}\otimes E_{jj}  + \dfrac{2}{\lambda^{1/2}-\lambda^{-1/2}} \sum\limits_{i\neq j} \lambda^{-\frac{1}{2}s_{ij}}E_{ij}\otimes E_{ji}.
\end{equation}
A classical chain of trigonometric type can be defined by the monodromy operator
\be
\label{eq:spinchaincurve}
T(\mu)=L_N(\mu/\mu_N)\ldots L_1(\mu/\mu_1)\in \mathrm{End}(\mathbb{C}^M)
\ee
where $M$ is called 'rank' of the chain. Integrability is guaranteed by classical
RTT-relation
\begin{equation}
\{T(\lambda)\otimes T(\mu)\}=\k[r(\lambda/\mu),T(\lambda)\otimes T(\mu)]
\end{equation}
for the monodromy operator that follows from \rf{rLLintro}, and gives rise to the integrals of motion, which can be extracted
from the spectral curve equation \rf{SC} given explicitly by the formula
\begin{equation}
\label{fTcurve}
f_\Delta(\lambda,\mu)=\det(\lambda Q - T(\mu))=0.
\end{equation}
where $Q$ - diagonal twist matrix with the constant entities. Relativistic Toda system can be considered as certain degenerate case of generic $\XXZ$ chain of rank $M=2$ (of length $N$ for $N$ particles). 
}

\paragraph{Examples of Newton polygons}

In what follows we mostly consider cluster integrable systems, corresponding to the Newton polygons of the following types:\\

\begin{figure}[h!]
\begin{center}

\scalebox{0.8}{
\begin{tikzpicture}

\tikzmath{\Lx=3;\Ly=1;\xs=-9.5;\ys=1.5;\d=1.5;}


\begin{scope}[thick]
	\draw (\xs,\ys)--(\xs,\ys+\d);
	\draw (\xs,\ys+\d)--(\xs+\d*3,\ys);
	\draw (\xs+\d*3,\ys)--(\xs+\d*3,\ys-\d);
	\draw (\xs+\d*3,\ys-\d)--(\xs,\ys);
\end{scope}

\foreach \x in {0,...,\Lx}
	\foreach \y in {-1,...,\Ly}
		\draw[fill] (\xs+\d*\x,\ys+\d*\y) circle[radius=0.05];

\tikzmath{\Lx=3;\Ly=2;\xs=-2.75;\ys=0;\d=1.5;}


\begin{scope}[thick]
	\draw (\xs,\ys)--(\xs,\ys+\d*2);
	\draw (\xs,\ys+\d*2)--(\xs+\d*3,\ys+\d*2);
	\draw (\xs+\d*3,\ys+\d*2)--(\xs+\d*3,\ys);
	\draw (\xs+\d*3,\ys)--(\xs,\ys);
\end{scope}

\foreach \x in {0,...,\Lx}
	\foreach \y in {0,...,\Ly}
		\draw[fill] (\xs+\d*\x,\ys+\d*\y) circle[radius=0.05];



\draw[white] (-2,0)--(-2,1);
\draw[white] (6.5,0)--(6.5,1);

\tikzmath{\Lx=3;\Ly=2;\xs=4;\ys=0;\d=1.5;}


\begin{scope}[thick]
	\draw (\xs,\ys)--(\xs+\d,\ys+\d*2);
	\draw (\xs+\d,\ys+\d*2)--(\xs+\d*4,\ys+\d*2);
	\draw (\xs+\d*4,\ys+\d*2)--(\xs+\d*3,\ys);
	\draw (\xs+\d*3,\ys)--(\xs,\ys);
\end{scope}

\foreach \x in {0,...,4}
	\foreach \y in {0,...,\Ly}
		\draw[fill] (\xs+\d*\x,\ys+\d*\y) circle[radius=0.05];


\end{tikzpicture}
}

\end{center}
\caption{From left to right Newton polygons for: Toda chain on three sites, $\mathfrak{gl}_2$ $\XXZ$ spin chain on three sites, $\mathfrak{gl}_2$ spin chain on three sites with cyclic twist matrix.}
\label{fig:polyexamples}
\end{figure}
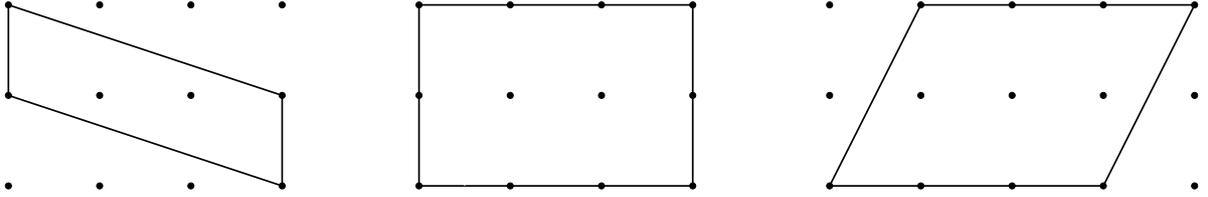

\begin{itemize}
  \item Quadrangles with four boundary points, where all internal points are located along the same straight line, as on Fig.~\ref{fig:polyexamples}, left. This is the case of relativistic Toda chains, studied in \cite{BGM:2017}. The corresponding gauge theory is 5d $\mathcal{N}=1$ Yang-Mills theory with $SU(N)$ gauge group (for $N-1$ internal points) without matter multiplets, possibly with the Chern-Simons term of level $|k|\leq N$ -- in such case quadrangle is not a parallelogram.
  \item ``Big'' rectangles (modulo $SA(2,\mathbb{Z})$ transform). For the $N\times M$ rectangle (see Fig.~\ref{fig:polyexamples}, center) this can be alternatively described as a $\mathfrak{gl}_N$ spin chain on $M$ sites
       (cf. with \cite{BPTY:2012}), or vice versa. The corresponding 5d gauge theories are given by linear quivers theories with the $SU(N)$ gauge group at each of $M-1$ nodes: see Fig. \ref{fig:gaugequiver}.

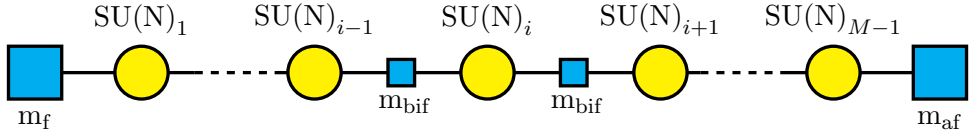
\begin{figure}[!h]

\begin{center}

\begin{tikzpicture}

\tikzmath{
	\w=1.25;
	\s=1;
}

\begin{scope}[scale = 0.7]

\draw[line width = \w] (-4.5,0) -- (-1.5,0);
\draw[line width = \w, dashed] (-1.5,0) -- (0,0);
\draw[line width = \w] (0,0) -- (8,0);
\draw[line width = \w, dashed] (8,0) -- (9.5,0);
\draw[line width = \w] (9.5,0) -- (12.5,0);

\fill[color = cyan] (12,-0.5) rectangle (13, 0.5);
\draw[line width = \w] (12,-0.5) rectangle (13, 0.5);
\node[scale = \s] at (12.5,-0.9) {$\mathrm{m_{af}}$};

\fill[color = yellow] (10.5,0) circle (0.5);
\draw[line width = \w] (10.5,0) circle (0.5);
\node[scale = \s] at (10.5,1) {$\mathrm{SU(N)}_{M-1}$};

\fill[color = yellow] (7.25,0) circle (0.5);
\draw[line width = \w] (7.25,0) circle (0.5);
\node[scale = \s] at (7.25,1) {$\mathrm{SU(N)}_{i+1}$};

\fill[color = cyan] (5.37, -0.25) rectangle (5.87, 0.25);
\draw[line width = \w] (5.37, -0.25) rectangle (5.87, 0.25);
\node[scale = \s] at (5.69,-0.6) {$\mathrm{m_{bif}}$};

\fill[color = yellow] (4,0) circle (0.5);
\draw[line width = \w] (4,0) circle (0.5);
\node[scale = \s] at (4,1) {$\mathrm{SU(N)}_{i}$};

\fill[color = cyan] (2.13, -0.25) rectangle (2.63, 0.25);
\draw[line width = \w] (2.13, -0.25) rectangle (2.63, 0.25);
\node[scale = \s] at (2.45,-0.6) {$\mathrm{m_{bif}}$};

\fill[color = yellow] (0.75,0) circle (0.5);
\draw[line width = \w] (0.75,0) circle (0.5);
\node[scale = \s] at (0.75,1) {$\mathrm{SU(N)}_{i-1}$};

\fill[color = yellow] (-2.5,0) circle (0.5);
\draw[line width = \w] (-2.5,0) circle (0.5);
\node[scale = \s] at (-2.5,1) {$\mathrm{SU(N)}_{1}$};

\fill[color = cyan] (-4,-0.5) rectangle (-5, 0.5);
\draw[line width = \w] (-4,-0.5) rectangle (-5, 0.5);
\node[scale = \s] at (-4.5,-0.9) {$\mathrm{m_{f}}$};

\end{scope}

\end{tikzpicture}
\end{center}
\caption{Linear quiver which defines multiplets for $\mathcal{N}=1$ gauge theory. Circles are for gauge vector multiplets, boxes are for hypermultiplets.}
\label{fig:gaugequiver}
\end{figure}

  \item ``Twisted rectangles'', or just the parallelograms, which are not $SA(2,\mathbb{Z})$-equivalent to the
  previous class (see Fig.~\ref{fig:polyexamples}, right), they can be alternatively formulated as spin chains with nontrivial twists. Gauge theory counterpart for this class of polygons is not yet known, except for the twisted $\mathfrak{gl}_N$ chain on one site, leading back to the basic class of Toda chains.
\end{itemize}

For all these families the spectral curve of an integrable system, determined by equation \rf{SC} is endowed with a pair of meromorphic differentials $\left(\frac{d\lambda}{\lambda},\frac{d\mu}{\mu}\right)$ with the fixed $2\pi i\mathbb{Z}$-valued periods. One can also use this pair to introduce (the $SL(2,\mathbb{Z})$-invariant for our family) 2-form $\frac{d\lambda}{\lambda}\wedge\frac{d\mu}{\mu}$
on $\mathbb{C}^\times\times\mathbb{C}^\times$, whose 'pre-symplectic' form is the SW differential.

\paragraph{Structure of the paper}

The main aim is to extend the correspondence between 5d theories and cluster integrable systems to wider class of models. We find isomorphism between the classes of $\mathfrak{gl}_N$ $\XXZ$-like spin chains on $M$ sites, corresponding to $5d$ $SU(N)$ linear quiver gauge theories (see Fig.~\ref{fig:gaugequiver}) \cite{BPTY:2012}, and cluster integrable systems with $N\times M$ rectangular Newton polygons.

We start from the brief overview of classical $\XXZ$ spin chains. We illustrate with the simple example of relativistic Toda chain, how Lax operators naturally arise from the Dirac-Kasteleyn operator of cluster integrable system. Then we do this for the general case of $\XXZ$ spin chain of arbitrary length and rank. Spectral (or fiber-base) duality arises as an obvious consequence of the structure of considered bipartite graph. Spin chains with additional cyclic permutation twist matrix arise in the cluster context naturally as well.

Then we explain structure of large subgroup of cluster mapping class group $\Gq$. We show that in case of general rank and length of chain it contains subgroup (\ref{eq:GqGeneralNM}) which act in autonomous $q=1$ limit by permutations of inhomogeneities and diagonal twist parameters of spin chain. We also discuss issue of deautonomization and propose a way to define action of $\Gq$ on zig-zags in $q\neq 1$ case. Then we derive bilinear equations for the action of generators of $\Gq$ on A-cluster variables.

\section{Spin chains}

\subsection{Relativistic Toda chain}

\begin{figure}[h!]
\begin{center}
\scalebox{0.9}{
\begin{tikzpicture}
\tikzmath{
	int \Lx,\Ly;
	\Lx=3;\Ly=1;
	\scale=1;
	\dofset=0.1;
	\cofset=0;
}

\begin{scope}[scale=\scale]

\clip(-\cofset,-\cofset) rectangle (2*\Lx + \cofset, 4*\Ly + \cofset);

\tikzmath{
		for \i in {0,...,\Lx}{
			{			
				\draw  (\i*2,0) pic[scale=\scale] {toda_one_text};
			};
		};
};

\end{scope}

\begin{scope}[scale=\scale]

\tikzmath{
	for \i in {0,...,\Lx-1}{
		\xc = int(round(\i+1));
		{	
			\node at (0.7+\i*2, 1) {$_\xc$};	
			\node at (0.7+\i*2, 3) {$_\xc$};	
			\node[styleTextEdges, anchor=east] at (\i*2+1, 2) {$\xi_{\xc}$};		
			\node[styleTextEdges, anchor=east] at (\i*2+1.5, 0.5) {$\eta_{\xc}$};
		};
  	};
	for \i in {0,...,\Lx-2}{
		\xc = int(round(\i+1));
		{	
  			\node at (2+\i*2, 1) {$\yabold_{\xc}$};
			\node at (2+\i*2, 3) {$\ybbold_{\xc}$};
			\node[styleTextEdges] at (\i*2+1.9, 1.5) {$\sqrt{\xi_{\xc}\eta_{\xc}}$};		
			\node[styleTextEdges] at (\i*2+1.9, 3.5) {$\sqrt{\xi_{\xc}\eta_{\xc}}$};
		};
  	};
  	\xc = int(round(\Lx));
	{  	
  	\node at (\Lx*2-0.3, 0.7) {$\yabold_{\xc}$};
	\node at (\Lx*2-0.3, 2.7) {$\ybbold_{\xc}$};
	\node[styleTextEdges] at (\Lx*2-0.3, 1.1) {$\sqrt{\xi_{\xc}\eta_{\xc}}\varkappa_1$};
	\node[styleTextEdges] at (\Lx*2-0.2, 3.1) {$\sqrt{\xi_{\xc}\eta_{\xc}}/\varkappa_2$};
	};
};

\draw[dotted, thick] (-\dofset, -\dofset) -- (2*\Lx+\dofset,-\dofset) -- (2*\Lx+\dofset,4*\Ly+\dofset) -- (-\dofset,4*\Ly+\dofset) -- (-\dofset,-\dofset);

\draw[thin, color = \colB, opacity=1, styleArrowShifted] (0.1-\cofset,-\dofset) -- (0.1+\cofset,4*\Ly+\dofset);
\draw[thin, color = \colA, opacity=1, styleArrowShifted] (\dofset+2*\Lx, 4*\Ly-0.1-\cofset) -- (-\dofset,4*\Ly-0.1-\cofset);

\end{scope}
\end{tikzpicture}
\begin{tikzpicture}
\tikzmath{
	int \Lx,\Ly;
	\Lx=3;\Ly=1;
	\scale=1;
	\dofset=0.1;
	\cofset=0;
}

\begin{scope}[scale=\scale]

\clip(-\cofset,-\cofset) rectangle (2*\Lx + \cofset, 4*\Ly + \cofset);

\tikzmath{
		for \i in {0,...,\Lx}{
			{			
				\draw  (\i*2,0) pic[scale=\scale] {toda_casi_hor};
			};
		};
};

\end{scope}

\begin{scope}[scale=\scale]

\tikzmath{
	for \i in {0,...,\Lx-1}{
		\xc = int(round(\i+1));
		{	
			\node at (0.7+\i*2, 1) {$_\xc$};	
			\node at (0.7+\i*2, 3) {$_\xc$};	
		};
  	};
};

\node at (0.3, 2.8) {$\alpha$};
\node at (0.3, 0.8) {$\beta$};

\draw[dotted, thick] (-\dofset, -\dofset) -- (2*\Lx+\dofset,-\dofset) -- (2*\Lx+\dofset,4*\Ly+\dofset) -- (-\dofset,4*\Ly+\dofset) -- (-\dofset,-\dofset);

\end{scope}
\end{tikzpicture}
\begin{tikzpicture}
\tikzmath{
	int \Lx,\Ly;
	\Lx=3;\Ly=1;
	\scale=1;
	\dofset=0.1;
	\cofset=0;
}

\begin{scope}[scale=\scale]

\clip(-\cofset,-\cofset) rectangle (2*\Lx + \cofset, 4*\Ly + \cofset);

\tikzmath{
		for \i in {0,...,\Lx}{
			{			
				\draw  (\i*2,0) pic[scale=\scale] {toda_casi_hor_twist};
			};
		};
};

\end{scope}

\begin{scope}[scale=\scale]

\tikzmath{
	for \i in {0,...,\Lx-1}{
		\xc = int(round(\i+1));
		{	
			\node at (0.7+\i*2, 1) {$_\xc$};	
			\node at (0.7+\i*2, 3) {$_\xc$};	
		};
  	};
};

\node at (0.3, 2.8) {$\gamma$};
\node at (0.3, 0.8) {$\delta$};

\draw[dotted, thick] (-\dofset, -\dofset) -- (2*\Lx+\dofset,-\dofset) -- (2*\Lx+\dofset,4*\Ly+\dofset) -- (-\dofset,4*\Ly+\dofset) -- (-\dofset,-\dofset);

\end{scope}
\end{tikzpicture}
}
\caption{Left: Bipartite graph for the Toda chain. Center, right: zig-zag paths $\alpha, \beta,\gamma,\delta$.}
\label{fig:toda1}
\end{center}
\end{figure}
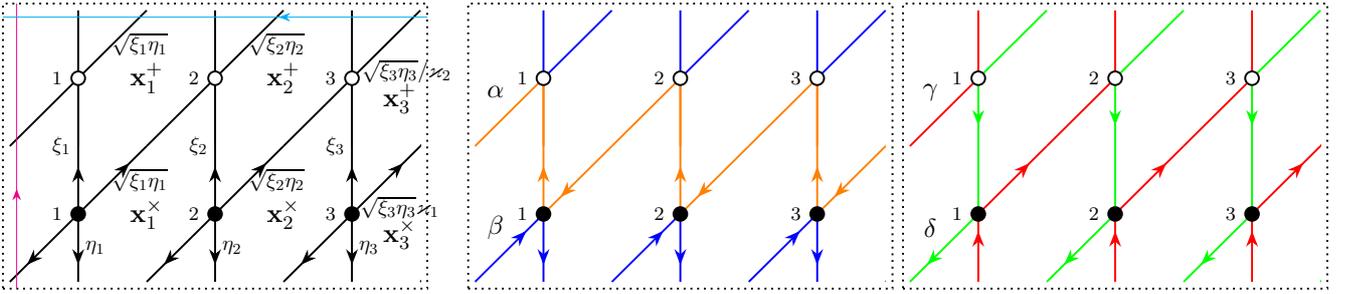

Let us start with the case of relativistic Toda chain, which is known to be related to Seiberg-Witten theory in $5d$ without matter \cite{N:1996}. Relativistic Toda chains arise naturally on Lie groups \cite{FM:1997}, and therefore have cluster description. A typical bipartite graph of affine relativistic Toda is shown in Fig.~\ref{fig:toda1}. For the Toda system with $N$ particles it has $2N$ vertices, $4N$ edges and $2N$ faces. Corresponding Newton polygon is shown in Fig.~\ref{fig:polyexamples}, left.

The cluster Poisson bracket \rf{PB} for the Toda face variables is
\be
\{\ya_i, \ya_j\}=\{\yb_i,\yb_j\}=0,\;\;\; \{\ya_i, \yb_j\} = (\delta_{i,j+1}+\delta_{i+1,j}-2\delta_{i,j}) \ya_i \yb_j,\ \ \ i,j\in \mathbb{Z}/N\mathbb{Z}
\ee
where in the non-vanishing r.h.s. one can immediately recognize the Cartan matrix of $\widehat{sl_N}$. This Poisson
bracket has obviously two Casimir functions, which can be chosen, say, as\footnote{Only the ratio of $\varkappa$'s is actually independent Casimir, but we introduce both of them for convenience in what follows.}
\be
q = \prod_j (\ya_j\yb_j),\ \ \ \ \varkappa_1/\varkappa_2 = \prod_j \yb_j.
\ee
However, in what follows we are going to use the edge variables (see Appendix~\ref{ss:cluster} for details), which do not have any canonical Poisson bracket,
e.g. since they are not gauge invariant, when treated as elements of $\mathbb{C}^\times$-valued gauge connection on the graph. Hence, following \cite{M:2012}, we fix the gauge and parameterize all edges by $2N$ exponentiated Darboux variables $\xi_k, \eta_k$
\begin{equation}
\label{xiet}
\{\xi_i,\eta_j\}=\delta_{ij}\xi_i \eta_j,~~~ \{\xi_i,\varkappa_a\}=\{\eta_i,\varkappa_a\}=0,
\end{equation}
so that the face variables are expressed, as a products of oriented edge variables (see Fig.~\ref{fig:toda1}, left) by
\begin{equation}
\ya_i = \dfrac{\xi_{i+1}}{\xi_i} (\varkappa_2/\varkappa_1)^{\delta_{iN}},\;\;
\yb_i = \dfrac{\eta_i}{\eta_{i+1}} (\varkappa_1/\varkappa_2 )^{\delta_{iN}}.
\end{equation}
In terms of the edge variables \rf{xiet} the monodromies over zig-zag paths (see Fig.~\ref{fig:toda1}, middle, right) can be expressed as follows
\begin{equation}
\alpha = \zeta/\varkappa_1,\;\;\;
\beta = \varkappa_2/\zeta,\;\;\;
\gamma = \varkappa_1\zeta,\;\;\;
\delta = 1/\varkappa_2\zeta,~~~ \zeta=\prod\limits_{k=1}^{N}\sqrt{\dfrac{\xi_k}{\eta_k}}
\end{equation}
In the autonomous limit $q=1$, there is a single independent Casimir -- diagonal twist of monodromy operator $\varkappa_1/\varkappa_2$ or coupling of the affine Toda chain. Reduction from four zig-zags $\alpha,\beta,\gamma,\delta$ to single Casimir $\varkappa_1/\varkappa_2$ is a reminiscence of the freedom $\lambda \to a \lambda,~ \mu \to b \mu$ and the fact that $\alpha \beta \gamma \delta=1$.

The Dirac-Kasteleyn operator here can be read of the left picture at Fig.~\ref{fig:toda1}, and is given by $N\times N$ matrix\footnote{The spectral parameters or quasimomenta $\lambda$ and $\mu$ appear due to intersection of the edge with the blue and purple cycles in $\H1(\T2,\mathbb{Z})$, and minuses arise due to discrete spin structure.}:
\be
\label{DKTo}
\Kast(\lambda,\mu) = \sum_{i=1}^N \left( (\xi_i+\mu^{-1}\eta_i) E_{ii}- \varkappa_1^{\delta_{iN}}\sqrt{\xi_i \eta_i} E_{i,i+1}+\varkappa_2^{-\delta_{iN}}\mu^{-1}\sqrt{\xi_i \eta_i} E_{i+1,i}\right)
\ee
where we have additionally defined
\begin{equation}
E_{N,N+1} = \lambda E_{N,1},\;\;\; E_{N+1,N} = \lambda^{-1} E_{1,N}
\end{equation}
and it almost coincides here \cite{EFS} with the standard $N\times N$ formalism for the spectral curve of relativistic Toda chain
\begin{equation}
\label{spectralToda}
\det\,\Kast(\lambda,\mu)=0\; \Leftrightarrow\; \exists\, \; \Kast(\lambda,\mu)\psi=0
\end{equation}
with Baker-Akhiezer function $\psi \in \mathbb{C}^{N}$.

\begin{figure}[h!]
\begin{center}
\begin{tikzpicture}
\tikzmath{
	int \Lx,\Ly;
	\Lx=3;\Ly=1;
	\scale=1;
	\dofset=0.1;
	\cofset=0;
}

\begin{scope}[scale=\scale]

\clip(-\cofset,-\cofset) rectangle (4*\Lx + \cofset, 4*\Ly + \cofset);

\tikzmath{
		for \i in {0,...,\Lx}{
			{			
				\draw  (\i*4,0) pic[scale=\scale] {toda_two_text};
			};
		};
};

\end{scope}

\begin{scope}[scale=\scale]

\tikzmath{
	for \i in {0,...,\Lx-1}{
		\xc = int(round(\i+1));
		{	
			\node at (0.6+\i*4, 1) {$_{\xc,2}$};	
			\node at (0.6+\i*4, 3) {$_{\xc,1}$};
			\node at (3.4+\i*4, 1) {$_{\xc,1}$};	
			\node at (3.4+\i*4, 3) {$_{\xc,2}$};
			\node at (1.8+\i*4, 1.2) {$1$};
			\node[styleTextEdges, anchor=west] at (\i*4+2.2, 2) {$\xi_{\xc}$};		
			\node[styleTextEdges, anchor=west] at (\i*4+2.7, 0.5) {$\eta_{\xc}$};
		};
  	};
	for \i in {0,...,\Lx-2}{
		\xc = int(round(\i+1));
		{	
  			\node at (4+\i*4, 1) {$\yabold_{\xc}$};
			\node at (4+\i*4, 3) {$\ybbold_{\xc}$};
			\node at (0.8+4+\i*4, 0.5) {$1$};
			\node[styleTextEdges] at (\i*4+3.3, 2) {$\sqrt{\xi_{\xc}\eta_{\xc}}$};		
			\node[styleTextEdges] at (\i*4+2.2, 3.3) {$\sqrt{\xi_{\xc}\eta_{\xc}}$};
		};
  	};
  	\xc = int(round(\Lx));
	{  	
  	\node at (\Lx*4-0.3, 0.7) {$\yabold_{\xc}$};
	\node at (\Lx*4-0.3, 2.7) {$\ybbold_{\xc}$};
	\node[styleTextEdges] at (0.8, 2.1) {$\varkappa_1\sqrt{\xi_{\xc}\eta_{\xc}}$};		
	\node[styleTextEdges] at (0.7, 0.5) {$\varkappa_2$};
	\node[styleTextEdges] at (\Lx*4-4+2.2, 3.3) {$\sqrt{\xi_{\xc}\eta_{\xc}}$}; 	
	};
};

\draw[dotted, thick] (-\dofset, -\dofset) -- (4*\Lx+\dofset,-\dofset) -- (4*\Lx+\dofset,4*\Ly+\dofset) -- (-\dofset,4*\Ly+\dofset) -- (-\dofset,-\dofset);

\draw[thin, color = \colB, opacity=1, styleArrowShifted] (0.1-\cofset,-\dofset) -- (0.1+\cofset,4*\Ly+\dofset);
\draw[thin, color = \colA, opacity=1, styleArrowShifted] (\dofset+4*\Lx, 4*\Ly-0.1-\cofset) -- (-\dofset,4*\Ly-0.1-\cofset);

\end{scope}
\end{tikzpicture}
\caption{Extended and deformed bipartite graph for the Toda chain.}
\label{fig:toda2}
\end{center}
\end{figure}
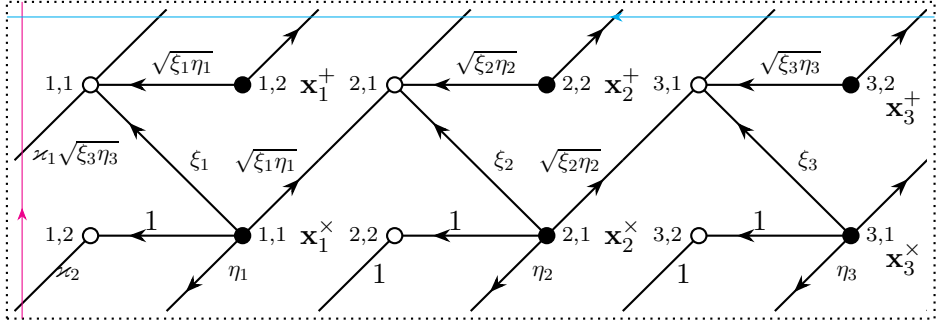

Now, to illustrate what is going to be done for the spin chains, let us rewrite this equation in terms of the well-known $2\times 2$ formalism for Toda chains, but not quite in a standard way. In order to do that, we first add an additional black (white) vertex to each top (bottom) edge in left Fig.~\ref{fig:toda1}, and draw it in deformed way as in Fig.~\ref{fig:toda2}. Such operation obviously does not change the set of dimer configurations, and new dimer partition function differs from the old one only by total nonvanishing factor.

The Dirac-Kasteleyn matrix, read from the Fig.~\ref{fig:toda2}, can be written in the block form
\be
\label{DKTe}
\Kast(\lambda,\mu) = \sum_{i=1}^N\left( E_{ii}\otimes A_i + E_{i,i+1}\otimes C_i Q^{\delta_{i,N}}\right)=
\\
=\sum_{i=1}^N \left( (\xi_i+\mu^{-1}\eta_i) E_{ii}\otimes E_{11}+E_{ii}\otimes E_{12}+\sqrt{\xi_i\eta_i} E_{ii}\otimes E_{21} - \right.
\\
\left.-\varkappa_1^{\delta_{i,N}}\sqrt{\xi_i \eta_i} E_{i,i+1}\otimes E_{11}-\mu\varkappa_2^{\delta_{i,N}} E_{i,i+1}\otimes E_{22}\right)
\ee
with
\begin{equation}
A_i =
\left(
\begin{array}{cc}
\xi_i+\mu^{-1}\eta_i & 1\\
\sqrt{\xi_i\eta_i} & 0
\end{array}
\right),\;\;\;
C_i =
\left(
\begin{array}{cc}
-\sqrt{\xi_i\eta_i} & 0\\
0 & -\mu
\end{array}
\right),\;\;\;
Q =
\left(
\begin{array}{cc}
\varkappa_1 & 0\\
0 & \varkappa_2
\end{array}
\right).
\end{equation}
The first factor in the tensor product corresponds to the number of the particle (or of the 'site'), arising naturally in the framework of $2\times 2$ formalism for Toda systems and spin chains below, while the second -- to position of a vertex inside the 'site'. For the 'extended' (compare to \rf{DKTo}) operator \rf{DKTe} one gets the same equation
(\ref{spectralToda}), but now with $\psi \in \mathbb{C}^{2N}$, which can be written as:
\begin{equation}
\psi=\sum_{i=1}^N e_i\otimes \left(\begin{array}{c} \psi_{i,1}\\ \psi_{i,2} \end{array}\right) = \sum_{i=1}^N e_i\otimes \psi_i .
\end{equation}
For the coefficients of this expansion (\ref{spectralToda}) gives
\begin{equation}
\label{spectralToda2}
\left\{
\begin{array}{l}
\psi_{k+1}= L_k(\mu) \psi_k \\
\psi_{N+1}=\lambda Q \psi_1
\end{array}
\right.
\end{equation}
or the system of finite-difference equations on Baker-Akhiezer functions with the quasi-periodic boundary conditions, where the $2\times 2$ Lax matrix 
\begin{equation}
L_i(\mu) = -C_i^{-1}(\mu)A_i(\mu) =
\mu^{-\frac{1}{2}}\left(
\begin{array}{cc}
\mu^{\frac{1}{2}} \sqrt{\dfrac{\xi_i}{\eta_i}}+\mu^{-\frac{1}{2}}\sqrt{\dfrac{\eta_i}{\xi_i}} & \dfrac{\mu^{\frac{1}{2}}}{\sqrt{\eta_i \xi_i}}\\
\mu^{-\frac{1}{2}}\sqrt{\xi_i\eta_i} & 0
\end{array}
\right)
\end{equation}
is equivalent to the standard Lax matrix for relativistic Toda chain (see e.g. \cite{M:2012}) up to conjugation by permutation matrix, and redefinition of the variables
\begin{equation}
\xi \mapsto \eta,~~~ \eta \mapsto \xi^{-1}, ~~~ \mu \mapsto \mu^{-1}.
\end{equation}
This Lax operator satisfies classical $\RLL$ relation
\begin{equation}
\label{rLL}
\{L_i(\lambda)\otimes L_j(\mu)\} = \delta_{ij}[r(\lambda/\mu),L_i(\lambda)\otimes L_j(\mu)]
\end{equation}
with the classical (trigonometric) $r$-matrix (\ref{eq:rmat1})~\footnote{Up to numeric rescaling, see Appendix \ref{ss:limit} for discussion.}. Compatibility condition of (\ref{spectralToda2}) gives spectral curve equation in the form
\begin{equation}
\det(\lambda Q - L_N(\mu)...L_1(\mu))=0
\end{equation}
where $Q=\diag(\varkappa_1, \varkappa_2)$ is extra twist matrix\footnote{Note that constant diagonal matrices $Q$ satisfy $[r,Q \otimes Q]=0$, and therefore can be also used in construction of monodromy operators.}, and inhomogeneities $\{ \mu_i\}$, which appear in the case of generic $\XXZ$ chain, are absorbed here into redefinition of dynamical variables.

\subsection{Spin chains of $\XXZ$ type}
\label{ss:XXZ}

Let us now apply the same arguments, which we used for the Toda chain, to the following class of chains: the rank $M$ chains on $N$ cites of $\XXZ$-type, which means that the Poisson structure \rf{rLL} is defined by trigonometric $r$-matrix. Such systems naturally arise in $q\to 1$ limit of $\UqM$, see Appendix~\ref{ss:limit}. We claim that such classical spin chain can be alternatively described as cluster integrable systems, constructed from 'big rectangles' of the size  $N\times M$.

\begin{figure}[!h]

\begin{center}

\begin{tikzpicture}

\tikzmath{\Lx=3;\Ly=2;\xs=-10;\ys=1;\d=2;}

\draw[dotted,step=2,shift={(0,-1)}] (\xs,\ys+1) grid (\xs+\d*\Lx, \ys+\d*\Ly+1);

\begin{scope}[thick]
	\draw[styleArrowShort, green] (\xs,\ys+\d)--(\xs,\ys);
	\draw[styleArrowShort, green] (\xs,\ys+2*\d)--(\xs,\ys+\d);	
	
	\draw[styleArrowShort, orange] (\xs+\d,\ys+\d*\Ly)--(\xs,\ys+\d*\Ly);
	\draw[styleArrowShort, orange] (\xs+2*\d,\ys+\d*\Ly)--(\xs+\d,\ys+\d*\Ly);	
	\draw[styleArrowShort, orange] (\xs+3*\d,\ys+\d*\Ly)--(\xs+2*\d,\ys+\d*\Ly);	

	\draw[styleArrowShort, red] (\xs+\d*\Lx,\ys)--(\xs+\d*\Lx,\ys+\d);
	\draw[styleArrowShort, red] (\xs+\d*\Lx,\ys+\d)--(\xs+\d*\Lx,\ys+\d*2);
	
	\draw[styleArrowShort, blue] (\xs,\ys)--(\xs+\d,\ys);
	\draw[styleArrowShort, blue] (\xs+\d,\ys)--(\xs+2*\d,\ys);
	\draw[styleArrowShort, blue] (\xs+2*\d,\ys)--(\xs+3*\d,\ys);

\end{scope}

\foreach \x in {0,...,\Lx}
	\foreach \y in {0,...,\Ly}
		\draw[fill] (\xs+\d*\x,\ys+\d*\y) circle[radius=0.05];

\node at (\xs-0.7,\ys) {(0,0)};
\node at (\xs+\d*\Lx+0.7,\ys) {(N,0)};
\node at (\xs-0.7, \ys+\d*\Ly) {(0,M)};
\node at (\xs+\d*\Lx+0.7, \ys+\d*\Ly) {(N,M)};


\tikzmath{
	int \N,\M;
	\N=3;\M=2;
	\scale=1;
	\dofset=0;
	\cofset=0.8;
	\inshift=0;
}

\begin{scope}[scale=\scale]

\clip(-\cofset+\inshift,-\cofset-\inshift) rectangle (2*\M + \cofset+\inshift, 2*\N + \cofset-\inshift);

\tikzmath{
		for \i in {-1,...,\N+1}{
			for \j in {-1,...,\M+1}{
				{			
					\draw (2*\i,2*\j) pic[scale=\scale] {quiver_cross_skew};
				};
			};
		};
};

\end{scope}

\tikzmath{
		for \i in {1,...,\N}{
			for \j in {1,...,\M}{
				\xc = int(round(\j));
				\yc = int(round(\i));
				{			
					\node at (2*\j-1.5,2*\i-1) {$\ybbold_{\yc \xc}$};
					\node at (2*\j-0.5,2*\i-2) {$\yabold_{\yc \xc}$};
				};
			};
		};
};

\draw[thick, dashed] (-\dofset+\inshift, -1)--(-\dofset+\inshift, 1+2*\N);
\draw[thick, dashed] (2*\M+\dofset+\inshift, -1)--(2*\M+\dofset+\inshift, 1+2*\N);

\draw[thick, dashed] (-1, -\dofset-\inshift)--(2*\M+\dofset+\inshift+1, -\dofset-\inshift);
\draw[thick, dashed] (-1, 2*\N+\dofset+\inshift)--(2*\M+\dofset+\inshift+1, 2*\N+\dofset+\inshift);


\end{tikzpicture}

\end{center}
\caption{Left: Newton polygon for $(N,M)=(3,2)$. Zig-zags from Fig.~\ref{fig:bipNM} as elements of torus first homology group are drawn by colored arrows. Right: Poisson quiver. It is drawn on the torus, so vertices lying on left-right and up-down sides have to be identified.}
\label{fig:quivNM}
\end{figure}
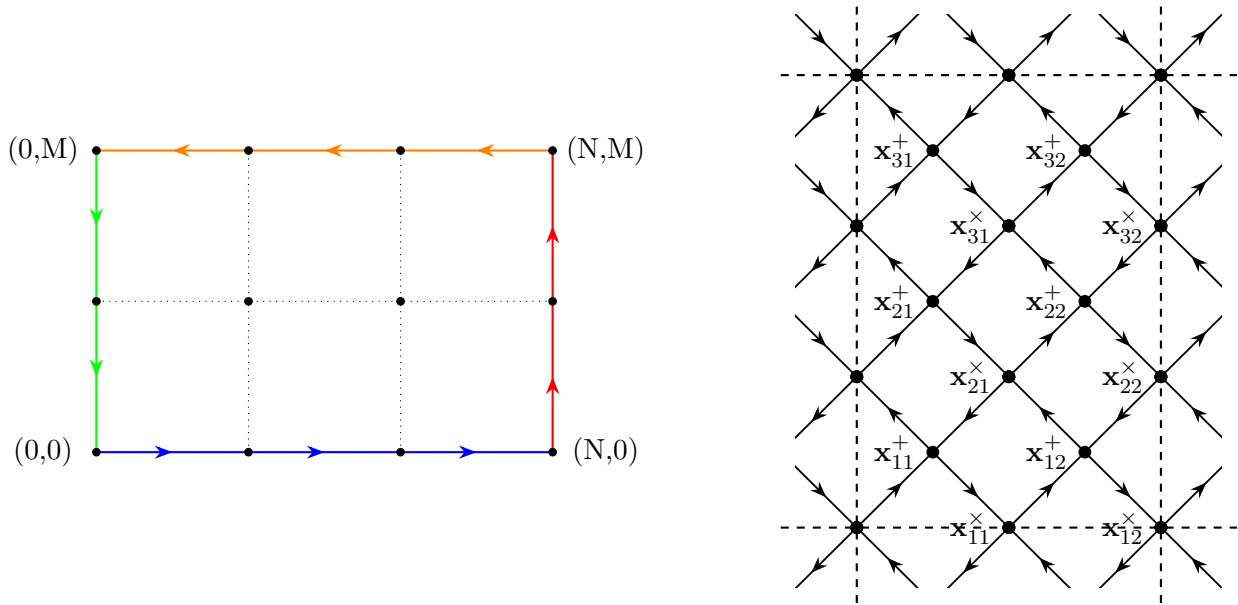

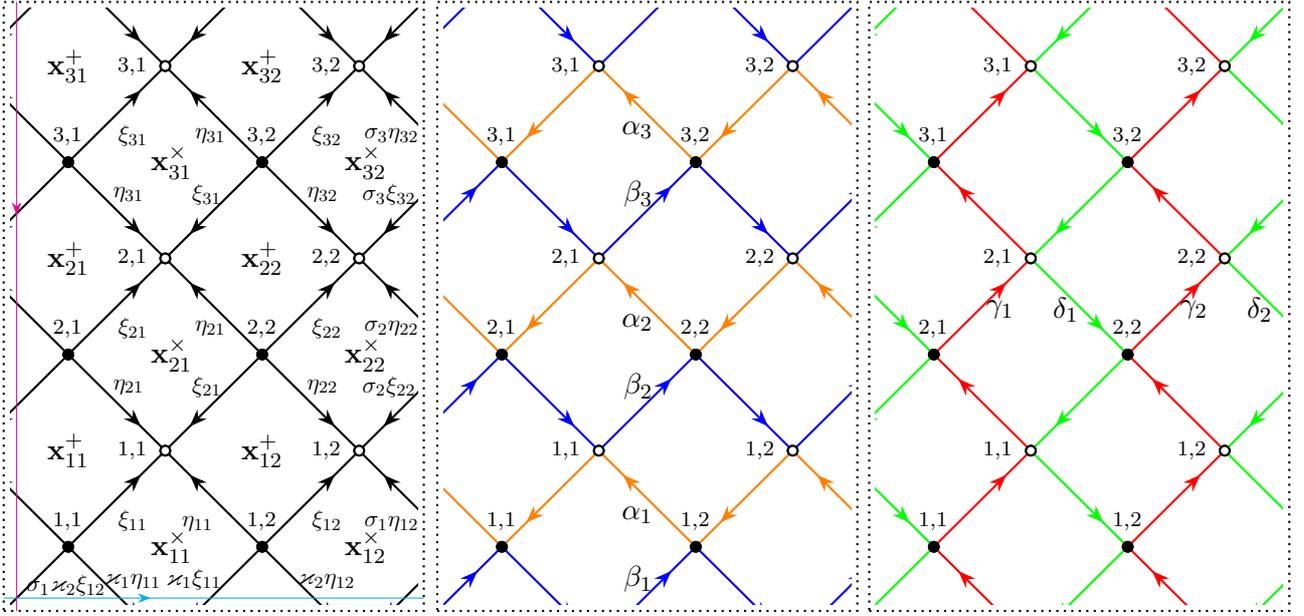
\begin{figure}[h!]
\begin{center}

\begin{tikzpicture}
\tikzmath{
	int \Lx,\Ly;
	\Lx=2;\Ly=3;
	\scale=0.85;
	\dofset=0.5;
	\cofset=0.6;
}

\begin{scope}[scale=\scale]

\clip(1.5 + \cofset,3 + \cofset) rectangle (3*\Lx+3 - \cofset, 3*\Ly+4.5 - \cofset);

\tikzmath{
	for \z in {1,...,\Ly+1}{
		for \i in {0,...,\Lx}{
			{			
				\draw  (\i*3,\z*3) pic[scale=\scale] {bipart_text};
			};
		};
  	};
};

\end{scope}

\begin{scope}[scale=\scale]

\tikzmath{
	for \z in {1,...,\Ly}{
		for \i in {1,...,\Lx}{
			\xc = int(round(\i));
    		\yc = int(round(\z));
			{	
				\node at (\i*3, \z*3+1.9) {$_{\yc,\xc}$};	
				\node at (\i*3+1.5-0.5, \z*3+3) {$_{\yc,\xc}$};			
				\node at (\i*3+1.6, \z*3+1.5) {$\yabold_{\yc\xc}$};
				\node at (\i*3, \z*3+3) {$\ybbold_{\yc\xc}$ };
				\node[styleTextEdges, anchor=north] at (\i*3+1, \z*3+2.2) {$\xi_{\yc\xc}$ };
			};
			if (((\i-\Lx)*(\z-1))!=0) then{
				{
					\node[styleTextEdges, anchor=north] at (\i*3+2.2, \z*3+2.15) {$\eta_{\yc\xc}$ };
					\node[styleTextEdges, anchor=east] at (\i*3+2.5, \z*3+1) {$\xi_{\yc\xc}$ };
					\node[styleTextEdges, anchor=east] at (\i*3+1.3, \z*3+1) {$\eta_{\yc\xc}$ };
				};
			};
  		};
	};
	for \z in {2,...,\Ly}{
		\xc = int(round(\Lx));
    	\yc = int(round(\z));
		{
			\node[styleTextEdges, anchor=east] at (\Lx*3+1.3, \z*3+1) {$\eta_{\yc\xc}$ };
			\node[styleTextEdges, anchor=north] at (3*\Lx+2, \z*3+2.15) {$\sigma_{\yc}\eta_{\yc\xc}$ };
			\node[styleTextEdges, anchor=east] at (3*\Lx+2.5, \z*3+1) {$\sigma_{\yc}\xi_{\yc\xc}$ };
		};
	};
	for \i in {1,...,\Lx-1}{
		\xc = int(round(\i));
    	\yc = int(round(1));
		{
			\node[styleTextEdges, anchor=north] at (\i*3+1, 3+1.2) {$\varkappa_{\xc}\eta_{\yc\xc}$ };
			\node[styleTextEdges, anchor=north] at (\i*3+2, 3+2.1) {$\eta_{\yc\xc}$ };
			\node[styleTextEdges, anchor=east] at (\i*3+2.5, 3+1) {$\varkappa_{\xc}\xi_{\yc\xc}$ };
		};
	};
	{
		\node[styleTextEdges, anchor=north] at (\Lx*3+1, 3+1.2) {$\varkappa_{\Lx}\eta_{1\Lx}$ };
		\node[styleTextEdges, anchor=north] at (\Lx*3+2, 3+2.1) {$\sigma_{1}\eta_{1\Lx}$ };
		\node[styleTextEdges, anchor=east] at (3.7, 3+0.9) {$\sigma_{1}\varkappa_{\Lx}\xi_{1\Lx}$ };
	};
}

\draw[dotted, thick] (1.5+\dofset,3+\dofset) -- (3*\Lx+3-\dofset,3+\dofset) -- (3*\Lx+3-\dofset,3*\Ly+4.5-\dofset) -- (1.5+\dofset,3*\Ly+4.5-\dofset) -- (1.5+\dofset,3+\dofset);

\draw[thin, color = \colB, opacity=1, styleArrowShifted] (1.7+\dofset,3*\Ly+4.5-\dofset) -- (1.7+\dofset,3+\dofset);
\draw[thin, color = \colA, opacity=1, styleArrowShifted] (1.5+\dofset,0.2+3+\dofset) -- (3*\Lx+3-\dofset,0.2+3+\dofset);

\end{scope}
\end{tikzpicture}
\begin{tikzpicture}
\tikzmath{
	int \Lx,\Ly;
	\Lx=2;\Ly=3;
	\scale=0.85;
	\dofset=0.5;
	\cofset=0.6;
}

\begin{scope}[scale=\scale]

\clip(1.5 + \cofset,3 + \cofset) rectangle (3*\Lx+3 - \cofset, 3*\Ly+4.5 - \cofset);

\tikzmath{
	for \z in {1,...,\Ly+1}{
		for \i in {0,...,\Lx}{
			{			
				\draw  (\i*3,\z*3) pic[scale=\scale] {bipart_casi_hor};
			};
		};
  	};
};

\end{scope}

\begin{scope}[scale=\scale]

\draw[dotted, thick] (1.5+\dofset,3+\dofset) -- (3*\Lx+3-\dofset,3+\dofset) -- (3*\Lx+3-\dofset,3*\Ly+4.5-\dofset) -- (1.5+\dofset,3*\Ly+4.5-\dofset) -- (1.5+\dofset,3+\dofset);

\tikzmath{
	for \z in {1,...,\Ly}{
		for \i in {1,...,\Lx}{
			\xc = int(round(\i));
    		\yc = int(round(\z));
			{	
				\node at (\i*3, \z*3+1.9) {$_{\yc,\xc}$};	
				\node at (\i*3+1.5-0.5, \z*3+3) {$_{\yc,\xc}$};			
			};
  		};
	};
	for \z in {1,...,\Ly}{
    	\yc = int(round(\z));
		{
			\node[anchor=east] at (3*\Lx/2+2.5, \z*3+1) {$\beta_{\yc}$ };
			\node[anchor=east] at (3*\Lx/2+2.5, \z*3+2) {$\alpha_{\yc}$ };
		};
	};
}

\end{scope}

\end{tikzpicture}
\begin{tikzpicture}
\tikzmath{
	int \Lx,\Ly;
	\Lx=2;\Ly=3;
	\scale=0.85;
	\dofset=0.5;
	\cofset=0.6;
}

\begin{scope}[scale=\scale]

\clip(1.5 + \cofset,3 + \cofset) rectangle (3*\Lx+3 - \cofset, 3*\Ly+4.5 - \cofset);

\tikzmath{
	for \z in {1,...,\Ly+1}{
		for \i in {0,...,\Lx}{
			{			
				\draw  (\i*3,\z*3) pic[scale=\scale] {bipart_casi_vert};
			};
		};
  	};
};

\end{scope}

\begin{scope}[scale=\scale]

\draw[dotted, thick] (1.5+\dofset,3+\dofset) -- (3*\Lx+3-\dofset,3+\dofset) -- (3*\Lx+3-\dofset,3*\Ly+4.5-\dofset) -- (1.5+\dofset,3*\Ly+4.5-\dofset) -- (1.5+\dofset,3+\dofset);

\tikzmath{
	for \z in {1,...,\Ly}{
		for \i in {1,...,\Lx}{
			\xc = int(round(\i));
    		\yc = int(round(\z));
			{	
				\node at (\i*3, \z*3+1.9) {$_{\yc,\xc}$};	
				\node at (\i*3+1.5-0.5, \z*3+3) {$_{\yc,\xc}$};			
			};
  		};
	};
	for \i in {1,...,\Lx}{
		\xc = int(round(\i));
    	\yc = int(round(\Ly));
		{
			\node[anchor=east] at (\i*3+1.4, 3+0.7+3*\Ly/2) {$\gamma_{\xc}$ };
			\node[anchor=east] at (\i*3+2.4, 3+0.7+3*\Ly/2) {$\delta_{\xc}$ };			
		};
	};
}

\end{scope}

\end{tikzpicture}

\end{center}
\caption{Left: bipartite graphs with labeled edges and faces: each edge, crossing purple cycle has to be multiplied by $\mu$, each edge, crossing blue cycle -- by $\lambda$. Center: horizontal zig-zag paths. Right: vertical zig-zag paths.}
\label{fig:bipNM}
\end{figure}

For a cluster integrable system with such Newton polygon (see Fig.~\ref{fig:quivNM}, left) one gets a bipartite graph, drawn at Fig.~\ref{fig:bipNM}. According to \cite{GK:2011} this graph is drawn on torus $\T2$, i.e. left side is glued with the right side, and top - with the bottom, we will call such graphs as $N\times M$ 'fence nets'.

The cluster coordinates $\ya_{ia}, \yb_{ia}$, now associated with the faces of graph at Fig.~\ref{fig:bipNM}, satisfy the following Poisson bracket relations
\be
\label{PBfn}
\{\ya_{ia},\yb_{jb}\}=(-\delta_{ij}\delta_{ab}+\delta_{i,j+1}\delta_{ab}+\delta_{ij}\delta_{a+1,b}-
\delta_{i,j+1}\delta_{a+1,b})\ya_{ia}\yb_{jb},
\\
 \{\ya_{ia},\ya_{jb}\}=\{\yb_{ia},\yb_{jb}\}=0,\ \ \ \ i,j \in \mathbb{Z}/N\mathbb{Z},\ \ \ \ a,b\in \mathbb{Z}/M\mathbb{Z}
\ee
with two kinds of indices living 'on circles': $i,j$ enumerating rows of bipartite graph and $a,b$ enumerating columns. Corresponding quiver is drawn at Fig.~\ref{fig:quivNM}, right. As in Toda case, 'fixing' a gauge, we pass now to the edge variables
\begin{equation}
\label{cledge}
\ya_{ia}=\dfrac{\eta_{ia}^2}{\xi_{ia}^2}, \;\;\;
\yb_{ia}=\dfrac{\xi_{ia}\xi_{i+1,a-1}}{\eta_{i+1,a} \eta_{i,a-1}} (\sigma_{i+1}/\sigma_{i})^{\delta_{a,1}}(\varkappa_{a-1}/\varkappa_{a})^{\delta_{i,N}}.
\end{equation}
with the
Poisson bracket
\begin{equation}
\{\xi_{ia},\eta_{jb}\} = \frac{1}{2}\delta_{ij}\delta_{ab} \xi_{ia}\eta_{jb},\ \ \ \
i,j \in \mathbb{Z}/N\mathbb{Z},\ \ \ \ a,b\in \mathbb{Z}/M\mathbb{Z}
\end{equation}	
Extra parameters in \rf{cledge} are the Casimir functions of the bracket \rf{PBfn}, together with
\be
\zeta^h_i=\prod\limits_{b=1}^{M}\dfrac{\xi_{ib}}{\eta_{ib}},~~~ \zeta^v_a=\prod\limits_{j=1}^{N}\dfrac{\xi_{ja}}{\eta_{ja}}, ~~~\{\ya,\zeta^{h,v}\}=\{\yb,\zeta^{h,v}\}=0.
\ee
It is useful to re-express them via the zig-zag variables (see the zig-zag paths on Fig.~\ref{fig:bipNM}, middle and right)
\be
\alpha_i = \sigma_i/\zeta^h_i,\;\;\;
\beta_i = 1/\zeta^h_i\sigma_i,\;\;\; i=1,\ldots,N
\\
\gamma_a = \zeta^v_a/\varkappa_a,\;\;\;
\delta_a = \zeta^v_a \varkappa_a,\ \ \ a=1,\ldots,M
\ee
These formulas relate convenient generators of the center of cluster Poisson algebra with inhomogeneities
$\{\mu_k = 1/\sigma_k \zeta^h_k = \beta_k\}$, twists $\{\kappa_a\}$, 'on-site' Casimirs $\zeta_i^{h}=(\alpha_i \beta_i)^{\frac{1}{2}}$ and 'projections of spins'\footnote{Notice that spin's projections are not originally the Casimir functions for spin's brackets, but rather 'trivial' integrals of motion -- like the total momentum of particles in Toda chains.} $\zeta_a^{v}=(\gamma_a \delta_a)^{\frac{1}{2}}$ of the chain.

Our main statement here is that the classical spin variables (for definition see Appendix A) associated with single site of the chain could also be expressed via the edge variables $\xi,\eta$ by
\be
\label{Szt}
e^{S_a^0}= z_a^2,~~~ S_{ab} = \frac{1}{2} z_b^{-2} (z_a^2+z_a^{-2})\dfrac{\tau_a}{\tau_b},~ a<b, ~~~ S_{ab} = -\frac{1}{2} z_a^{2} (z_a^2+z_a^{-2})\dfrac{\tau_a}{\tau_b}, ~  a>b,
\ee
where\footnote{This is basically standard bosonization formulas for the spin variables, cf. for example
with \cite{BS},\cite{MMRZZ}.}
\be
\label{convvars}
z_{a} = \sqrt{\xi_{a}/\eta_{a}},~~~ \tau_{a}=\sqrt{\xi_{a} \eta_{a}}\prod\limits_{b=1}^{M}z_{b}^{\mathrm{sgn}(b-a)}
\ee
and the 'site index' $i=1,\ldots,N$ is omitted here. Spin-variables cannot be directly expressed through the cluster variables in a natural way, but rather as a product of edge variables over some non-closed paths. However it is possible to express cluster variables via the spin variables on two adjacent sites by
\begin{equation}
\ya_{i,a} = e^{-2(S_{a}^0)_i}, \ \ \
\yb_{i,a} = -(S_{a-1,a})_{i+1}(S_{a,a-1})_{i}\dfrac{e^{(S_a^0)_{i+1}+(S_{a-1}^0)_{i}}}{\cosh\, (S^0_{a-1})_{i+1}\cosh\, (S^0_{a})_{i}}\left(\dfrac{\sigma_{i+1}}{\sigma_{i}}\right)^{\delta_{a,1}}\left(\dfrac{\varkappa_{a-1}}{\varkappa_{a}}\right)^{\delta_{i,N}}
\end{equation}
where index outside brackets of spin variables enumerates number of site. This relation will be discussed in details elsewhere \cite{SZ}.

The spectral curve again can be given by determinant of the Dirac-Kasteleyn operator, which is the weighted adjacency matrix of the bipartite graph. For generic $(N,M)$ system it has the form:
\begin{equation}
\label{DKNM}
\begin{array}{c}
\Kast(\lambda,\mu)=\sum\limits_{i=1}^{N}\sum\limits_{a=1}^{M} \xi_{ia} (E_{i,i}\otimes E_{a,a} - \varkappa_a ^{\delta_{i,1}}\sigma_i^{\delta_{M,a}} E_{i,i-1}\otimes E_{a+1,a})+ \\
+\eta_{ia} (\varkappa_a^{\delta_{1,i}} E_{i,i-1}\otimes E_{a,a} + \sigma_i^{\delta_{M,a}}  E_{i,i}\otimes E_{a+1,a})
\end{array}
\end{equation}
where the summand $E_{ij}\otimes E_{ab}$ is corresponding to the edge between black and white vertices\footnote{Signs '$-$' in $\Kast$ arise in a standard way \cite{GK:2011} due to choice of Kasteleyn marking or discrete spin structure on $\T2$.} $(i,a)\to (j,b)$, and those matrices $E_{ij}$ which get out of fundamental domain are promoted to the elements of the 'loop algebra', with the 'loop' parameters $(\lambda,\mu)$:
\be
\label{tEdef}
E_{1,0}\equiv \lambda E_{1,N},\ \ \ \ \ \E_{M+1,M}\equiv \mu E_{1,M}.
\ee
\Remark{ The operator \rf{DKNM} as an element of $\mathrm{End}(\mathbb{C}^N) \lbr\lbr  \lambda^{-1} \rbr\rbr \otimes \mathrm{End}(\mathbb{C}^M) \lbr\lbr  \mu^{-1} \rbr\rbr$ can be naturally embedded into tensor product of evaluation representations of the loop algebras $\widetilde{\mathfrak{gl}}_N\otimes \widetilde{\mathfrak{gl}}_M$, i.e.
\begin{equation}
\begin{array}{c}
\Kast(\lambda,\mu)=\sum\limits_{i=1}^{N}\sum\limits_{a=1}^{M} \xi_{ia} (h_i\otimes h_a - \varkappa_a ^{\delta_{i,1}}\sigma_i^{\delta_{M,a}} f_{i-1}\otimes f_{a}) + \eta_{ia} (\varkappa_a^{\delta_{1,i}} f_{i-1}\otimes h_{a} + \sigma_i^{\delta_{M,a}}  h_{i} \otimes f_{a})
\end{array}
\end{equation}
for two evaluation representations $\widetilde{\mathfrak{gl}}_K \to \mathrm{End}(\mathbb{C}^K) \lbr\lbr  \zeta \rbr\rbr$:
\begin{equation}
\begin{array}{lll}
e_i = E_{i,i+1}, &  1\leq i \leq K - 1, &
e_0 = e_K = \zeta E_{K,1}\\
f_i = E_{i+1,i}, & 1\leq i \leq K - 1, &
f_0 = f_K = \zeta^{-1} E_{1,K}\\
h_i = E_{ii}, & 1\leq i \leq K.
\end{array}
\end{equation}
}

Let us now, breaking $M\leftrightarrow N$ symmetry, collect the terms, corresponding to $E_{ii}$ and $E_{i,i-1}$ in the first tensor factor, i.e. rewrite \rf{DKNM} as:
\begin{equation}
\label{Kblock}
\Kast(\lambda,\mu)=\sum\limits_{i=1}^{N} E_{i,i}\otimes A_i + E_{i,i-1}\otimes C_i (Q)^{\delta_{1,i}}
\end{equation}
with
\begin{equation}
A_i = \sum\limits_{b=1}^{M}\left( \xi_{ib} E_{b,b}+ \eta_{ib} \sigma_i^{\delta_{M,b}} E_{b+1,b}\right),\;\;\;  C_i = \sum\limits_{b=1}^{M} \left(\eta_{ib} E_{b,b}-\xi_{ib} \sigma_i^{\delta_{M,b}} E_{b+1,b}\right),\;\;\; Q = \sum\limits_{b=1}^{M} \varkappa_b E_{bb}
\end{equation}
From the spectral curve equation $\det \Kast(\lambda,\mu)=0$ one finds for
\begin{equation}
\label{spectpsi}
\psi=\sum\limits_{i=1}^{N}\psi_ie_i=\sum\limits_{i=1}^{N}\sum\limits_{a=1}^{M} \psi_{ia}e_i\otimes e_a\in \mathbb{C}^{MN}:\; \Kast(\lambda,\mu)\psi=0.
\end{equation}
that
\begin{equation}
A_{i} \psi_{i} + C_i (Q)^{\delta_{i,1}}\psi_{i-1}=0,\;\;\; i=1,\ldots,N, ~~~ \psi_0 \equiv \lambda\psi_N.
\end{equation}
Solving these equations recursively for the vectors $\psi_{i} = \sum\limits_{a=1}^{M} \psi_{ia} e_a$, one
finally gets
\begin{equation}
\label{spectpsi1}
\left(\lambda Q - (-1)^{N}C^{-1}_1 A_1...C^{-1}_N A_N\right)\psi_N=0
\end{equation}
with consistency condition
\begin{equation}
\label{eq:spectralxxz}
\det\left( \lambda \,Q - L_1\left(\sigma_1 \zeta^h_1 \mu\right)...L_N\left(\sigma_N \zeta^h_N \mu\right) \right)=0
\end{equation}
of the form \rf{fTcurve}, with the Lax matrices
\begin{equation}
\label{LCA}
L_k\left(\sigma_k \zeta^h_k \mu\right)=-C^{-1}_k A_k,\ \ \ \ k=1,\ldots,N.
\end{equation}
Hence, the spectral curve $\det \Kast(\lambda,\mu)=0$ is represented in the form \rf{eq:spinchaincurve}, common for the classical integrable chains with inhomogeneities $\mu_k = 1/\sigma_k \zeta^h_k = \beta_k$ and twist $Q=\sum_a \varkappa_a E_{aa} =\sum_a \sqrt{\delta_a/\gamma_a} E_{aa}$. There are also two sets of Casimirs related to spin variables: total projections of spin $\zeta^{v}_a = \prod_i e^{S_{ia}^0}$ and single non-trivial on-site Casimirs $\zeta^h_i$. The Lax operators \rf{LCA} on different sites satisfy classical $\RLL$-relations
\begin{equation}
\label{RTTClusterMain}
\{L_{n}(\mu)\otimes L_{n'}(\mu')\} = \dfrac{1}{2}\delta_{nn'}[r(\mu/\mu'),L_{n}(\mu)\otimes L_{n'}(\mu')]
\end{equation}
which coincide with (\ref{r2}) arising from the classical limit of $\UqM$ with $q=e^{-\hbar}$ and $\k=\frac{1}{2}$ in (\ref{eq:commprescr}), see Appendix \ref{ss:RLL} for details. In such way one gets explicit formulas (with the sign-factors \rf{sifa})
\begin{equation}
\label{Lclust}
(L_n)_{ab}(\mu) =  \dfrac{1}{\mu^{\frac{1}{2}}-\mu^{-\frac{1}{2}}}
\left\{
\begin{array}{ll}
a=b, & \mu^{\frac{1}{2}}z_{na}^{-2}+\mu^{-\frac{1}{2}}z_{na}^2 \\
a\neq b, & \mu^{-\frac{s_{ab}}{2}} (z_{nb}^2+z_{nb}^{-2})\dfrac{\tau_{nb}}{\tau_{na}}
\end{array}
\right. ,~~~
\end{equation}
for the Lax operators \rf{LCA} on the sites $n\in 1,...,N$ in terms of variables introduced in \rf{convvars}.

Comparing $L$-operator (\ref{Lclust}) with (\ref{Lspinminus}) one comes to the formulas \rf{Szt}, expressing the 'spin operators' on each site in terms of the edge variables.
Expressions \rf{Szt} satisfy all the relations of the classical limit of $\UqM$ with $\k=\frac{1}{2}$. Note that this Lax operator is belonging to the lowest rank Kirillov orbit. 

\Remark{\label{rem:group}An equivalent construction of the cluster integrable systems is based on the Poisson submanifolds or double Bruhat cells in $\GLb$, endowed with the usual $r$-matrix Poisson structure \cite{FG:2005, FM:2014}. For the family of systems we consider here, given by the $SA(2,\mathbb{Z})$-orbit of rectangular $N\times M$ Newton polygons, one gets in such way a double Bruhat cell of $\GLb(N+M)$, given by the word
\begin{equation}
\label{uNM}
u=(s_M \overline{s}_{M}\, ...\, s_{1}\overline{s}_{1}\Lambda)^N
\end{equation}
in the co-extended double Weyl group $\Wext(A_{K}^{(1)}\times A_{K}^{(1)})$ (here with $K=N+M$) with the generators $s_i,s_{\overline{i}},\Lambda$ satisfying relations
\begin{equation}
\label{eq:doubleWeylGroup}
\begin{array}{l}
s_i^2=1, \ \ \ (s_i s_{i+1})^3=1, \ \ \ s_i s_{j} = s_{j} s_{i},\ \ \mathrm{for~~}|i-j|>1 \\
\bar{s}_i^2=1, \ \ \ (\bar{s}_i \bar{s}_{i+1})^3=1, \ \ \ \bar{s}_i \bar{s}_{j} = \bar{s}_{j} \bar{s}_{i},\ \ \mathrm{for~~}|i-j|>1 \\
\Lambda^{K}=1, \ \ \ \Lambda s_{i+1} = s_{i} \Lambda, \ \ \ \Lambda \bar{s}_{i+1} = \bar{s}_{i} \Lambda
\end{array}
~~~ i,j=1,\ldots,K
\end{equation}

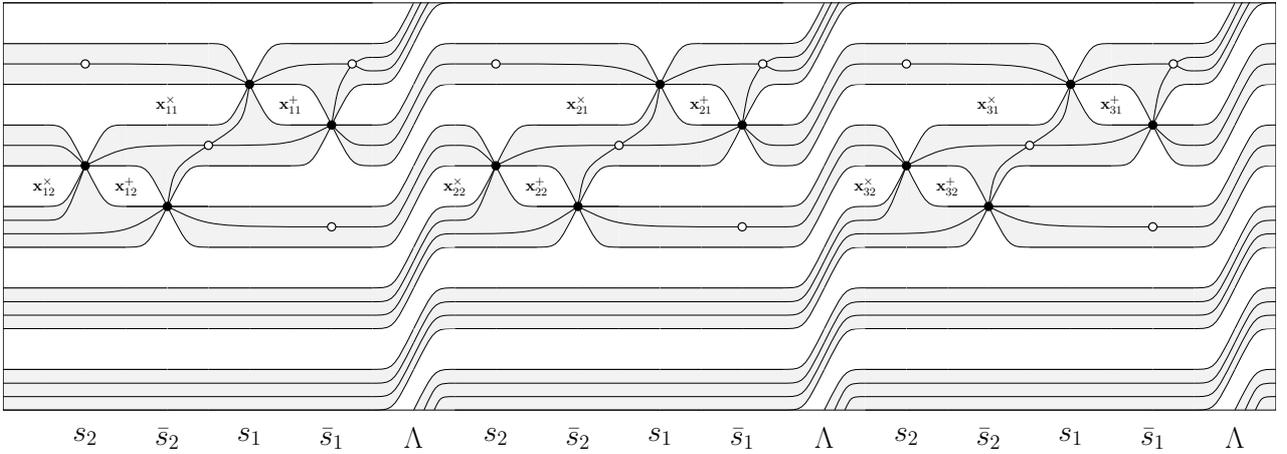
\begin{figure}[!h]
\begin{center}

\scalebox{0.54}{
\begin{tikzpicture}

\clip(-1,-1) rectangle (30,10);

\draw  (-1,0) pic {Wt_triv_gr};
\draw  (-1,1) pic {Wt_triv_wh};
\draw  (-1,2) pic {Wt_triv_gr};
\draw  (-1,3) pic {Wt_triv_wh};
\draw  (-1,4) pic {Wt_triv_gr};
\draw  (-1,5) pic {Wt_triv_wh};
\draw  (-1,6) pic {Wt_triv_gr};
\draw  (-1,7) pic {Wt_triv_wh};
\draw  (-1,8) pic {Wt_triv_gr};
\draw  (-1,9) pic {Wt_triv_wh};
\tikzmath{\x=0;}
\node[anchor=north, font = \huge] at (\x+1, -0.3) {$s_{2}$};
\draw  (\x,0) pic {Wt_triv_gr};
\draw  (\x,1) pic {Wt_triv_wh};
\draw  (\x,2) pic {Wt_triv_gr};
\draw  (\x,3) pic {Wt_triv_wh};
\draw  (\x,7) pic {Wt_triv_wh_no_bot};
\draw  (\x,4) pic {Wt_triv_gr_no_top};
\draw  (\x,5) pic {Wt_lower_left_wh};
\draw  (\x,6) pic {Wt_upper_left_gr};
\draw  (\x,8) pic {Wt_triv_gr};
\draw  (\x,9) pic {Wt_triv_wh};
\tikzmath{\x=1;}
\draw  (\x,0) pic {Wt_triv_gr};
\draw  (\x,1) pic {Wt_triv_wh};
\draw  (\x,2) pic {Wt_triv_gr};
\draw  (\x,3) pic {Wt_triv_wh};
\draw  (\x,7) pic {Wt_triv_wh_no_bot};
\draw  (\x,4) pic {Wt_triv_gr_no_top};
\draw  (\x,5) pic {Wt_lower_right_wh};
\draw  (\x,6) pic {Wt_upper_right_gr};
\draw  (\x,8) pic {Wt_triv_gr};
\draw  (\x,9) pic {Wt_triv_wh};
\tikzmath{\x=2;}
\node[anchor=north, font = \huge] at (\x+1, -0.3) {$\bar{s}_{2}$};
\draw  (\x,0) pic {Wt_triv_gr};
\draw  (\x,1) pic {Wt_triv_wh};
\draw  (\x,2) pic {Wt_triv_gr};
\draw  (\x,6) pic {Wt_triv_gr_no_bot};
\draw  (\x,3) pic {Wt_triv_wh_no_top};
\draw  (\x,4) pic {Wt_lower_left_gr};
\draw  (\x,5) pic {Wt_upper_left_wh};	
\draw  (\x,7) pic {Wt_triv_wh};
\draw  (\x,8) pic {Wt_triv_gr};
\draw  (\x,9) pic {Wt_triv_wh};
\tikzmath{\x=3;}
\draw  (\x,0) pic {Wt_triv_gr};
\draw  (\x,1) pic {Wt_triv_wh};
\draw  (\x,2) pic {Wt_triv_gr};
\draw  (\x,7) pic {Wt_triv_wh};
\draw  (\x,6) pic {Wt_triv_gr_no_bot};
\draw  (\x,3) pic {Wt_triv_wh_no_top};
\draw  (\x,4) pic {Wt_lower_right_gr};
\draw  (\x,5) pic {Wt_upper_right_wh};
\draw  (\x,8) pic {Wt_triv_gr};
\draw  (\x,9) pic {Wt_triv_wh};
\tikzmath{\x=4;}
\node[anchor=north, font = \huge] at (\x+1, -0.3) {$s_{1}$};
\draw  (\x,0) pic {Wt_triv_gr};
\draw  (\x,1) pic {Wt_triv_wh};
\draw  (\x,2) pic {Wt_triv_gr};
\draw  (\x,3) pic {Wt_triv_wh};
\draw  (\x,4) pic {Wt_triv_gr};
\draw  (\x,5) pic {Wt_triv_wh};
\draw  (\x,9) pic {Wt_triv_wh_no_bot};
\draw  (\x,6) pic {Wt_triv_gr_no_top};
\draw  (\x,7) pic {Wt_lower_left_wh};
\draw  (\x,8) pic {Wt_upper_left_gr};
\tikzmath{\x=5;}
\draw  (\x,0) pic {Wt_triv_gr};
\draw  (\x,1) pic {Wt_triv_wh};
\draw  (\x,2) pic {Wt_triv_gr};
\draw  (\x,3) pic {Wt_triv_wh};
\draw  (\x,4) pic {Wt_triv_gr};
\draw  (\x,5) pic {Wt_triv_wh};
\draw  (\x,9) pic {Wt_triv_wh_no_bot};
\draw  (\x,6) pic {Wt_triv_gr_no_top};
\draw  (\x,7) pic {Wt_lower_right_wh};
\draw  (\x,8) pic {Wt_upper_right_gr};
\tikzmath{\x=6;}
\node[anchor=north, font = \huge] at (\x+1, -0.3) {$\bar{s}_{1}$};
\draw  (\x,0) pic {Wt_triv_gr};
\draw  (\x,1) pic {Wt_triv_wh};
\draw  (\x,2) pic {Wt_triv_gr};
\draw  (\x,3) pic {Wt_triv_wh};
\draw  (\x,4) pic {Wt_triv_gr};
\draw  (\x,8) pic {Wt_triv_gr_no_bot};
\draw  (\x,5) pic {Wt_triv_wh_no_top};
\draw  (\x,6) pic {Wt_lower_left_gr};
\draw  (\x,7) pic {Wt_upper_left_wh};	
\draw  (\x,9) pic {Wt_triv_wh};
\tikzmath{\x=7;}
\draw  (\x,0) pic {Wt_triv_gr};
\draw  (\x,1) pic {Wt_triv_wh};
\draw  (\x,2) pic {Wt_triv_gr};
\draw  (\x,3) pic {Wt_triv_wh};
\draw  (\x,4) pic {Wt_triv_gr};
\draw  (\x,8) pic {Wt_triv_gr_no_bot};
\draw  (\x,5) pic {Wt_triv_wh_no_top};
\draw  (\x,6) pic {Wt_lower_right_gr};
\draw  (\x,7) pic {Wt_upper_right_wh};
\draw  (\x,9) pic {Wt_triv_wh};
\tikzmath{\x=8;}
\node[anchor=north, font = \huge] at (\x+1, -0.3) {$\Lambda$};
\draw  (\x,0) pic {Wt_LambdaL_gr};
\draw  (\x,1) pic {Wt_LambdaL_wh};
\draw  (\x,2) pic {Wt_LambdaL_gr};
\draw  (\x,3) pic {Wt_LambdaL_wh};
\draw  (\x,4) pic {Wt_LambdaL_gr};
\draw  (\x,5) pic {Wt_LambdaL_wh};
\draw  (\x,6) pic {Wt_LambdaL_gr};
\draw  (\x,7) pic {Wt_LambdaL_wh};
\draw  (\x,8) pic {Wt_LambdaL_gr};
\draw  (\x,9) pic {Wt_LambdaL_wh};
\tikzmath{\x=9;}
\draw  (\x,0) pic {Wt_LambdaU_wh};
\draw  (\x,1) pic {Wt_LambdaU_gr};
\draw  (\x,2) pic {Wt_LambdaU_wh};
\draw  (\x,3) pic {Wt_LambdaU_gr};
\draw  (\x,4) pic {Wt_LambdaU_wh};
\draw  (\x,5) pic {Wt_LambdaU_gr};
\draw  (\x,6) pic {Wt_LambdaU_wh};
\draw  (\x,7) pic {Wt_LambdaU_gr};
\draw  (\x,8) pic {Wt_LambdaU_wh};
\draw  (\x,9) pic {Wt_LambdaU_gr};
\tikzmath{\x=10;}
\node[anchor=north, font = \huge] at (\x+1, -0.3) {$s_{2}$};
\draw  (\x,0) pic {Wt_triv_gr};
\draw  (\x,1) pic {Wt_triv_wh};
\draw  (\x,2) pic {Wt_triv_gr};
\draw  (\x,3) pic {Wt_triv_wh};
\draw  (\x,7) pic {Wt_triv_wh_no_bot};
\draw  (\x,4) pic {Wt_triv_gr_no_top};
\draw  (\x,5) pic {Wt_lower_left_wh};
\draw  (\x,6) pic {Wt_upper_left_gr};
\draw  (\x,8) pic {Wt_triv_gr};
\draw  (\x,9) pic {Wt_triv_wh};
\tikzmath{\x=11;}
\draw  (\x,0) pic {Wt_triv_gr};
\draw  (\x,1) pic {Wt_triv_wh};
\draw  (\x,2) pic {Wt_triv_gr};
\draw  (\x,3) pic {Wt_triv_wh};
\draw  (\x,7) pic {Wt_triv_wh_no_bot};
\draw  (\x,4) pic {Wt_triv_gr_no_top};
\draw  (\x,5) pic {Wt_lower_right_wh};
\draw  (\x,6) pic {Wt_upper_right_gr};
\draw  (\x,8) pic {Wt_triv_gr};
\draw  (\x,9) pic {Wt_triv_wh};
\tikzmath{\x=12;}
\node[anchor=north, font = \huge] at (\x+1, -0.3) {$\bar{s}_{2}$};
\draw  (\x,0) pic {Wt_triv_gr};
\draw  (\x,1) pic {Wt_triv_wh};
\draw  (\x,2) pic {Wt_triv_gr};
\draw  (\x,6) pic {Wt_triv_gr_no_bot};
\draw  (\x,3) pic {Wt_triv_wh_no_top};
\draw  (\x,4) pic {Wt_lower_left_gr};
\draw  (\x,5) pic {Wt_upper_left_wh};	
\draw  (\x,7) pic {Wt_triv_wh};
\draw  (\x,8) pic {Wt_triv_gr};
\draw  (\x,9) pic {Wt_triv_wh};
\tikzmath{\x=13;}
\draw  (\x,0) pic {Wt_triv_gr};
\draw  (\x,1) pic {Wt_triv_wh};
\draw  (\x,2) pic {Wt_triv_gr};
\draw  (\x,7) pic {Wt_triv_wh};
\draw  (\x,6) pic {Wt_triv_gr_no_bot};
\draw  (\x,3) pic {Wt_triv_wh_no_top};
\draw  (\x,4) pic {Wt_lower_right_gr};
\draw  (\x,5) pic {Wt_upper_right_wh};
\draw  (\x,8) pic {Wt_triv_gr};
\draw  (\x,9) pic {Wt_triv_wh};
\tikzmath{\x=14;}
\node[anchor=north, font = \huge] at (\x+1, -0.3) {$s_{1}$};
\draw  (\x,0) pic {Wt_triv_gr};
\draw  (\x,1) pic {Wt_triv_wh};
\draw  (\x,2) pic {Wt_triv_gr};
\draw  (\x,3) pic {Wt_triv_wh};
\draw  (\x,4) pic {Wt_triv_gr};
\draw  (\x,5) pic {Wt_triv_wh};
\draw  (\x,9) pic {Wt_triv_wh_no_bot};
\draw  (\x,6) pic {Wt_triv_gr_no_top};
\draw  (\x,7) pic {Wt_lower_left_wh};
\draw  (\x,8) pic {Wt_upper_left_gr};
\tikzmath{\x=15;}
\draw  (\x,0) pic {Wt_triv_gr};
\draw  (\x,1) pic {Wt_triv_wh};
\draw  (\x,2) pic {Wt_triv_gr};
\draw  (\x,3) pic {Wt_triv_wh};
\draw  (\x,4) pic {Wt_triv_gr};
\draw  (\x,5) pic {Wt_triv_wh};
\draw  (\x,9) pic {Wt_triv_wh_no_bot};
\draw  (\x,6) pic {Wt_triv_gr_no_top};
\draw  (\x,7) pic {Wt_lower_right_wh};
\draw  (\x,8) pic {Wt_upper_right_gr};
\tikzmath{\x=16;}
\node[anchor=north, font = \huge] at (\x+1, -0.3) {$\bar{s}_{1}$};
\draw  (\x,0) pic {Wt_triv_gr};
\draw  (\x,1) pic {Wt_triv_wh};
\draw  (\x,2) pic {Wt_triv_gr};
\draw  (\x,3) pic {Wt_triv_wh};
\draw  (\x,4) pic {Wt_triv_gr};
\draw  (\x,8) pic {Wt_triv_gr_no_bot};
\draw  (\x,5) pic {Wt_triv_wh_no_top};
\draw  (\x,6) pic {Wt_lower_left_gr};
\draw  (\x,7) pic {Wt_upper_left_wh};	
\draw  (\x,9) pic {Wt_triv_wh};
\tikzmath{\x=17;}
\draw  (\x,0) pic {Wt_triv_gr};
\draw  (\x,1) pic {Wt_triv_wh};
\draw  (\x,2) pic {Wt_triv_gr};
\draw  (\x,3) pic {Wt_triv_wh};
\draw  (\x,4) pic {Wt_triv_gr};
\draw  (\x,8) pic {Wt_triv_gr_no_bot};
\draw  (\x,5) pic {Wt_triv_wh_no_top};
\draw  (\x,6) pic {Wt_lower_right_gr};
\draw  (\x,7) pic {Wt_upper_right_wh};
\draw  (\x,9) pic {Wt_triv_wh};
\tikzmath{\x=18;}
\node[anchor=north, font = \huge] at (\x+1, -0.3) {$\Lambda$};
\draw  (\x,0) pic {Wt_LambdaL_gr};
\draw  (\x,1) pic {Wt_LambdaL_wh};
\draw  (\x,2) pic {Wt_LambdaL_gr};
\draw  (\x,3) pic {Wt_LambdaL_wh};
\draw  (\x,4) pic {Wt_LambdaL_gr};
\draw  (\x,5) pic {Wt_LambdaL_wh};
\draw  (\x,6) pic {Wt_LambdaL_gr};
\draw  (\x,7) pic {Wt_LambdaL_wh};
\draw  (\x,8) pic {Wt_LambdaL_gr};
\draw  (\x,9) pic {Wt_LambdaL_wh};
\tikzmath{\x=19;}
\draw  (\x,0) pic {Wt_LambdaU_wh};
\draw  (\x,1) pic {Wt_LambdaU_gr};
\draw  (\x,2) pic {Wt_LambdaU_wh};
\draw  (\x,3) pic {Wt_LambdaU_gr};
\draw  (\x,4) pic {Wt_LambdaU_wh};
\draw  (\x,5) pic {Wt_LambdaU_gr};
\draw  (\x,6) pic {Wt_LambdaU_wh};
\draw  (\x,7) pic {Wt_LambdaU_gr};
\draw  (\x,8) pic {Wt_LambdaU_wh};
\draw  (\x,9) pic {Wt_LambdaU_gr};
\tikzmath{\x=20;}
\node[anchor=north, font = \huge] at (\x+1, -0.3) {$s_{2}$};
\draw  (\x,0) pic {Wt_triv_gr};
\draw  (\x,1) pic {Wt_triv_wh};
\draw  (\x,2) pic {Wt_triv_gr};
\draw  (\x,3) pic {Wt_triv_wh};
\draw  (\x,7) pic {Wt_triv_wh_no_bot};
\draw  (\x,4) pic {Wt_triv_gr_no_top};
\draw  (\x,5) pic {Wt_lower_left_wh};
\draw  (\x,6) pic {Wt_upper_left_gr};
\draw  (\x,8) pic {Wt_triv_gr};
\draw  (\x,9) pic {Wt_triv_wh};
\tikzmath{\x=21;}
\draw  (\x,0) pic {Wt_triv_gr};
\draw  (\x,1) pic {Wt_triv_wh};
\draw  (\x,2) pic {Wt_triv_gr};
\draw  (\x,3) pic {Wt_triv_wh};
\draw  (\x,7) pic {Wt_triv_wh_no_bot};
\draw  (\x,4) pic {Wt_triv_gr_no_top};
\draw  (\x,5) pic {Wt_lower_right_wh};
\draw  (\x,6) pic {Wt_upper_right_gr};
\draw  (\x,8) pic {Wt_triv_gr};
\draw  (\x,9) pic {Wt_triv_wh};
\tikzmath{\x=22;}
\node[anchor=north, font = \huge] at (\x+1, -0.3) {$\bar{s}_{2}$};
\draw  (\x,0) pic {Wt_triv_gr};
\draw  (\x,1) pic {Wt_triv_wh};
\draw  (\x,2) pic {Wt_triv_gr};
\draw  (\x,6) pic {Wt_triv_gr_no_bot};
\draw  (\x,3) pic {Wt_triv_wh_no_top};
\draw  (\x,4) pic {Wt_lower_left_gr};
\draw  (\x,5) pic {Wt_upper_left_wh};	
\draw  (\x,7) pic {Wt_triv_wh};
\draw  (\x,8) pic {Wt_triv_gr};
\draw  (\x,9) pic {Wt_triv_wh};
\tikzmath{\x=23;}
\draw  (\x,0) pic {Wt_triv_gr};
\draw  (\x,1) pic {Wt_triv_wh};
\draw  (\x,2) pic {Wt_triv_gr};
\draw  (\x,7) pic {Wt_triv_wh};
\draw  (\x,6) pic {Wt_triv_gr_no_bot};
\draw  (\x,3) pic {Wt_triv_wh_no_top};
\draw  (\x,4) pic {Wt_lower_right_gr};
\draw  (\x,5) pic {Wt_upper_right_wh};
\draw  (\x,8) pic {Wt_triv_gr};
\draw  (\x,9) pic {Wt_triv_wh};
\tikzmath{\x=24;}
\node[anchor=north, font = \huge] at (\x+1, -0.3) {$s_{1}$};
\draw  (\x,0) pic {Wt_triv_gr};
\draw  (\x,1) pic {Wt_triv_wh};
\draw  (\x,2) pic {Wt_triv_gr};
\draw  (\x,3) pic {Wt_triv_wh};
\draw  (\x,4) pic {Wt_triv_gr};
\draw  (\x,5) pic {Wt_triv_wh};
\draw  (\x,9) pic {Wt_triv_wh_no_bot};
\draw  (\x,6) pic {Wt_triv_gr_no_top};
\draw  (\x,7) pic {Wt_lower_left_wh};
\draw  (\x,8) pic {Wt_upper_left_gr};
\tikzmath{\x=25;}
\draw  (\x,0) pic {Wt_triv_gr};
\draw  (\x,1) pic {Wt_triv_wh};
\draw  (\x,2) pic {Wt_triv_gr};
\draw  (\x,3) pic {Wt_triv_wh};
\draw  (\x,4) pic {Wt_triv_gr};
\draw  (\x,5) pic {Wt_triv_wh};
\draw  (\x,9) pic {Wt_triv_wh_no_bot};
\draw  (\x,6) pic {Wt_triv_gr_no_top};
\draw  (\x,7) pic {Wt_lower_right_wh};
\draw  (\x,8) pic {Wt_upper_right_gr};
\tikzmath{\x=26;}
\node[anchor=north, font = \huge] at (\x+1, -0.3) {$\bar{s}_{1}$};
\draw  (\x,0) pic {Wt_triv_gr};
\draw  (\x,1) pic {Wt_triv_wh};
\draw  (\x,2) pic {Wt_triv_gr};
\draw  (\x,3) pic {Wt_triv_wh};
\draw  (\x,4) pic {Wt_triv_gr};
\draw  (\x,8) pic {Wt_triv_gr_no_bot};
\draw  (\x,5) pic {Wt_triv_wh_no_top};
\draw  (\x,6) pic {Wt_lower_left_gr};
\draw  (\x,7) pic {Wt_upper_left_wh};	
\draw  (\x,9) pic {Wt_triv_wh};
\tikzmath{\x=27;}
\draw  (\x,0) pic {Wt_triv_gr};
\draw  (\x,1) pic {Wt_triv_wh};
\draw  (\x,2) pic {Wt_triv_gr};
\draw  (\x,3) pic {Wt_triv_wh};
\draw  (\x,4) pic {Wt_triv_gr};
\draw  (\x,8) pic {Wt_triv_gr_no_bot};
\draw  (\x,5) pic {Wt_triv_wh_no_top};
\draw  (\x,6) pic {Wt_lower_right_gr};
\draw  (\x,7) pic {Wt_upper_right_wh};
\draw  (\x,9) pic {Wt_triv_wh};
\tikzmath{\x=28;}
\node[anchor=north, font = \huge] at (\x+1, -0.3) {$\Lambda$};
\draw  (\x,0) pic {Wt_LambdaL_gr};
\draw  (\x,1) pic {Wt_LambdaL_wh};
\draw  (\x,2) pic {Wt_LambdaL_gr};
\draw  (\x,3) pic {Wt_LambdaL_wh};
\draw  (\x,4) pic {Wt_LambdaL_gr};
\draw  (\x,5) pic {Wt_LambdaL_wh};
\draw  (\x,6) pic {Wt_LambdaL_gr};
\draw  (\x,7) pic {Wt_LambdaL_wh};
\draw  (\x,8) pic {Wt_LambdaL_gr};
\draw  (\x,9) pic {Wt_LambdaL_wh};
\tikzmath{\x=29;}
\draw  (\x,0) pic {Wt_LambdaU_wh};
\draw  (\x,1) pic {Wt_LambdaU_gr};
\draw  (\x,2) pic {Wt_LambdaU_wh};
\draw  (\x,3) pic {Wt_LambdaU_gr};
\draw  (\x,4) pic {Wt_LambdaU_wh};
\draw  (\x,5) pic {Wt_LambdaU_gr};
\draw  (\x,6) pic {Wt_LambdaU_wh};
\draw  (\x,7) pic {Wt_LambdaU_gr};
\draw  (\x,8) pic {Wt_LambdaU_wh};
\draw  (\x,9) pic {Wt_LambdaU_gr};


\draw (-1,0.33)..controls(0,0.33)..(0,0.33);
\draw (-1,0.66)..controls(0,0.66)..(0,0.66);
\draw (-1,2.33)..controls(0,2.33)..(0,2.33);
\draw (-1,2.66)..controls(0,2.66)..(0,2.66);
\draw (-1,4.33)..controls(0,4.33)..(0,4.33);
\draw (-1,4.66)..controls(0,4.66)..(0,4.66);
\draw (-1,6.5)..controls(0,6.5)..(0,6.5);
\draw (-1,8.5)..controls(0,8.5)..(0,8.5);

\tikzmath{\x=0;}

\draw (\x+1,8.5)..controls(\x+1,8.5)..(\x,8.5);
\draw (\x+1,6)..controls(\x+0.5,6.5)..(\x,6.5);
\draw (\x+1,6)..controls(\x+0.5,4.66)..(\x,4.66);
\draw (\x+3,5)..controls(\x+2,4.33)..(\x,4.33);

\draw (\x+1,6)..controls(\x+2,6.5)..(\x+4,6.5);
\draw (\x+3,5)..controls(\x+3.15,5.95)..(\x+4,6.5);

\draw (\x+7,7)..controls(\x+6,6.5)..(\x+4,6.5);
\draw (\x+5,8)..controls(\x+4.85,7.05)..(\x+4,6.5);

\draw (\x+5,8)..controls(\x+6,8.5)..(\x+7.5,8.5);
\draw (\x+7,7)..controls(\x+7.15,7.95)..(\x+7.5,8.5);

\draw (\x+1,8.5)..controls(\x+4,8.5)..(\x+5,8);
\draw (\x+7,4.5)..controls(\x+4,4.5)..(\x+3,5);

\draw (\x+7.5,8.5)..controls(\x+7.8,8.66)..(\x+8,8.66);
\draw (\x+7.5,8.5)..controls(\x+7.8,8.33)..(\x+8,8.33);
\draw (\x+7,7)..controls(\x+7.5,6.5)..(\x+8,6.5);

\draw (\x+7,4.5)..controls(\x+8,4.5)..(\x+8,4.5);

\draw (\x,2.66)..controls(\x,2.66)..(\x+8,2.66);
\draw (\x,2.33)..controls(\x,2.33)..(\x+8,2.33);

\draw (\x,0.66)..controls(\x,0.66)..(\x+8,0.66);
\draw (\x,0.33)..controls(\x,0.33)..(\x+8,0.33);

\draw (\x+8,8.66)..controls(\x+8.5,8.66)..(\x+9,9.66);
\draw (\x+10,10.66)..controls(\x+9.5,10.66)..(\x+9,9.66);

\draw (\x+8,8.33)..controls(\x+8.5,8.33)..(\x+9,9.33);
\draw (\x+10,10.33)..controls(\x+9.5,10.33)..(\x+9,9.33);

\draw (\x+8,6.5)..controls(\x+8.5,6.5)..(\x+9,7.5);
\draw (\x+10,8.5)..controls(\x+9.5,8.5)..(\x+9,7.5);

\draw (\x+8,4.5)..controls(\x+8.5,4.5)..(\x+9,5.5);
\draw (\x+10,6.5)..controls(\x+9.5,6.5)..(\x+9,5.5);

\draw (\x+8,2.66)..controls(\x+8.5,2.66)..(\x+9,3.66);
\draw (\x+10,4.66)..controls(\x+9.5,4.66)..(\x+9,3.66);

\draw (\x+8,2.33)..controls(\x+8.5,2.33)..(\x+9,3.33);
\draw (\x+10,4.33)..controls(\x+9.5,4.33)..(\x+9,3.33);

\draw (\x+8,0.66)..controls(\x+8.5,0.66)..(\x+9,1.66);
\draw (\x+10,2.66)..controls(\x+9.5,2.66)..(\x+9,1.66);

\draw (\x+8,0.33)..controls(\x+8.5,0.33)..(\x+9,1.33);
\draw (\x+10,2.33)..controls(\x+9.5,2.33)..(\x+9,1.33);

\draw (\x+10,0.66)..controls(\x+9.55,0.66)..(\x+9.25,0);
\draw (\x+10,0.33)..controls(\x+9.65,0.33)..(\x+9.5,0);

\draw[blackCircle] (\x+1,6) circle;
\draw[blackCircle] (\x+3,5) circle;
\draw[whiteCircle] (\x+4,6.5) circle;
\draw[blackCircle] (\x+5,8) circle;
\draw[blackCircle] (\x+7,7) circle;
\draw[whiteCircle] (\x+7.5,8.5) circle;
\draw[whiteCircle] (\x+7,4.5) circle;
\draw[whiteCircle] (\x+1,8.5) circle;


\tikzmath{\x=10;}

\draw (\x+1,8.5)..controls(\x+1,8.5)..(\x,8.5);
\draw (\x+1,6)..controls(\x+0.5,6.5)..(\x,6.5);
\draw (\x+1,6)..controls(\x+0.5,4.66)..(\x,4.66);
\draw (\x+3,5)..controls(\x+2,4.33)..(\x,4.33);

\draw (\x+1,6)..controls(\x+2,6.5)..(\x+4,6.5);
\draw (\x+3,5)..controls(\x+3.15,5.95)..(\x+4,6.5);

\draw (\x+7,7)..controls(\x+6,6.5)..(\x+4,6.5);
\draw (\x+5,8)..controls(\x+4.85,7.05)..(\x+4,6.5);

\draw (\x+5,8)..controls(\x+6,8.5)..(\x+7.5,8.5);
\draw (\x+7,7)..controls(\x+7.15,7.95)..(\x+7.5,8.5);

\draw (\x+1,8.5)..controls(\x+4,8.5)..(\x+5,8);
\draw (\x+7,4.5)..controls(\x+4,4.5)..(\x+3,5);

\draw (\x+7.5,8.5)..controls(\x+7.8,8.66)..(\x+8,8.66);
\draw (\x+7.5,8.5)..controls(\x+7.8,8.33)..(\x+8,8.33);
\draw (\x+7,7)..controls(\x+7.5,6.5)..(\x+8,6.5);

\draw (\x+7,4.5)..controls(\x+8,4.5)..(\x+8,4.5);

\draw (\x,2.66)..controls(\x,2.66)..(\x+8,2.66);
\draw (\x,2.33)..controls(\x,2.33)..(\x+8,2.33);

\draw (\x,0.66)..controls(\x,0.66)..(\x+8,0.66);
\draw (\x,0.33)..controls(\x,0.33)..(\x+8,0.33);

\draw (\x+8,8.66)..controls(\x+8.5,8.66)..(\x+9,9.66);
\draw (\x+10,10.66)..controls(\x+9.5,10.66)..(\x+9,9.66);

\draw (\x+8,8.33)..controls(\x+8.5,8.33)..(\x+9,9.33);
\draw (\x+10,10.33)..controls(\x+9.5,10.33)..(\x+9,9.33);

\draw (\x+8,6.5)..controls(\x+8.5,6.5)..(\x+9,7.5);
\draw (\x+10,8.5)..controls(\x+9.5,8.5)..(\x+9,7.5);

\draw (\x+8,4.5)..controls(\x+8.5,4.5)..(\x+9,5.5);
\draw (\x+10,6.5)..controls(\x+9.5,6.5)..(\x+9,5.5);

\draw (\x+8,2.66)..controls(\x+8.5,2.66)..(\x+9,3.66);
\draw (\x+10,4.66)..controls(\x+9.5,4.66)..(\x+9,3.66);

\draw (\x+8,2.33)..controls(\x+8.5,2.33)..(\x+9,3.33);
\draw (\x+10,4.33)..controls(\x+9.5,4.33)..(\x+9,3.33);

\draw (\x+8,0.66)..controls(\x+8.5,0.66)..(\x+9,1.66);
\draw (\x+10,2.66)..controls(\x+9.5,2.66)..(\x+9,1.66);

\draw (\x+8,0.33)..controls(\x+8.5,0.33)..(\x+9,1.33);
\draw (\x+10,2.33)..controls(\x+9.5,2.33)..(\x+9,1.33);

\draw (\x+10,0.66)..controls(\x+9.55,0.66)..(\x+9.25,0);
\draw (\x+10,0.33)..controls(\x+9.65,0.33)..(\x+9.5,0);

\draw[blackCircle] (\x+1,6) circle;
\draw[blackCircle] (\x+3,5) circle;
\draw[whiteCircle] (\x+4,6.5) circle;
\draw[blackCircle] (\x+5,8) circle;
\draw[blackCircle] (\x+7,7) circle;
\draw[whiteCircle] (\x+7.5,8.5) circle;
\draw[whiteCircle] (\x+7,4.5) circle;
\draw[whiteCircle] (\x+1,8.5) circle;

\tikzmath{\x=20;}

\draw (\x+1,8.5)..controls(\x+1,8.5)..(\x,8.5);
\draw (\x+1,6)..controls(\x+0.5,6.5)..(\x,6.5);
\draw (\x+1,6)..controls(\x+0.5,4.66)..(\x,4.66);
\draw (\x+3,5)..controls(\x+2,4.33)..(\x,4.33);

\draw (\x+1,6)..controls(\x+2,6.5)..(\x+4,6.5);
\draw (\x+3,5)..controls(\x+3.15,5.95)..(\x+4,6.5);

\draw (\x+7,7)..controls(\x+6,6.5)..(\x+4,6.5);
\draw (\x+5,8)..controls(\x+4.85,7.05)..(\x+4,6.5);

\draw (\x+5,8)..controls(\x+6,8.5)..(\x+7.5,8.5);
\draw (\x+7,7)..controls(\x+7.15,7.95)..(\x+7.5,8.5);

\draw (\x+1,8.5)..controls(\x+4,8.5)..(\x+5,8);
\draw (\x+7,4.5)..controls(\x+4,4.5)..(\x+3,5);

\draw (\x+7.5,8.5)..controls(\x+7.8,8.66)..(\x+8,8.66);
\draw (\x+7.5,8.5)..controls(\x+7.8,8.33)..(\x+8,8.33);
\draw (\x+7,7)..controls(\x+7.5,6.5)..(\x+8,6.5);

\draw (\x+7,4.5)..controls(\x+8,4.5)..(\x+8,4.5);

\draw (\x,2.66)..controls(\x,2.66)..(\x+8,2.66);
\draw (\x,2.33)..controls(\x,2.33)..(\x+8,2.33);

\draw (\x,0.66)..controls(\x,0.66)..(\x+8,0.66);
\draw (\x,0.33)..controls(\x,0.33)..(\x+8,0.33);

\draw (\x+8,8.66)..controls(\x+8.5,8.66)..(\x+9,9.66);
\draw (\x+10,10.66)..controls(\x+9.5,10.66)..(\x+9,9.66);

\draw (\x+8,8.33)..controls(\x+8.5,8.33)..(\x+9,9.33);
\draw (\x+10,10.33)..controls(\x+9.5,10.33)..(\x+9,9.33);

\draw (\x+8,6.5)..controls(\x+8.5,6.5)..(\x+9,7.5);
\draw (\x+10,8.5)..controls(\x+9.5,8.5)..(\x+9,7.5);

\draw (\x+8,4.5)..controls(\x+8.5,4.5)..(\x+9,5.5);
\draw (\x+10,6.5)..controls(\x+9.5,6.5)..(\x+9,5.5);

\draw (\x+8,2.66)..controls(\x+8.5,2.66)..(\x+9,3.66);
\draw (\x+10,4.66)..controls(\x+9.5,4.66)..(\x+9,3.66);

\draw (\x+8,2.33)..controls(\x+8.5,2.33)..(\x+9,3.33);
\draw (\x+10,4.33)..controls(\x+9.5,4.33)..(\x+9,3.33);

\draw (\x+8,0.66)..controls(\x+8.5,0.66)..(\x+9,1.66);
\draw (\x+10,2.66)..controls(\x+9.5,2.66)..(\x+9,1.66);

\draw (\x+8,0.33)..controls(\x+8.5,0.33)..(\x+9,1.33);
\draw (\x+10,2.33)..controls(\x+9.5,2.33)..(\x+9,1.33);

\draw (\x+10,0.66)..controls(\x+9.55,0.66)..(\x+9.25,0);
\draw (\x+10,0.33)..controls(\x+9.65,0.33)..(\x+9.5,0);

\draw[blackCircle] (\x+1,6) circle;
\draw[blackCircle] (\x+3,5) circle;
\draw[whiteCircle] (\x+4,6.5) circle;
\draw[blackCircle] (\x+5,8) circle;
\draw[blackCircle] (\x+7,7) circle;
\draw[whiteCircle] (\x+7.5,8.5) circle;
\draw[whiteCircle] (\x+7,4.5) circle;
\draw[whiteCircle] (\x+1,8.5) circle;

\node at (0, 5.5) {$\yabold_{12}$};
\node at (3, 7.5) {$\yabold_{11}$};
\node at (2, 5.5) {$\ybbold_{12}$};
\node at (6, 7.5) {$\ybbold_{11}$};

\node at (10, 5.5) {$\yabold_{22}$};
\node at (13, 7.5) {$\yabold_{21}$};
\node at (12, 5.5) {$\ybbold_{22}$};
\node at (16, 7.5) {$\ybbold_{21}$};

\node at (20, 5.5) {$\yabold_{32}$};
\node at (23, 7.5) {$\yabold_{31}$};
\node at (22, 5.5) {$\ybbold_{32}$};
\node at (26, 7.5) {$\ybbold_{31}$};

\draw[thin] (-1,-0) -- (30,-0) -- (30,10) -- (-1,10) -- (-1,-0);

\end{tikzpicture}
}
\caption{Thurston diagram in the $(3,2)$ case, which appears from $u=(s_2\bar{s}_2 s_1\bar{s}_1\Lambda)^{3}$.}
\label{fig:thurst}
\end{center}
\end{figure}

We are not going to repeat here all steps of the construction in detail, and just present the main ingredient --
the Thurston diagram for \rf{uNM}, drawn for $(N,M)=(3,2)$ at  Fig.~\ref{fig:thurst}. The corresponding bipartite graph (see Fig.~\ref{fig:thurst}) differs from the discussed above 'fence-net' by additional horizontal twist of the cylinder by $2\pi$, which does not affect an integrable system, since it corresponds to the $SL(2,\mathbb{Z})$ transformation of the spectral parameters $(\lambda,\mu)\to (\lambda,\mu\lambda^{-1})$.

\paragraph{Example. SU(2) theory with $N_f=4$}{
The most well-known case of the system we consider here corresponds to the five-dimensional supersymmetric gauge theory with the $SU(2)$ gauge group and $N_f=4$ fundamental multiplets. The corresponding Newton polygon is a square with sides of length $N=M=2$ (see Fig.~\ref{fig:polygon2x2}), and as a spin chain this is just common $\XXZ$-model on two sites with the Lax operator\footnote{This form is slightly different from (\ref{Lax2x2Uq}) arising from the classical limit of $U_q(\mathfrak{gl}_2)$. However, in $2\times 2$ case these two forms are equivalent.}  (see e.g. \cite{MM:1997})
\begin{equation}
\label{Lax2x2}
L(\mu) =
\left(
\begin{array}{cc}
\mu e^{S^z}-\mu^{-1}e^{-S^z} & 2S^{-}\\
2S^{+} & \mu e^{-S^z}-\mu^{-1}e^{S^z}
\end{array}
\right),~~~~~
Q =
\left(
\begin{array}{cc}
\varkappa & 0\\
0 & \varkappa^{-1}
\end{array}
\right).
\end{equation}
Spectral curve for the system is given by
\begin{equation}
\label{eq:spectralxxzsl2}
\mathrm{det}\left(L\left(\mu/\mu_1\right)L\left(\mu/\mu_2\right)Q-\lambda\right)=0.
\end{equation}
The Poisson brackets of spin operators are given by classical trigonometric $r$-matrix and written as:
\begin{equation}
\label{eq:spinsl2}
\{S^z,\;S^{\pm}\}=\pm \,S^{\pm}, \;\; \{S^{+},\;S^{-}\}=\sinh\, 2S^z
\end{equation}
for the $S$-variables on the same site, and zero for the variables on the different sites. Such bracket has one natural Casimir function
\begin{equation}
K=-\zeta^h-(\zeta^h)^{-1}=\frac{1}{2}\cosh 2S^z + S^+ S^-.
\end{equation}

\begin{figure}[!h]
\begin{center}
\begin{tikzpicture}

\tikzmath{\Lx=2;\Ly=2;\xs=0;\ys=0;\d=2;}

\draw[dotted,step=2,shift={(0,0)}] (\xs,\ys) grid (\xs+\d*\Lx, \ys+\d*\Ly);

\begin{scope}[thick]
	\draw (\xs,\ys+\d)--(\xs,\ys);
	\draw (\xs,\ys+2*\d)--(\xs,\ys+\d);	
	
	\draw (\xs+\d,\ys+\d*\Ly)--(\xs,\ys+\d*\Ly);
	\draw (\xs+2*\d,\ys+\d*\Ly)--(\xs+\d,\ys+\d*\Ly);	
	\draw (\xs+2*\d,\ys+\d*\Ly)--(\xs+2*\d,\ys+\d*\Ly);	

	\draw (\xs+\d*\Lx,\ys)--(\xs+\d*\Lx,\ys+\d);
	\draw (\xs+\d*\Lx,\ys+\d)--(\xs+\d*\Lx,\ys+\d*2);
	
	\draw (\xs,\ys)--(\xs+\d,\ys);
	\draw (\xs+\d,\ys)--(\xs+2*\d,\ys);
	\draw (\xs+2*\d,\ys)--(\xs+2*\d,\ys);

\end{scope}

\foreach \x in {0,...,\Lx}
	\foreach \y in {0,...,\Ly}
		\draw[fill] (\xs+\d*\x,\ys+\d*\y) circle[radius=0.05];

\end{tikzpicture}

\end{center}
\caption{Newton polygon for $(N,M)=(2,2)$.}
\label{fig:polygon2x2}
\end{figure}
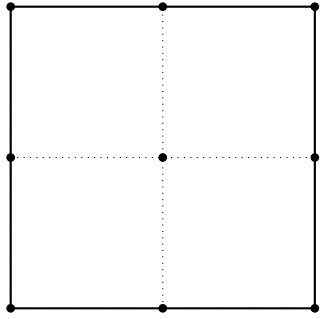

As a cluster integrable system it lives on X-variety with the quiver corresponding to $A_3^{(1)}$-type system from Fig.~2 in \cite{BGM:2017}, and its deautonomization leads to the Painlev\'e VI equation, solvable by conformal blocks, or equivalently topological strings amplitudes \cite{Jimbo:2017}. We derive Lax operator for this system from Kasteleyn operator in details in the next example, which is simply generalization of this example to three sites.
}
\paragraph{Example. SU(3) theory with $N_f=6$.}{
\label{spin:ex2}

This case is corresponding to the word $u=(2\bar{2}1\bar{1}\Lambda)^3$ in double Weyl group of $\GLb(5)$. Bipartite graph is drawn on Fig.~\ref{fig:bipNM}. Kasteleyn operator is $6\times 6$ matrix
\begin{equation}
\label{K322x2}
\Kast=
\begin{array}{l|cccccc}
bw & 11 & 12 & 21 & 22 & 31 & 32 \\
\hline
11 & \xi_{11} & \mu \sigma_1 \eta_{12} & 0 & 0 & \lambda\varkappa_1\eta_{11} & -\lambda\mu\varkappa_2\sigma_1 \xi_{12} \\
12 & \eta_{11} & \xi_{12} & 0 & 0 & -\lambda\varkappa_1 \xi_{11} & \lambda\varkappa_2\eta_{12}\\
21 & \eta_{21} & -\mu\sigma_2 \xi_{22} & \xi_{21} & \mu \sigma_2 \eta_{22} & 0 & 0 \\
22 & -\xi_{21} & \eta_{22} & \eta_{21} & \xi_{22} & 0 & 0 \\
31 & 0 & 0 & \eta_{31} & -\mu\sigma_2 \xi_{32} & \xi_{31} & \mu \sigma_2 \eta_{32}\\
32 & 0 & 0 & -\xi_{31} & \eta_{32} & \eta_{31} & \xi_{32}
\end{array}=
\left(
\begin{array}{ccc}
A_1 & 0 & \lambda C_1 Q\\
C_2 & A_2 & 0\\
0 & C_3 & A_3
\end{array}
\right).
\end{equation}
Spectral curve is given by condition
\begin{equation}
\det \Kast(\lambda,\mu)=0 ~~ \Leftrightarrow ~~ \exists~\psi=
\left(\begin{array}{c}
\psi_1\\
\psi_2\\
\psi_3
\end{array}\right):~~ \Kast(\lambda,\mu)\psi=0 ~~~
~~ \Leftrightarrow ~~
\left\{
\begin{array}{c}
\lambda Q\psi_3 = L_1(\sigma_1 \zeta^h_1\mu) \psi_1 \\
\psi_1=L_2(\sigma_2 \zeta^h_2\mu)\psi_2\\
\psi_2=L_3(\sigma_3 \zeta^h_3\mu)\psi_3
\end{array}
\right.
\end{equation}
\begin{equation}
\label{eq:Lax2x2clusterex}
L_k(\mu)=
\dfrac{1}{\mu^{\frac{1}{2}}-\mu^{-\frac{1}{2}}}
\left(
\def\arraystretch{2.2}
\begin{array}{cc}
\mu^{-\frac{1}{2}}\dfrac{\xi_{k1}}{\eta_{k1}}+\mu^{\frac{1}{2}}\dfrac{\eta_{k1}}{\xi_{k1}} &
\mu^{\frac{1}{2}}\dfrac{\eta_{k2}}{\xi_{k1}}\left(\dfrac{\xi_{k2}}{\eta_{k2}}+\dfrac{\eta_{k2}}{\xi_{k2}}\right) \\
\mu^{-\frac{1}{2}}\dfrac{\xi_{k1}}{\eta_{k2}}\left(\dfrac{\xi_{k1}}{\eta_{k1}}+\dfrac{\eta_{k1}}{\xi_{k1}}\right) &
\mu^{-\frac{1}{2}}\dfrac{\xi_{k2}}{\eta_{k2}}+\mu^{\frac{1}{2}}\dfrac{\eta_{k2}}{\xi_{k2}}
\end{array}
\right)
\end{equation}
$$\zeta_k^h=\dfrac{\xi_{k1}\xi_{k2}}{\eta_{k1}\eta_{k2}}
,~~~ Q =
\left(
\begin{array}{cc}
\varkappa_1 & 0\\
0 & \varkappa_2
\end{array}
\right)$$
which could be rewritten using monodromy operator
\begin{equation}
\label{specT3s2x2}
\begin{array}{c}
\left(\lambda Q - T_{3}^{2 \times 2}(\mu)\right)\psi_3=0 ~~ \Leftrightarrow ~~  \det\left(\lambda Q - T_{3}^{2 \times 2}(\mu)\right)=0,\\
T_{3}^{2 \times 2}(\mu)=L_1(\sigma_1 \zeta^h_1\mu)L_2(\sigma_2 \zeta^h_2\mu)L_3(\sigma_3 \zeta^h_3\mu).
\end{array}
\end{equation}
Lax operator (\ref{eq:Lax2x2clusterex}) is of $\mathfrak{gl}_2$ type, so can be mapped to (\ref{Lax2x2Uq}). To transform it in $\mathfrak{sl}_2$ form (\ref{Lax2x2}) we have to apply transformations like (\ref{eq:2x2transfroms})
\begin{equation}
\mu \mapsto -\mu\,\dfrac{\xi_1 \xi_2}{\eta_1 \eta_2}, ~~\mathrm{then} ~~
L(\mu) \mapsto
\left(\sqrt{\dfrac{\xi_1 \xi_2}{\eta_1 \eta_2}}\mu^{\frac{1}{2}}-\sqrt{\dfrac{\eta_1 \eta_2}{\xi_1 \xi_2}}\mu^{-\frac{1}{2}}\right)
\left(
\begin{array}{cc}
\mu^{-1/2} & 0\\
0 & 1
\end{array}
\right)
\cdot L(\mu) \cdot
\left(
\begin{array}{cc}
\mu^{1/2} & 0\\
0 & 1
\end{array}
\right)
\end{equation}
so it becomes
\begin{equation}
\label{eq:lax2x2exclusttransform}
L(\mu)=
\left(
\def\arraystretch{2.2}
\begin{array}{cc}
\mu^{\frac{1}{2}}\sqrt{\dfrac{\eta_1 \xi_2}{\xi_1 \eta_2}}  - \mu^{-\frac{1}{2}}\sqrt{\dfrac{\xi_1 \eta_2}{\eta_1 \xi_2}} &
\sqrt{\dfrac{\xi_2 \eta_2}{\xi_1\eta_1}}\left(\dfrac{\xi_{2}}{\eta_{2}}+\dfrac{\eta_{2}}{\xi_{2}}\right) \\
-\sqrt{\dfrac{\xi_1 \eta_1}{\xi_2 \eta_2}}\left(\dfrac{\xi_{1}}{\eta_{1}}+\dfrac{\eta_{1}}{\xi_{1}}\right) &
\mu^{\frac{1}{2}}\sqrt{\dfrac{\xi_1 \eta_2}{\eta_1 \xi_2}}-\mu^{-\frac{1}{2}}\sqrt{\dfrac{\eta_1 \xi_2}{\xi_1 \eta_2}}
\end{array}
\right).
\end{equation}
Defining classical $\mathfrak{sl}_2$ spin variables by
\begin{equation}
\begin{array}{lll}
S^{-} = \dfrac{1}{2}\sqrt{\dfrac{\xi_2\eta_2}{\xi_1\eta_1}}\left(\dfrac{\xi_{2}}{\eta_{2}}+\dfrac{\eta_{2}}{\xi_{2}}\right), &
S^{+} = -\dfrac{1}{2}\sqrt{\dfrac{\xi_1\eta_1}{\eta_2 \xi_2}}\left(\dfrac{\xi_{1}}{\eta_{1}}+\dfrac{\eta_{1}}{\xi_{1}}\right), &
e^{S^{z}} = \sqrt{\dfrac{\xi_1\eta_2}{\eta_1 \xi_2}}
\end{array}
\end{equation}
we see that Lax operator (\ref{eq:lax2x2exclusttransform}) coincides with the (\ref{Lax2x2}) up to replacement $\mu^{1/2}\to \mu$ and $S^z \mapsto -S^{z}$. The latter is a consequence of the fact that (\ref{eq:lax2x2exclusttransform}) is coming from $q=e^{-\hbar}$ prescription, but $(\ref{Lax2x2})$ - from the usual $q=e^{\hbar}$. Poisson brackets of spin variables coming from edge variables bracket $\{\xi_i,\eta_j\} = \frac{1}{2}\delta_{ij}\xi_i\eta_j$ are
\begin{equation}
\{S^z,\;S^{\pm}\}=\pm \frac{1}{2}\,S^{\pm}, \;\; \{S^{+},\;S^{-}\}=\frac{1}{2}\sinh\, 2S^z
\end{equation}
which differs from (\ref{eq:spinsl2}) by factor $1/2$, appearing from $\k=\frac{1}{2}$ in the prescription for the classical limit of commutators (\ref{eq:commprescr}). For details see Appendix \ref{ss:limit}.

Spectral curve (\ref{eq:spectralxxzsl2}) could be obtained from (\ref{eq:spectralxxz}) by transformation $\lambda \mapsto \lambda(\varkappa_1\varkappa_2)^{-\frac{1}{2}}$ with identification of parameters $\varkappa=(\varkappa_1/\varkappa_2)^{\frac{1}{2}}$, $\mu_i = (\varkappa_1\varkappa_2)^{\frac{1}{2}} (\sigma_i \zeta^{h}_i)^{-1}$.
}

\section{Dualities and twists}

\subsection{Spectral duality}
For some integrable chains special kind of duality could be observed both on the classical and on the quantum level: namely system with $N$-dimensional auxiliary space on $M$ sites share Hamiltonians with some other system with $M$-dimensional auxiliary space on $N$ sites. Under duality spectral parameter which monodromy operator depends on, and spectral parameter of characteristic equation exchange, so this duality is often called spectral duality (however, sometimes referred as 'level-rank' or 'fiber-base' duality, see \cite{MMRZZ} and references therein).

In the case of our interest, system doesn't change its type: $\XXZ$ classical spin chain of $\mathfrak{gl}_M$ type on $N$ sites is dual to the $\XXZ$ chain of the $\mathfrak{gl}_N$ type on $M$ sites \cite{MMRZZ}, \cite{BS}. Looking at $M\times N$ fence-net bipartite graph, it becomes obvious: graph keeps its structure under $90$-degree rotation. On the level of Kasteleyn operator, this corresponds to exchange of factors in tensor product, and using different expressions for spin variables.

\paragraph{SU(2) theory with $N_f=4$ and one bi-fundamental multiplet.}{
 We start discussion of spectral duality in our context from simplest non-trivial example. Let us consider $\mathfrak{gl}_3$ spin chain on two sites, which is dual to $\mathfrak{gl}_2$ chain on three sites, considered in section \ref{ss:XXZ}. To derive dual Lax operators, we should permute some rows and columns of Kasteleyn operator (\ref{K322x2}), which is exchanging of factors in tensor product $\mathrm{End}(\mathbb{C}^2 \otimes \mathbb{C}^3) = \mathrm{End}(\mathbb{C}^3 \otimes \mathbb{C}^2) $:
\begin{equation}
\Kast=
\begin{array}{l|cccccc}
 & 11 & 21 & 31 & 12 & 22 & 32 \\
\hline
11 & \xi_{11} & 0 & \lambda\varkappa_1\eta_{11}  &  \mu \sigma_1 \eta_{12} & 0 & -\lambda\mu\varkappa_2\sigma_1 \xi_{12} \\
21 & \eta_{21} & \xi_{21} & 0 & -\mu\sigma_2 \xi_{22} & \mu \sigma_2 \eta_{22} & 0 \\
31 & 0 & \eta_{31} & \xi_{31} & 0 & -\mu\sigma_3 \xi_{32} & \mu \sigma_3 \eta_{32}\\
12 & \eta_{11} & 0 & -\lambda\varkappa_1 \xi_{11} & \xi_{12} & 0 & \lambda\varkappa_2\eta_{12}\\
22 & -\xi_{21} & \eta_{21} & 0 & \eta_{22} & \xi_{22} & 0 \\
32 & 0 & -\xi_{31} & \eta_{31} & 0 & \eta_{32} & \xi_{32}
\end{array}=
\left(
\begin{array}{cc}
\tilde{A}_1 & \mu \tilde{Q}\tilde{C}_2\\
\tilde{C}_1 & \tilde{A}_2
\end{array}
\right)
\end{equation}
Spectral curve is given by condition
\begin{equation}
\det \Kast(\lambda,\mu)=0 ~~ \Leftrightarrow ~~ \exists~\tilde{\psi}=
\left(\tilde{\psi}_1 ~~ \tilde{\psi}_2 \right):~~ \tilde{\psi}\Kast(\lambda,\mu)=0 ~~~
~~ \Leftrightarrow ~~
\left\{
\begin{array}{c}
\tilde{\psi}_2=\tilde{\psi}_1 \tilde{L}_1(\varkappa_1\zeta^v_1 \lambda)\\
\mu \tilde{\psi}_1 \tilde{Q}=\tilde{\psi}_2 \tilde{L}_2(\varkappa_2\zeta^v_2 \lambda)
\end{array}
\right.
\end{equation}
\begin{equation}
\tilde{L}_k(\lambda)=
\dfrac{1}{\lambda^{\frac{1}{2}}-\lambda^{-\frac{1}{2}}}
\left(
\def\arraystretch{2.2}
\begin{array}{ccc}
\lambda^{-\frac{1}{2}}\dfrac{\xi_{1k}}{\eta_{1k}}+\lambda^{\frac{1}{2}}\dfrac{\eta_{1k}}{\xi_{1k}} &
\lambda^{\frac{1}{2}}\dfrac{\eta_{1k}}{\xi_{2k}}\left(\dfrac{\xi_{1k}}{\eta_{1k}}+\dfrac{\eta_{1k}}{\xi_{1k}}\right) &
\lambda^{\frac{1}{2}}\dfrac{\eta_{1k}\eta_{2k}}{\xi_{2k}\xi_{3k}}\left(\dfrac{\xi_{1k}}{\eta_{1k}}+\dfrac{\eta_{1k}}{\xi_{1k}}\right) \\
\lambda^{-\frac{1}{2}}\dfrac{\xi_{2k}}{\eta_{1k}}\left(\dfrac{\xi_{2k}}{\eta_{2k}}+\dfrac{\eta_{2k}}{\xi_{2k}}\right) &
\lambda^{-\frac{1}{2}}\dfrac{\xi_{2k}}{\eta_{2k}}+\lambda^{\frac{1}{2}}\dfrac{\eta_{2k}}{\xi_{2k}} &
\lambda^{\frac{1}{2}}\dfrac{\eta_{2k}}{\xi_{3k}}\left(\dfrac{\xi_{2k}}{\eta_{2k}}+\dfrac{\eta_{2k}}{\xi_{2k}}\right) \\
\lambda^{-\frac{1}{2}}\dfrac{\xi_{2k}\xi_{3k}}{\eta_{1k}\eta_{2k}}\left(\dfrac{\xi_{3k}}{\eta_{3k}}+\dfrac{\eta_{3k}}{\xi_{3k}}\right) &
\lambda^{-\frac{1}{2}}\dfrac{\xi_{3k}}{\eta_{2k}}\left(\dfrac{\xi_{3k}}{\eta_{3k}}+\dfrac{\eta_{3k}}{\xi_{3k}}\right) &
\lambda^{-\frac{1}{2}}\dfrac{\xi_{3k}}{\eta_{3k}}+\lambda^{\frac{1}{2}}\dfrac{\eta_{3k}}{\xi_{3k}}
\end{array}
\right)
\end{equation}
\begin{equation}
\zeta_k^v=\dfrac{\xi_{1k}\xi_{2k}\eta_{3k}}{\eta_{1k}\eta_{2k}\eta_{3k}}
,~~~ \tilde{Q} =
\left(
\begin{array}{ccc}
\sigma_1 & 0 & 0\\
0 & \sigma_2 & 0\\
0 & 0 & \sigma_3
\end{array}
\right)
\end{equation}
which could be rewritten using monodromy operator
\begin{equation}
\tilde{\psi}_1 \left(\mu \tilde{Q} - \tilde{T}_{2}^{3 \times 3}(\lambda)\right)=0 ~~ \Leftrightarrow ~~  \det\left(\mu \tilde{Q} - \tilde{T}_{2}^{3 \times 3}(\lambda)\right)=0, ~~~  \tilde{T}_{2}^{3 \times 3}(\lambda)=\tilde{L}_1(\varkappa_1\zeta^v_1 \lambda) \tilde{L}_2(\varkappa_2\zeta^v_2 \lambda).
\end{equation}
It is indeed spectral dual to the curve (\ref{specT3s2x2}). One can check by direct calculation that
\begin{equation}
\begin{array}{lr}
(1-\varkappa_1 \zeta^v_1\lambda)(1-\varkappa_2 \zeta^v_2\lambda)\det\left(\mu \tilde{Q} - \tilde{T}_{2}^{3 \times 3}(\lambda)\right) =
\\
= (1-\sigma_1 \zeta^h_1\mu)(1-\sigma_2 \zeta^h_2\mu)(1-\sigma_3 \zeta^h_3\mu)\det\left(\lambda Q - T_{3}^{2 \times 2}(\mu)\right).
\end{array}
\end{equation}
}
\paragraph{General case.}{If the order of factors in tensor product in (\ref{Kblock}) had been chosen in the other way, we would get $M$ matrices $A_k$ and $C_k$ of size $N\times N$:
\begin{equation}
\Kast(\lambda,\mu)=\sum\limits_{m=1}^{M} \tilde{A}_m \otimes E_{m,m} +  (\tilde{Q})^{\delta_{M,m}} \tilde{C}_m\otimes E_{m+1,m}
\end{equation}
\begin{equation}
\label{ACdefs}
\tilde{A}_m = \sum\limits_{n=1}^{N} \xi_{nm} E_{n,n} +\eta_{nm}  \varkappa_m^{\delta_{1,n}} E_{n,n-1} ,\;\;\;  \tilde{C}_m = \sum\limits_{n=1}^{N} \eta_{nm} E_{n,n} - \xi_{nm} \varkappa_m ^{\delta_{n,1}} E_{n,n-1},\;\;\; \tilde{Q} = \sum\limits_{n=1}^{N} \sigma_n E_{nn}.
\end{equation}
Again, we present spectral curve as condition
\begin{equation}
\exists\, \tilde{\psi}=\sum\limits_{n=1}^{N}\sum\limits_{m=1}^{M} \tilde{\psi}_{nm}e_n\otimes e_m\in \mathbb{C}^{MN}:\; \tilde{\psi}\Kast(\lambda,\mu)=0
\end{equation}
which gives for the spectral curve
\begin{equation}
\det(\tilde{L}_1(\varkappa_1 \zeta_1^{v} \lambda) ... \tilde{L}_M(\varkappa_M \zeta_M^{v} \lambda) - \mu\tilde{Q} )=0,~~~
\tilde{L}_k(\varkappa_k \zeta_k^{v} \lambda) = -\tilde{A}_k \tilde{C}_k^{-1}.
\end{equation}
Using variables (\ref{convvars}) we can write dual Lax operator
\begin{equation}
(\tilde{L}_m)_{ij}(\lambda) =  \dfrac{1}{\lambda^{\frac{1}{2}}-\lambda^{-\frac{1}{2}}}
\left\{
\begin{array}{ll}
i\neq j, & \lambda^{-\frac{s_{ij}}{2}} (z_{im}^2+z_{im}^{-2})\dfrac{\tilde{\tau}_{im}}{\tilde{\tau}_{jm}}\\
i=j, & \lambda^{\frac{1}{2}}z_{im}^{-2}+\lambda^{-\frac{1}{2}}z_{im}^2
\end{array}
\right. ,~~~
\tilde{\tau}_{nm}=w_{nm}\prod\limits_{i=1}^{N}z_{im}^{-s_{in}}~.
\end{equation}
We can relate them to $L$-operators (\ref{Lclust}) of the same size
\begin{equation}
L(z,w,\mu)=\tilde{L}(z \to z^{-1}, w, \lambda \to \mu^{-1})^{\top}.
\end{equation}
Noting that for the classical $r$-matrix
\begin{equation}
r(a^{-1})^{\top}=-r(a)
\end{equation}
where transposition is taken in each tensor multiplier, we can deduce from (\ref{RTTClusterMain}) that
\begin{equation}
\{\tilde{L}(\lambda)\otimes \tilde{L}(\mu)\} = \frac{1}{2}[\tilde{L}(\lambda)\otimes \tilde{L}(\mu),r(\lambda/\mu)].
\end{equation}
To obtain explicit relation for the dual spectral curves, we have to come back to the Kasteleyn operator of the system, and consider its determinant. In terms of $M\times M$ blocks $A_k, C_k$ defined by ($\ref{ACdefs}$) spectral curve is given by
\begin{equation}
\det\, \Kast(\lambda,\mu)=
\begin{vmatrix}
A_1 & 0 & ... & 0 & \lambda C_1 Q \\
C_2 & A_2 & ... & 0 & 0 \\
... & ... & ... & ... & ... \\
0 & 0 & ...& A_{N-1} & 0 \\
0 & 0 & ... & C_{N} & A_{N} \\
\end{vmatrix}
=
\prod\limits_i \left(\det  C_i\right)\cdot
\begin{vmatrix}
C_1^{-1} A_1 & 0 & ... & 0 & \lambda Q \\
\M1 & C_2^{-1}A_2 & ... & 0 & 0 \\
... & ... & ... & ... & ... \\
0 & 0 & ...& C_{N-1}^{-1} A_{N-1} & 0 \\
0 & 0 & ... & \M1 & C_{N}^{-1}A_{N} \\
\end{vmatrix}=
\end{equation}
$$
=...=\prod\limits_i \left(\det  C_i\right) \cdot
\begin{vmatrix}
\M1 & 0 & ... & 0 & \lambda Q \\
\M1 & \M1 & ... & 0 & 0 \\
... & ... & ... & ... & ... \\
0 & 0 & ...& \M1 & 0 \\
0 & 0 & ... & \M1 & (-1)^{N}T_{N}^{M\times M} \\
\end{vmatrix}, ~~~
T_{N}^{M\times M}=L_1...L_N,~~~ L_k=-C_k^{-1}A_k,
$$
and subtracting consequentially lines from first to last
\begin{equation}
\det\, \Kast(\lambda,\mu)= (-1)^{NM}\det \left(C_1 ... C_N\right) \det(T_{N}^{M\times M}(\mu)-\lambda Q).
\end{equation}
Acting in the same way, we get for the dual spectral curve
\begin{equation}
\det\, \Kast(\lambda,\mu)=(-1)^{NM}\det \left(\tilde{C}_1...\tilde{C}_M\right) \det(\tilde{T}_{M}^{N\times N}(\lambda)-\mu \tilde{Q}), ~~~ \tilde{T}_{M}^{N\times N}=\tilde{L}_1...\tilde{L}_M,~~~ \tilde{L}_k=-\tilde{A}_k \tilde{C}_k^{-1}
\end{equation}
so, precise relation between curves is
\begin{equation}
\det \left(C_1...C_N\right) \det(T_{N}^{M\times M}(\mu)-\lambda Q) = \det \left(\tilde{C}_1...\tilde{C}_M\right) \det(\tilde{T}_{M}^{N\times N}(\lambda)-\mu \tilde{Q})
\end{equation}
Note that the relation of pre-factors is Casimir of the bracket
\begin{equation}
\dfrac{\det \left(C_1...C_N\right)}{\det \left(\tilde{C}_1...\tilde{C}_N\right)}=\dfrac{\mu^{\frac{N}{2}}}{\lambda^{\frac{M}{2}}}\left(\dfrac{\sigma_1...\sigma_N}{\varkappa_1...\varkappa_M}\right)^{1/2}\dfrac{\prod\limits_{n=1}^{N}(\sigma_n \zeta^h_n \mu)^{-1/2}-(\sigma_n \zeta^h_n \mu)^{1/2}}{\prod\limits_{m=1}^{M}(\varkappa_m \zeta^v_m \lambda)^{-1/2}-(\varkappa_m \zeta^v_m \mu)^{1/2}}.
\end{equation}}
\subsection{Twisted chains \label{ss:twT}}

A diagonal twist matrix is not the only one, commuting with $r$-matrices. A cyclic twist
\begin{equation}
Q_{\Lambda}(\lambda)=\sum\limits_{i=1}^{N}  E_{i+1,i} =\sum\limits_{i=1}^{N-1}  \E_{i+1,i} + \lambda \E_{1,N}
\end{equation}
also satisfies $[r(\lambda/\mu),Q_{\Lambda}(\lambda)\otimes Q_{\Lambda}(\mu)]=0$.
In terms of bipartite graphs it corresponds to the twist on a cycle of the torus, where the bipartite graph is drawn on, or the gluing condition for the sides of fundamental domain, see Fig.~\ref{fig:bipNM}. Such twist also changes a Poisson quiver, even though the edge variables are not affected themselves.

The twist of a bipartite graph results further in change of the zig-zag's structure. Several parallel zig-zags now join into 'longer sequences' with non-trivial winding so that rectangle Newton polygon undergoes a 'shear shift' -- see examples on Fig.~\ref{fig:twisted}.

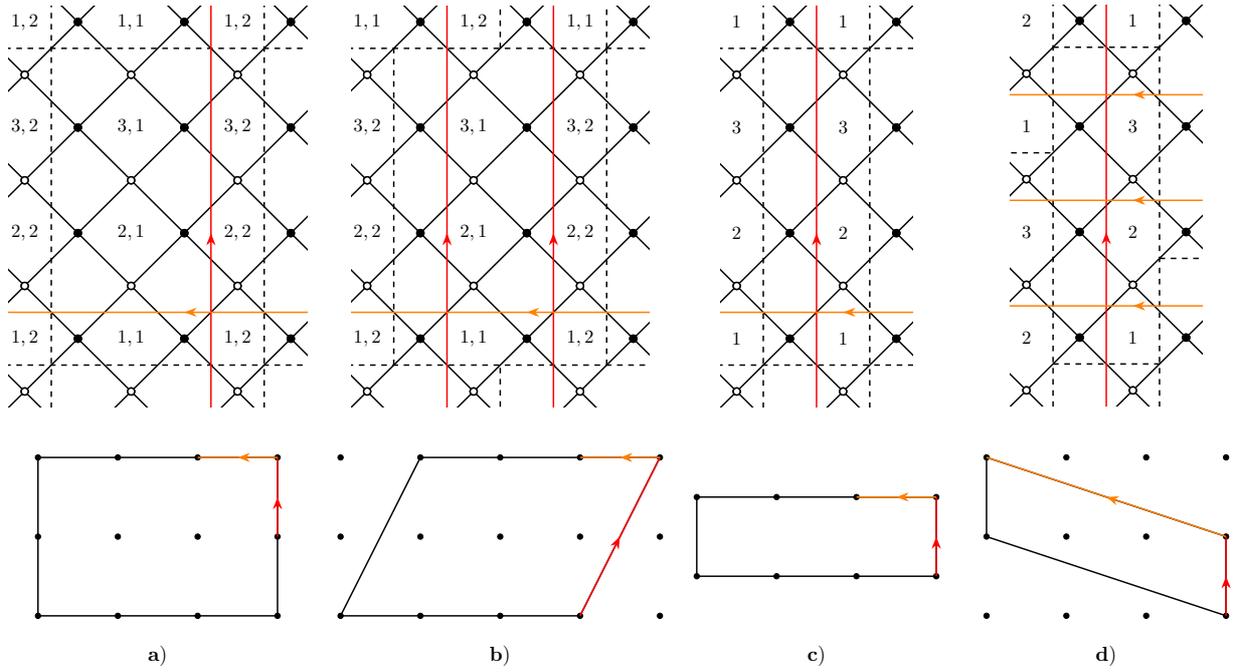
\begin{figure}[h!]
\begin{center}

\scalebox{0.7}{
\begin{tikzpicture}

\tikzmath{\Lx=3;\Ly=2;\xs=0.25;\ys=-1.25;\d=1.5;}

\node at (\xs+1.5*\d, \ys-0.5*\d) {$\mathbf{a)}$};

\begin{scope}[thick]
	\draw (\xs,\ys)--(\xs,\ys+\d*2);
	\draw (\xs,\ys+\d*2)--(\xs+\d*3,\ys+\d*2);
	\draw (\xs+\d*3,\ys+\d*2)--(\xs+\d*3,\ys);
	\draw (\xs+\d*3,\ys)--(\xs,\ys);
\end{scope}

\foreach \x in {0,...,\Lx}
	\foreach \y in {0,...,\Ly}
		\draw[fill] (\xs+\d*\x,\ys+\d*\y) circle[radius=0.05];

\draw[red, thick, styleArrowShort] (\xs+\d*3,\ys+\d*1)--(\xs+\d*3,\ys+\d*2);
\draw[orange, thick, styleArrowShort] (\xs+\d*3,\ys+\d*2)--(\xs+\d*2,\ys+\d*2);



\tikzmath{
	int \Lx,\Ly;
	\Lx=2;\Ly=3;
	\scale=1;
	\dofset=0;
	\cofset=-1.3;
}

\begin{scope}[scale=\scale]

\clip(\cofset+1,\cofset+4) rectangle (2*\Lx - \cofset, 3+2*\Ly - \cofset);


\draw[thick, dashed] (0.5,1)--(0.5,11);
\draw[thick, dashed] (4.5,1)--(4.5,11);
\draw[thick, dashed] (-1,3.5)--(7,3.5);
\draw[thick, dashed] (-1,9.5)--(7,9.5);

\tikzmath{
	for \z in {0,...,\Ly+2}{
		for \i in {-1,...,\Lx+1}{
			{			
				\draw  (\i*2,\z*2) pic {bipart};
			};
		};
  	};
};

{
	\node[font=\small] at (0*2, 1*2) {$3,2$};	
	\node[font=\small] at (1*2, 1*2) {$3,1$};	
	\node[font=\small] at (2*2, 1*2) {$3,2$};	
	\node[font=\small] at (3*2, 1*2) {$3,1$};

	\node[font=\small] at (0*2, 2*2) {$1,2$};	
	\node[font=\small] at (1*2, 2*2) {$1,1$};	
	\node[font=\small] at (2*2, 2*2) {$1,2$};	
	\node[font=\small] at (3*2, 2*2) {$1,1$};
	
	\node[font=\small] at (0*2, 3*2) {$2,2$};	
	\node[font=\small] at (1*2, 3*2) {$2,1$};	
	\node[font=\small] at (2*2, 3*2) {$2,2$};	
	\node[font=\small] at (3*2, 3*2) {$2,1$};
	
	\node[font=\small] at (0*2, 4*2) {$3,2$};	
	\node[font=\small] at (1*2, 4*2) {$3,1$};	
	\node[font=\small] at (2*2, 4*2) {$3,2$};	
	\node[font=\small] at (3*2, 4*2) {$3,1$};
	
	\node[font=\small] at (0*2, 5*2) {$1,2$};	
	\node[font=\small] at (1*2, 5*2) {$1,1$};	
	\node[font=\small] at (2*2, 5*2) {$1,2$};	
	\node[font=\small] at (3*2, 5*2) {$1,1$};
}

\draw[red, thick, styleArrowShort] (3.5,1)--(3.5,11);
\draw[orange, thick, styleArrowShort] (7,4.5)--(-1,4.5);

\end{scope}

\end{tikzpicture}
\begin{tikzpicture}
\draw[white] (-1,0)--(-1,1);
\draw[white] (6,0)--(6,1);

\tikzmath{\Lx=3;\Ly=2;\xs=-0.5;\ys=-1.25;\d=1.5;}


\begin{scope}[thick]
	\draw (\xs,\ys)--(\xs+\d,\ys+\d*2);
	\draw (\xs+\d,\ys+\d*2)--(\xs+\d*4,\ys+\d*2);
	\draw (\xs+\d*4,\ys+\d*2)--(\xs+\d*3,\ys);
	\draw (\xs+\d*3,\ys)--(\xs,\ys);
\end{scope}
\node at (\xs+2*\d, \ys-0.5*\d) {$\mathbf{b)}$};

\foreach \x in {0,...,4}
	\foreach \y in {0,...,\Ly}
		\draw[fill] (\xs+\d*\x,\ys+\d*\y) circle[radius=0.05];

\draw[red, thick, styleArrowShort] (\xs+\d*3,\ys)--(\xs+\d*4,\ys+\d*2);
\draw[orange, thick, styleArrowShort] (\xs+\d*4,\ys+\d*2)--(\xs+\d*3,\ys+\d*2);



\tikzmath{
	int \Lx,\Ly;
	\Lx=2;\Ly=3;
	\scale=1;
	\dofset=0;
	\cofset=-1.3;
}

\begin{scope}[scale=\scale]

\clip(\cofset+1,\cofset+4) rectangle (2*\Lx - \cofset, 3+2*\Ly - \cofset);


\draw[thick, dashed] (2.5,9.5)--(2.5,10.5);
\draw[thick, dashed] (4.5,9.5)--(4.5,10.5);

\draw[thick, dashed] (0.5,3.5)--(0.5,9.5);
\draw[thick, dashed] (4.5,3.5)--(4.5,9.5);

\draw[thick, dashed] (2.5,2.5)--(2.5,3.5);
\draw[thick, dashed] (6.5,2.5)--(6.5,3.5);

\draw[thick, dashed] (-1,3.5)--(7,3.5);
\draw[thick, dashed] (-1,9.5)--(7,9.5);

\tikzmath{
	for \z in {0,...,\Ly+2}{
		for \i in {-1,...,\Lx+1}{
			{			
				\draw  (\i*2,\z*2) pic {bipart};
			};
		};
  	};
};

{
	\node[font=\small] at (0*2, 1*2) {$3,1$};	
	\node[font=\small] at (1*2, 1*2) {$3,2$};	
	\node[font=\small] at (2*2, 1*2) {$3,1$};	
	\node[font=\small] at (3*2, 1*2) {$3,2$};

	\node[font=\small] at (0*2, 2*2) {$1,2$};	
	\node[font=\small] at (1*2, 2*2) {$1,1$};	
	\node[font=\small] at (2*2, 2*2) {$1,2$};	
	\node[font=\small] at (3*2, 2*2) {$1,1$};
	
	\node[font=\small] at (0*2, 3*2) {$2,2$};	
	\node[font=\small] at (1*2, 3*2) {$2,1$};	
	\node[font=\small] at (2*2, 3*2) {$2,2$};	
	\node[font=\small] at (3*2, 3*2) {$2,1$};
	
	\node[font=\small] at (0*2, 4*2) {$3,2$};	
	\node[font=\small] at (1*2, 4*2) {$3,1$};	
	\node[font=\small] at (2*2, 4*2) {$3,2$};	
	\node[font=\small] at (3*2, 4*2) {$3,1$};
	
	\node[font=\small] at (0*2, 5*2) {$1,1$};	
	\node[font=\small] at (1*2, 5*2) {$1,2$};	
	\node[font=\small] at (2*2, 5*2) {$1,1$};	
	\node[font=\small] at (3*2, 5*2) {$1,2$};
}

\draw[red, thick, styleArrowShort] (5.5,1)--(5.5,11);
\draw[red, thick, styleArrowShort] (3.5,1)--(3.5,11);
\draw[red, thick, styleArrowShort] (1.5,1)--(1.5,11);
\draw[red, thick, styleArrowShort] (-0.5,1)--(-0.5,11);

\draw[orange, thick, styleArrowShort] (7,4.5)--(-1,4.5);

\end{scope}

\end{tikzpicture}
\begin{tikzpicture}

\draw[white] (4.5,3)--(4.5,4);
\draw[white] (1,-1.27)--(2,-1.27);

\tikzmath{\Lx=3;\Ly=1;\xs=-0.75;\ys=-0.5;\d=1.5;}


\begin{scope}[thick]
	\draw (\xs,\ys)--(\xs,\ys+\d);
	\draw (\xs,\ys+\d)--(\xs+\d*3,\ys+\d);
	\draw (\xs+\d*3,\ys+\d)--(\xs+\d*3,\ys);
	\draw (\xs+\d*3,\ys)--(\xs,\ys);
\end{scope}
\node at (\xs+1.5*\d, \ys-1*\d) {$\mathbf{c)}$};

\foreach \x in {0,...,\Lx}
	\foreach \y in {0,...,\Ly}
		\draw[fill] (\xs+\d*\x,\ys+\d*\y) circle[radius=0.05];

\draw[red, thick, styleArrowShort] (\xs+\d*3,\ys)--(\xs+\d*3,\ys+\d);
\draw[orange, thick, styleArrowShort] (\xs+\d*3,\ys+\d)--(\xs+2*\d,\ys+\d);



\tikzmath{
	int \Lx,\Ly;
	\Lx=1;\Ly=3;
	\scale=1;
	\dofset=0;
	\cofset=-1.3;
}

\begin{scope}[scale=\scale]

\clip(\cofset+1,\cofset+4) rectangle (2*\Lx - \cofset, 3+2*\Ly - \cofset);


\draw[thick, dashed] (0.5,1)--(0.5,11);
\draw[thick, dashed] (2.5,1)--(2.5,11);
\draw[thick, dashed] (-1,3.5)--(5,3.5);
\draw[thick, dashed] (-1,9.5)--(5,9.5);

\tikzmath{
	for \z in {0,...,\Ly+2}{
		for \i in {-1,...,\Lx+1}{
			{			
				\draw  (\i*2,\z*2) pic {bipart};
			};
		};
  	};
};

{

	\node[font=\small] at (0*2, 2*2) {$1$};	
	\node[font=\small] at (1*2, 2*2) {$1$};	
	\node[font=\small] at (2*2, 2*2) {$1$};	
	
	\node[font=\small] at (0*2, 3*2) {$2$};	
	\node[font=\small] at (1*2, 3*2) {$2$};	
	\node[font=\small] at (2*2, 3*2) {$2$};	
	
	\node[font=\small] at (0*2, 4*2) {$3$};	
	\node[font=\small] at (1*2, 4*2) {$3$};	
	\node[font=\small] at (2*2, 4*2) {$3$};	
	
	\node[font=\small] at (0*2, 5*2) {$1$};	
	\node[font=\small] at (1*2, 5*2) {$1$};	
	\node[font=\small] at (2*2, 5*2) {$1$};	
}

\draw[red, thick, styleArrowShort] (1.5,1)--(1.5,11);
\draw[orange, thick, styleArrowShort] (5,4.5)--(-1,4.5);

\end{scope}

\end{tikzpicture}
\begin{tikzpicture}

\draw[white] (1,-1.27)--(2,-1.27);

\tikzmath{\Lx=3;\Ly=1;\xs=-0.75;\ys=0.23;\d=1.5;}


\begin{scope}[thick]
	\draw (\xs,\ys)--(\xs,\ys+\d);
	\draw (\xs,\ys+\d)--(\xs+\d*3,\ys);
	\draw (\xs+\d*3,\ys)--(\xs+\d*3,\ys-\d);
	\draw (\xs+\d*3,\ys-\d)--(\xs,\ys);
\end{scope}
\node at (\xs+1.5*\d, \ys-1.5*\d) {$\mathbf{d)}$};

\foreach \x in {0,...,\Lx}
	\foreach \y in {-1,...,\Ly}
		\draw[fill] (\xs+\d*\x,\ys+\d*\y) circle[radius=0.05];

\draw[red, thick, styleArrowShort] (\xs+\d*3,\ys-\d)--(\xs+\d*3,\ys);
\draw[orange, thick, styleArrowShort] (\xs+\d*3,\ys)--(\xs,\ys+\d);



\tikzmath{
	int \Lx,\Ly;
	\Lx=1;\Ly=3;
	\scale=1;
	\dofset=0;
	\cofset=-1.3;
}

\begin{scope}[scale=\scale]

\clip(\cofset+1,\cofset+4) rectangle (2*\Lx - \cofset, 3+2*\Ly - \cofset);


\draw[thick, dashed] (0.5,1)--(0.5,11);
\draw[thick, dashed] (2.5,1)--(2.5,11);

\draw[thick, dashed] (2.5,5.5)--(3.5,5.5);
\draw[thick, dashed] (0.5,3.5)--(2.5,3.5);
\draw[thick, dashed] (0.5,9.5)--(2.5,9.5);
\draw[thick, dashed] (-0.5,7.5)--(0.5,7.5);

\tikzmath{
	for \z in {0,...,\Ly+2}{
		for \i in {-1,...,\Lx+1}{
			{			
				\draw  (\i*2,\z*2) pic {bipart};
			};
		};
  	};
};

{

	\node[font=\small] at (0*2, 2*2) {$2$};	
	\node[font=\small] at (1*2, 2*2) {$1$};	
	\node[font=\small] at (2*2, 2*2) {$3$};	
	
	\node[font=\small] at (0*2, 3*2) {$3$};	
	\node[font=\small] at (1*2, 3*2) {$2$};	
	\node[font=\small] at (2*2, 3*2) {$1$};	
	
	\node[font=\small] at (0*2, 4*2) {$1$};	
	\node[font=\small] at (1*2, 4*2) {$3$};	
	\node[font=\small] at (2*2, 4*2) {$2$};	
	
	\node[font=\small] at (0*2, 5*2) {$2$};	
	\node[font=\small] at (1*2, 5*2) {$1$};	
	\node[font=\small] at (2*2, 5*2) {$3$};	
}

\draw[red, thick, styleArrowShort] (1.5,1)--(1.5,11);

\tikzmath{\dz=0.1;}

\draw[orange, thick, styleArrowShort] (5,2.5+\dz)--(-1,2.5+\dz);
\draw[orange, thick, styleArrowShort] (5,4.5+\dz)--(-1,4.5+\dz);
\draw[orange, thick, styleArrowShort] (5,6.5+\dz)--(-1,6.5+\dz);
\draw[orange, thick, styleArrowShort] (5,8.5+\dz)--(-1,8.5+\dz);
\end{scope}

\end{tikzpicture}
}

\end{center}
\caption{Examples of twisted $\mathfrak{gl}_2$ chains. Dashed lines bound fundamental domains. We use different notations for zig-zags here, comparing to the pictures above. Edges crossed by red arrows belong to $\gamma_2$ zig-zag, orange arrows are for $\alpha_1$. a,b) $\XXZ$ chain of rank two and its twisted cousin. Note that the twisted twice chain is equivalent up to $\mathrm{SL}(2,\mathbb{Z})$ transformation $\lambda \to \lambda \mu$ to the untwisted chain, as $Q_{\Lambda}^2 = \mu\, \M1$, like in Remark \ref{rem:group}. c,d) Making Toda chain by twisting $\mathfrak{gl}_N$ chain dual to $\mathfrak{gl}_1$ chain.}
\label{fig:twisted}
\end{figure}

In the context of such transformations one can expect nontrivial consequences for spectral duality. Consider the trivial case of $\mathfrak{gl}_N$ chain on a single site, which is dual to rank $1$ chain on $N$ sites, and apply the cyclic twist along the longer side of a bipartite graph. In original picture this is just a multiplication of a single $N\times N$ Lax operator by cyclic permutation matrix. However in the dual setup, this results in passing from trivial $\mathfrak{gl}_1$ chain to the Toda chain on the same number of sites, which can be verified by comparing Fig.~\ref{fig:twisted} and Fig.~\ref{fig:toda1}. After such procedure the number of Casimirs drops by $2N-2$, while number of Hamiltonians jumps from $0$ to $N-1$.

For supersymmetric gauge theories such transformation turns the theory of a single $\mathrm{SU}(N)$ hypermultiplet with only $\mathrm{SU}(N)\times\mathrm{SU}(N)$  flavor symmetry into pure $\mathrm{SU}(N)$ gauge theory.

\section{Discrete dynamics}

The cluster mapping class group $\Gq$ consists of sequences of mutations and permutations of quiver vertices, which maps quiver to itself, but acts in general non-trivially to the cluster variables (see Appendix \ref{ss:cluster} for details). As a simplification one can restrict the action of $\Gq$ to the set of Casimirs of the Poisson bracket. Each monomial Casimir maps to the monomial in Casimir functions. When the necessary for integrability condition $\prod_i x_i=1$ is relaxed to $\prod_i x_i=q$ (which is called as deautonomization), these flows act on the set of Casimirs, inducing non-trivial $q-$dynamics.

In \cite{BGM:2017} the cluster mapping class groups for the quivers, corresponding to Newton polygons with a single internal point, were identified with the symmetry groups of q-Painlev\'e equations\footnote{Such relation for particular cases was  earlier mentioned in \cite{Hone:2014, Okubo:2015, BS:2016:1, Okubo:2017}.}. Passing from $X$-cluster to $A$-cluster variety, the q-Painlev\'e equations acquire bilinear form for the tau-functions, and can be solved via the dual Nekrasov partition functions for 5d supersymmetric SU(2) gauge theories \cite{BS:2016:1,BGM:2017,Jimbo:2017,BGT:2017}, which is a natural '5d uplift' of '4d' isomonodromic/CFT correspondence \cite{GIL:1207}. In \cite{BGM:2018} the cluster description was further applied to discrete dynamics of relativistic Toda chains of arbitrary lengths, where the solutions of non-autonomous versions are given by $SU(N)$ partition functions with the $|k|\leq N$ Chern-Simons terms. Recently, cluster realization of generalized q-Painlev\'e VI system was also observed in \cite{OS:2018}. Note that for $q=1$ case with trivial Casimirs solution of discrete dynamics for arbitrary bipartite graph can be written in terms of $\theta$-functions \cite{Fock}.

Below in this section we discuss the cluster mapping class groups and non-autonomous bilinear equations, arising for generic rectangle Newton polygons. We present their explicit construction in the example, which will illustrate the following results:

\paragraph{Structure of the group $\Gq$.}{\ \\
\emph{For the $SA(2,\mathbb{Z})$-class of $N\times M$ rectangular Newton polygon, the MCG $\Gq$ always contains a subgroup of the form}
\be
\label{eq:GqGeneralNM}
\Wext\left(A_{N-1}^{(1)} \times A_{N-1}^{(1)}\right) \times \Wext\left(A_{M-1}^{(1)}\times A_{M-1}^{(1)}\right) \rtimes \mathbb{Z} \subset \Gq.
\ee
\emph{where $\Wext\left(A_{N-1}^{(1)} \times A_{N-1}^{(1)}\right)$ is a co-extended double Weyl group
\rf{eq:doubleWeylGroup}. }


The generators of each subgroup are naturally labeled by intervals on sides of a Newton polygon, or subset of
'parallel' zig-zag paths (in the same homology class) on a bipartite graph:
\be
\label{genW}
\Wext\left(A_{N-1}^{(1)} \times A_{N-1}^{(1)}\right):\
\{ s_{\alpha_i,\alpha_{i+1}}\}, \ \ \ \{ s_{\beta_i,\beta_{i+1}}\},\ \ \ \ i\in \mathbb{Z}/N\mathbb{Z}
\\
\Wext\left(A_{M-1}^{(1)} \times A_{M-1}^{(1)}\right):\
\{ s_{\gamma_a,\gamma_{a+1}}\}, \ \ \ \{ s_{\delta_a,\delta_{a+1}}\},\ \ \ \ a\in \mathbb{Z}/M\mathbb{Z}
\ee
where subscripts $\alpha,\beta,\gamma,\delta$ label the corresponding group of paths, see Fig.~\ref{fig:bipNM} middle and right. The group being extended by the additional generator $\rho$ contains lattice of the rank $2N+2M-3$ of $q-$difference flows of integrable system.

Moreover, in special cases there is an obvious symmetry enhancement: for example, for $N=M$ an additional 'external' generator appears, which rotates the whole picture by $\pi/2$. However, sometimes this enhancement is more essential: if any of the sides is of length 2, two rest Weyl groups can be 'glued' together by additional permutation, so the known subgroup of $\Gq$ becomes
\begin{equation}
\Wext\left(A_{2N-1}^{(1)}\right) \times \Wext\left(A_{1}^{(1)}\times A_{1}^{(1)}\right) \subset \Gq
\end{equation}
This enhancement is closely related to the fact that spectral curves with the $N\times 2$ rectangular Newton polygon can be mapped to the curves with the triangular Newton polygon with the integer sides $2N\times 2\times 2$ (see e.g. (3.70) in \cite{Gaiotto}). If both $N=M=2$ one finds the extra enhancement from $\Wext(A_1^{(1)}\times A_1^{(1)}) \times \Wext(A_1^{(1)}\times A_1^{(1)})$ to $ \Wext(D_5^{(1)})$, see below.
}
\paragraph{Action on spin chain Casimirs.}{\ \\
\emph{Inhomogeneities, total spins, on-site Casimirs and twists of spin chain are permuted under the action of different components of $\Gq$.}\\

Inhomogeneities are given by single zig-zags $\mu_i = \beta_i$, while on-site Casimirs are given by products of zig-zags $\zeta_i^{h}=(\alpha_i \beta_i)^{\frac{1}{2}}$. So the well defined transformation of them, which 'permutes sites' of spin chain are products of primitive permutations
\begin{equation}
\sc_{\alpha_i,\alpha_{i+1}} \sc_{\beta_i,\beta_{i+1}}: ~~~ \mu_i \mapsto \mu_{i+1},~ \mu_{i+1} \mapsto \mu_{i}, ~~~ \zeta_i^{h}\mapsto \zeta_{i+1}^{h},~ \zeta_{i+1}^{h}\mapsto \zeta_{i}^{h}.
\end{equation}

Permutations of twists $\varkappa_a = (\delta_a/\gamma_a)^{\frac{1}{2}}$ and projections of spins $\zeta_a^{v}=(\gamma_a \delta_a)^{\frac{1}{2}}$ by products
\begin{equation}
\sc_{\gamma_a,\gamma_{a+1}}\sc_{\delta_a,\delta_{a+1}}:~~~ \varkappa_a \mapsto \varkappa_{a+1}, ~ \varkappa_{a+1} \mapsto \varkappa_{a}, ~~~ \zeta_a^{v} \mapsto \zeta_{a+1}^{v}, ~ \zeta_{a+1}^{v} \mapsto \zeta_{a}^{v}.
\end{equation}
can be viewed as an action of the Weyl group by permutations on the maximal torus of Lie group.
}

\paragraph{Bilinear equations.}{\ \\
\emph{Equations defining the action of each single generator of $\Gq$ on $A$-cluster variables $(\ta_{ij},\tb_{ij})$ could be rewritten in the form of bilinear equations. Evolution of coefficients can be encapsulated into the transformations of frozen variables $\{\c_{\alpha_i},\c_{\beta_i},\c_{\gamma_a},\c_{\delta_a}\}$, which are evolving in the same way as Casimirs in $\mathcal{X}$-variables.}\\

For example $\avars$-variables $\tba_{k,a}, \tbb_{k,a}$ transformed under the action of generator $\sc_{\beta_{i},\beta_{i+1}}$ satisfy bilinear equations
\begin{equation}
\label{eq:bilinearBetaGeneral}
{\def\arraystretch{2.2}
\begin{array}{l}
(\c_{\beta_{i+1}}-q^{\frac{1}{N}}\c_{\beta_{i}})(\c_{\delta}\c_{\gamma_a})^{\frac{1}{N}}\tb_{i-1,a}\ta_{i+1,a}
=
\c_{\beta_{i+1}}^{\frac{1}{M}}\tbb_{i,a}\ta_{i,a} - q^{\frac{1}{NM}}\c_{\beta_i}^{\frac{1}{M}}\tba_{i,a}\tb_{i,a}
\\
(\c_{\beta_{i+1}}-q^{\frac{1}{N}}\c_{\beta_i})(\c_{\delta}/\c_{\delta_a})^{\frac{1}{N}}\tb_{i-1,a+1}\ta_{i+1,a}
=
\c_{\alpha_i}^{-\frac{1}{M}}\tba_{i,a}\tb_{i,a+1} - q^{\frac{1}{NM}}\c_{\alpha_i}^{-\frac{1}{M}}\tbb_{i,a+1}\ta_{i,a}
\end{array}
}
\ \ \ \forall \, a \in \mathbb{Z}/M\mathbb{Z}
\end{equation}
where $\c_{\delta} = \prod_a \c_{\delta_a} $. Frozen variables are transforming as
\begin{equation}
\label{sbet}
\sc_{\beta_i,\beta_{i+1}}: \ \ \
\c_{\beta_i} \mapsto q^{-\frac{1}{N}}\c_{\beta_{i+1}}, \
\c_{\beta_{i+1}} \mapsto q^{\frac{1}{N}} \c_{\beta_i}.
\end{equation}
Bilinear equations for the action of generators $\sc_{\alpha_{i},\alpha_{i+1}},\sc_{\gamma_{a},\gamma_{a+1}},\sc_{\delta_{a},\delta_{a+1}}$ are similar.
}
\subsection{Structure of $\Gq$}
\label{ss:clusterexplicit}

Now we present generators of $\Gq$ in terms of the quiver mutations\footnote{For the definitions on cluster algebras see Appendix B.} $\{ \mua_{ij}, \mub_{ij}\}$ (in the vertices, initially assigned with $\{\ya_{ij},\yb_{ij}\}$) and permutations of the vertices $\{s^{\la,\lb}_{ij,kl}\}$. Consider for simplicity the $(3,2)$-example, which already illustrates how the explicit formulas look like in generic case. Here $2(N+M)=10$ generators \rf{genW} can be realized as
\begin{equation}
\begin{aligned}
\sc_{\beta_1, \beta_2}=s^{\la,\lb}_{12,12}\mub_{11}\mua_{11}\mua_{12}\mub_{12}\mua_{11}\mub_{11}
~~~~~
\sc_{\alpha_3, \alpha_1}=s^{\la,\lb}_{12,31}\mub_{32}\mua_{11}\mua_{12}\mub_{31}\mua_{11}\mub_{32}
\\
\sc_{\beta_2, \beta_3}=s^{\la,\lb}_{22,22}\mub_{21}\mua_{21}\mua_{22}\mub_{22}\mua_{21}\mub_{21}
~~~~~
\sc_{\alpha_1, \alpha_2}=s^{\la,\lb}_{22,11}\mub_{12}\mua_{21}\mua_{22}\mub_{11}\mua_{21}\mub_{12}
\\
\sc_{\beta_3, \beta_1}=s^{\la,\lb}_{32,32}\mub_{31}\mua_{31}\mua_{32}\mub_{32}\mua_{31}\mub_{31}
~~~~~
\sc_{\alpha_2, \alpha_3}=s^{\la,\lb}_{32,21}\mub_{22}\mua_{31}\mua_{32}\mub_{21}\mua_{31}\mub_{22}
\end{aligned}
\end{equation}
\begin{equation}
\begin{aligned}
\sc_{\delta_2, \delta_1}=s^{\la,\lb}_{31,31}\mub_{21}\mua_{21}\mub_{11}\mua_{11}\mua_{31}\mub_{31}\mua_{11}\mub_{11}\mua_{21}\mub_{21}
~~~~~
\sc_{\gamma_1, \gamma_2}=s^{\la,\lb}_{21,12}\mub_{22}\mua_{31}\mub_{32}\mua_{11}\mua_{21}\mub_{12}\mua_{11}\mub_{32}\mua_{31}\mub_{22}
\\
\sc_{\delta_1, \delta_2}=s^{\la,\lb}_{32,32}\mub_{22}\mua_{22}\mub_{12}\mua_{12}\mua_{32}\mub_{32}\mua_{12}\mub_{12}\mua_{22}\mub_{22}
~~~~~
\sc_{\gamma_2, \gamma_1}=s^{\la,\lb}_{22,11}\mub_{21}\mua_{32}\mub_{31}\mua_{12}\mua_{22}\mub_{11}\mua_{12}\mub_{31}\mua_{32}\mub_{21}
\end{aligned}
\end{equation}
which are sequences of mutations in the vertices along zig-zags in the forward and then backward directions.
One can check that each generator here is an involution i.e. $\sc^2=1$. On the $X$-cluster variables it acts by rational transformation, e.g. for $\sc_{\beta_2,\beta_3}=s^{\la,\lb}_{22,22}\mub_{21}\mua_{21}\mua_{22}\mub_{22}\mua_{21}\mub_{21}$ one can
explicitly write:
\be
\label{sbeta}
\ya_{31}\mapsto \ya_{31}\cdot \yb_{22} \ya_{21}\dfrac{\lbr \ya_{22},\yb_{21}, \ya_{21}\rbr}{\lbr \ya_{21},\yb_{22},\ya_{22}\rbr}, \ \ \
\ya_{32}\mapsto \ya_{32}\cdot \yb_{21}\ya_{22}\dfrac{\lbr \ya_{21},\yb_{22}, \ya_{22}\rbr}{\lbr \ya_{22},\yb_{21},\ya_{21}\rbr},
\\
\yb_{21}\mapsto \dfrac{1}{\ya_{21}}\cdot \dfrac{\lbr \yb_{21},\ya_{21}, \yb_{22}\rbr}{\lbr \yb_{22},\ya_{22},\yb_{21}\rbr},
\ \ \
\yb_{22}\mapsto \dfrac{1}{\ya_{22}}\cdot \dfrac{\lbr \yb_{22},\ya_{22}, \yb_{21}\rbr}{\lbr \yb_{21},\ya_{21},\yb_{22}\rbr},
\\
\ya_{21}\mapsto \dfrac{1}{\yb_{22}}\cdot \dfrac{\lbr \ya_{21},\yb_{22}, \ya_{22}\rbr}{\lbr \ya_{22},\yb_{21},\ya_{21}\rbr},
\ \ \
\ya_{22}\mapsto \dfrac{1}{\yb_{21}}\cdot \dfrac{\lbr \ya_{22},\yb_{21}, \ya_{21}\rbr}{\lbr \ya_{21},\yb_{22},\ya_{22}\rbr},
\\
\yb_{11}\mapsto \yb_{11}\cdot \ya_{21} \yb_{21}\dfrac{\lbr \yb_{22},\ya_{22}, \yb_{21}\rbr}{\lbr \yb_{21},\ya_{21},\yb_{22}\rbr}, \ \ \
\yb_{12}\mapsto \yb_{12}\cdot \ya_{22} \yb_{22}\dfrac{\lbr \yb_{21},\ya_{21}, \yb_{22}\rbr}{\lbr \yb_{22},\ya_{22},\yb_{21}\rbr},
\ee
while all the other variables remain unchanged. Here we have used the notation
\begin{equation}
\lbr x_1,x_2,..,x_n\rbr = 1+x_1+x_1 \cdot x_2+...+x_1\cdot ...\cdot x_n=1+x_1(1+x_2(....+x_{n-1}(1+x_n)...)).
\end{equation}
Notice also that the result of zig-zag mutation sequences actually do not depends on the point of the 'zig-zag strip' one starts with the first mutation and direction of the jumps along/across given zig-zag. Note that the $[~]$-function possesses nice 'inversion' property
\begin{equation}
\lbr x_1,...,x_n\rbr = x_1...x_n\cdot \lbr x_n^{-1},...,x_1^{-1}\rbr
\end{equation}
which allows to write equivalently, for example
\begin{equation}
\ya_{21}\mapsto \dfrac{1}{\yb_{22}}\cdot \dfrac{\lbr \ya_{21},\yb_{22}, \ya_{22}\rbr}{\lbr \ya_{22},\yb_{21},\ya_{21}\rbr} =
\dfrac{1}{\yb_{21}}\cdot \dfrac{\lbr (\ya_{22})^{-1},(\yb_{22})^{-1}, (\ya_{21})^{-1}\rbr}{\lbr (\ya_{21})^{-1},(\yb_{21})^{-1},(\ya_{22})^{-1}\rbr}.
\end{equation}
Each set of permutations $\sc_{\z_i,\z_{i+1}}$ with similar $\z$ constitute affine Weyl group of $A^{(1)}$-type. The groups for different $z$ are commuting, so they satisfy usual relations
\begin{equation}
\begin{array}{ll}
\left\{
\begin{array}{l}
\sc_{\z_i,\z_{i+1}}^2=1,\\
(\sc_{\z_i,\z_{i+1}}\sc_{\z_{i+1},\z_{i+2}})^{3}=1\\
\sc_{\z_i,\z_{i+1}}\sc_{\z_j,\z_{j+1}}=\sc_{\z_j,\z_{j+1}}\sc_{\z_i,\z_{i+1}},~~~|i-j|>1
\end{array}
\right.
&
\z=\alpha,\beta ~ \mathrm{with} ~ i,j\in \mathbb{Z}/3\mathbb{Z}
\\
~~~~ \sc_{\z_i,\z_{a+1}}^2=1
&
\z=\gamma,\delta ~ \mathrm{with} ~ i,j\in \mathbb{Z}/2\mathbb{Z}.
\\
~~~~ \sc_{\z_i,\z_{i+1}}\sc_{\z'_j,\z'_{j+1}} = \sc_{\z'_j,\z'_{j+1}}\sc_{\z_i,\z_{i+1}},
&
\z,\z' = \alpha,\beta,\gamma,\delta \mathrm{~such~that~} \z \neq \z'.
\end{array}
\end{equation}
There are two more 'external' automorphisms preserving bipartite graph
\begin{equation}
\begin{array}{lll}
\Lambda_h:~~~
& \ya_{ia} \mapsto \ya_{i,a-1},
&\yb_{ia} \mapsto \yb_{i,a-1}\\
\Lambda_v:~~~
& \ya_{ia} \mapsto \ya_{i-1,a},
& \yb_{ia} \mapsto \yb_{i-1,a}
\end{array}
\end{equation}
which satisfy obvious relations
\begin{equation}
\Lambda_h \Lambda_v = \Lambda_v \Lambda_h, ~~~
\Lambda_h^{2}=1,~~~
\Lambda_v^{3}=1,
\end{equation}
\begin{equation}
\begin{array}{cc}
\Lambda_h \sc_{\z_a,\z_{a+1}} = \sc_{\z_{a-1},\z_{a}} \Lambda_h,~~~ \mathrm{for~\z=\gamma,\delta}, ~~~ &
\Lambda_h \sc_{\z_i,\z_{i+1}} = \sc_{\z_{i},\z_{i+1}} \Lambda_h,~~~ \mathrm{for~\z=\alpha,\beta},
\end{array}
\end{equation}
\begin{equation}
\begin{array}{ll}
\Lambda_v \sc_{\z_i,\z_{i+1}} = \sc_{\z_{i-1},\z_{i}}\Lambda_v,~~~ \mathrm{for~\z=\alpha,\beta}~~~ &
\Lambda_v \sc_{\z_a,\z_{a+1}} = \sc_{\z_{a},\z_{a+1}} \Lambda_v,~~~ \mathrm{for~\z=\gamma,\delta}
\end{array},
\end{equation}
and promote affine Weyl groups to extended affine Weyl groups. There is also one more generator of infinite order
\begin{equation}
\begin{array}{lcc}
\rho = s^{\lb \la}\mu^{\lb}: &
\mu^{\lb}=\prod\limits_{i,a}\mu_{ia}^{\lb},&
s^{\lb \la}:~~~ \yb_{ia} \mapsto \ya_{ia}, ~~~
\ya_{ia} \mapsto \yb_{i-1,a+1},
\end{array}
\end{equation}
satisfying relations
\begin{equation}
\rho \, \sc_{\alpha_{i-1},\alpha_{i}} = \sc_{\alpha_i,\alpha_{i+1}} \, \rho , ~~~
\rho \, \sc_{\beta_i,\beta_{i+1}} = \sc_{\beta_i,\beta_{i+1}}\, \rho , ~~~
\rho \, \sc_{\gamma_i,\gamma_{i+1}} = \sc_{\gamma_{i-1},\gamma_{i}}\, \rho, ~~~
\rho \, \sc_{\delta_i,\delta_{i+1}} = \sc_{\delta_{i},\delta_{i+1}}\, \rho, ~~~
\end{equation}
so the cluster mapping class group contains
\begin{equation}
\Wext\left(A_{2}^{(1)} \times A_{2}^{(1)} \right) \times \Wext\left(A_{1}^{(1)}\times A_{1}^{(1)}\right) \rtimes \mathbb{Z} \subset \Gq.
\end{equation}

We conjecture that for general rectangular $N\times M$ Newton polygon, cluster mapping class group contains subgroup (\ref{eq:GqGeneralNM}). Construction of generators for general $N$ and $M$ is straightforward, by 'jumps over zig-zags' as in example.

In the case $N=M$ there is also one additional 'external' generator $R_{\pi/2}$ of order 4, which rotates bipartite graph by $\pi/2$
\begin{equation}
R_{\pi/2}:~ \ya_{i,a} \mapsto \yb_{-a,i}, ~~~
\yb_{i,a} \mapsto \ya_{1-a,i}.
\end{equation}

In the case $N=2K$ or $M=2K$ there is another additional  'external' generator, which flips the rectangle.
\paragraph{Discrete flows.}{
The group $\Gq$ contains lattice of discrete flows of rank $B-3$, where $B=2N+2M$ is the number of boundary integral points of Newton polygon. It consists of four pairwise commuting lattices contained in two copies of $W(A^{(1)}_{N-1})=\mathbb{Z}^{N-1} \rtimes W(A_{N-1})$ and two copies of $W(A^{(1)}_{M-1})=\mathbb{Z}^{M-1} \rtimes W(A_{M-1})$, and generator $(\rho)^{\mathrm{lcm}(N,M)}$ where $\mathrm{lcm}(N,M)$ is the least common multiple of $N$ and $M$. The lattice is generated by elements $\mathrm{T}_{\z_i,\z_{i+1}}$ which take pair of adjacent strands, wind them up in opposite directions over cylinder and put on the initial places, if one imagine $W(A^{(1)}_{N-1}),W(A^{(1)}_{M-1})$	 as a groups acting by permutations of strands on cylinder.
For $(3,2)$ example $\beta$-piece of $\Gq$ can be presented as $W(A^{(1)}_{2})=\mathbb{Z}^{2} \rtimes W(A_{2})$ with $\mathbb{Z}^2$ and $W(A_2)$ generated by
\begin{equation}
\label{Tbet}
\mathrm{T}_{\beta_1,\beta_2} = \sc_{\beta_1,\beta_2} \sc_{\beta_2,\beta_3} \sc_{\beta_3,\beta_1} \sc_{\beta_2,\beta_3},~~~ \mathrm{T}_{\beta_2,\beta_3} = \sc_{\beta_2,\beta_3} \sc_{\beta_3,\beta_1} \sc_{\beta_1,\beta_2} \sc_{\beta_3,\beta_1}
\end{equation}
and by
\begin{equation}
\sc_{\beta_1,\beta_2},\sc_{\beta_2,\beta_3}
\end{equation}
correspondingly.

One can find embedding of the lattice into the conjectured in \cite{FM:2014} group of discrete flows $\Gdp$. The group is defined to be a group of integral valued functions on boundary vertices of Newton polygon modulo sub-lattice $A$ generated by the restrictions from $\mathbb{Z}^2$ to the boundary of Newton polygon of affine functions $f(i,j) = a i +b j + c$, so $\Gdp = \mathbb{Z}^B/A$. For the case of rectangular Newton polygon one can easily find that $\Gdp = \mathbb{Z}^{B-3}\oplus \mathbb{Z}/N\mathbb{Z}\oplus \mathbb{Z}/M\mathbb{Z}$.

We are not going to stop here at the construction of embedding of generators of the lattice from $\Gq$ to $\Gdp$, interested reader can find it in Section 7.3 of \cite{FM:2014}. The computation comes from the consideration of the action of $\Gq$ on zig-zags presented in the next sub-section, and the result is that the image is $\Gd = \mathbb{Z}^{B-3}\oplus \mathbb{Z}/N\mathbb{Z}\oplus \mathbb{Z}/M\mathbb{Z}$, however the embedding is of the index $\Gdp/\Gd = \mathrm{lcm}(N,M)$. The non-trivial index is due to the generator $(\rho)^{\mathrm{lcm}(N,M)}$ which coincides with the generator $\tau$ from \cite{FM:2014}.
}
\subsection{Monomial dynamics of Casimirs}

According to \cite{GK:2011} the lattice of Casimir functions $x_\gamma$ is generated by zig-zag paths\footnote{For details on definitions see Appendix \ref{ss:cluster}.}
\begin{equation}
\mathsf{Z}=\{\gamma\in \H1(\Gamma,\mathbb{Z})~|~ \varepsilon(\gamma,\cdot)=0 \}.
\end{equation}
As the skew-symmetric form $\varepsilon$ is intersection form on dual surface, this condition is equivalent to being trivial in dual surface $\hat{S}$ homologies. In order to be expressed in terms of cluster variables $\{\ya_{ij},\yb_{ij}\}$ Casimir should be also trivial in torus homologies, i.e. we are interested in subset
\begin{equation}
\mathsf{C} = \{\gamma \in \H1(\Gamma,\mathbb{Z})~|~[\gamma]=0\in \H1(\hat{S},\mathbb{Z}),~~ [\gamma]=0\in \H1(\mathbb{T}^2,\mathbb{Z})\}.
\end{equation}
As zig-zags and faces are drawn on torus $\mathsf{Z},\mathsf{F}\subset \H1(\Gamma,\mathbb{Z})$, they are constrained by $\prod_i x_{\z_i}=1$, where the product goes over all zig-zag paths and $\prod_i x_{f_i}=1$, where the product goes over all faces of bipartite graph on torus. To obtain non-trivial $q$-dynamic these constraints have to be relaxed to $\prod_i x_{f_i}=q\neq 1$ so that $x_\gamma$ now is an element of extension $\H1(\tilde{\Gamma},\mathbb{Z}) = \H1(\Gamma,\mathbb{Z})\oplus \mathbb{Q}^2_{\langle \omega,\hat{\omega}\rangle}$ with the relations $\sum_i f_i = \omega,~\sum_i \z  _i = \hat{\omega}$. In multiplicative notations this reads
\begin{equation}
\prod_i x_{f_i}=q, ~ \prod_i x_{\z_i}=\hat{q}
\end{equation}
where we have additionally defined $q = x_{\omega}$, $\hat{q} = x_{\hat{\omega}}$. Introduction of $q\neq 1$ can be considered by lifting of bipartite graph to universal cover of $\mathbb{T}^2$ which is $\mathbb{R}^2$.

Any variable $x_{\gamma},\, \gamma\in\mathsf{C}$ can be expressed via face variables $x_{f_i}$, which are cluster variables, and can be mutated by usual rules (\ref{eq:mutations}). However, there is no generic rule for mutation of variable associated with a single zig-zag, except for mutation in four-valent vertex identified with a 'spider move' \cite{GK:2011}. We propose
here the generic rule for transformation of zig-zags~\footnote{We abuse notations, denoting $x_{\z} = \z$ for zig-zags.} under the action of generators \rf{sbet},
namely, for the $N\times M$ rectangle:
\begin{equation}
\label{eq:ZigZagDynNM}
\begin{array}{rll}
\sc_{\alpha_i,\alpha_{i+1}}: &
\alpha_i \mapsto q^{\frac{1}{N}}\alpha_{i+1}, &
\alpha_{i+1} \mapsto q^{-\frac{1}{N}}\alpha_i,\\
\sc_{\beta_i,\beta_{i+1}}: &
\beta_i \mapsto q^{-\frac{1}{N}}\beta_{i+1}, &
\beta_{i+1} \mapsto q^{\frac{1}{N}}\beta_i,\\
\sc_{\gamma_a,\gamma_{a+1}}: &
\gamma_a \mapsto q^{\frac{1}{M}}\gamma_{a+1}, &
\gamma_{a+1} \mapsto q^{-\frac{1}{M}}\gamma_a,\\
\sc_{\delta_a,\delta_{a+1}}: &
\delta_a \mapsto q^{-\frac{1}{M}}\delta_{a+1}, &
\delta_{a+1} \mapsto q^{\frac{1}{M}}\delta_a,\\
\end{array}
\end{equation}
where $i=1,\ldots,N$, $a=1,\ldots,M$. The group $\Gq$ acts on the elements of $\mathsf{C}$, embedded in multiplicative lattice generated by zig-zags, precisely as Coxeter groups of $A_{K-1}$-type act on the root lattices embedded into $\mathbb{Z}^K$ (c.f. \cite{OS:2018, IIO}).

These rules basically come just
from consistency with mutation transformations for the elements of $\mathsf{C}$. There is a two-parametric family of transformations for zig-zag variables
\begin{equation}
\z \mapsto \z a^{[\z]_A} b^{[\z]_B},\mathrm{~if~ [ \z ]=([\z]_A,[\z]_B) - class~of~} \z \mathrm{~in~\H1}(\mathbb{T}^2,\mathbb{Z})
\end{equation}
which do not affect $\mathsf{C}$, since $\mathsf{C}$ consists of the combinations of zig-zags with zero class in torus homology. This ambiguity is fixed using the 'locality assumption' that zig-zags not adjacent to the transformed faces are not changed.\\

Let us now demonstrate, how formulas \rf{eq:ZigZagDynNM} come for
$(N,M)=(3,2)$ from consistency with transformations of $\mathsf{C}$, where one can introduce the following over-determined set of generators
\begin{equation}
\label{eq:casiChor}
\begin{array}{lll}
\Z_{\beta_1, \alpha_1} = \ya_{11}\ya_{12},&
\Z_{\beta_2, \alpha_2} = \ya_{21}\ya_{22},&
\Z_{\beta_3, \alpha_3} = \ya_{31}\ya_{32},\\
\Z_{\alpha_1, \beta_2} = (\yb_{11}\yb_{12})^{-1},&
\Z_{\alpha_2, \beta_3} = (\yb_{21}\yb_{22})^{-1},&
\Z_{\alpha_3, \beta_1} = (\yb_{31}\yb_{32})^{-1}
\end{array}
\end{equation}
\begin{equation}
\label{eq:casiCver}
\begin{array}{ll}
\Z_{\gamma_1, \delta_1} = (\ya_{11}\ya_{21}\ya_{31})^{-1}, &
\Z_{\delta_1, \gamma_2} = \yb_{12}\yb_{22}\yb_{32},\\
\Z_{\gamma_2, \delta_2} = (\ya_{12}\ya_{22}\ya_{32})^{-1}, &
\Z_{\delta_2, \gamma_1} = \yb_{11}\yb_{21}\yb_{31}
\end{array}
\end{equation}
satisfying
\begin{equation}
\label{eq:constrC}
\begin{array}{l}
\Z_{\beta_1, \alpha_1}\Z_{\beta_2, \alpha_2}\Z_{\beta_3, \alpha_3}\Z_{\gamma_1, \delta_1}\Z_{\gamma_2, \delta_2} = 1 \\
\Z_{\alpha_1, \beta_2}\Z_{\alpha_2, \beta_3}\Z_{\alpha_3, \beta_1}\Z_{\delta_1, \gamma_2}\Z_{\delta_2, \gamma_1} = 1 \\
\Z_{\beta_1, \alpha_1}\Z_{\beta_2, \alpha_2}\Z_{\beta_3, \alpha_3}(\Z_{\alpha_1, \beta_2}\Z_{\alpha_2, \beta_3}\Z_{\alpha_3, \beta_1})^{-1}=q=1.
\end{array}
\end{equation}
so that the number of independent Casimirs is seven. In the autonomous limit, these Casimirs reduce to $\Z_{\z,\z'}=\z \cdot \z'$, where $\z,\z'$ correspond to zig-zags $\{\alpha,\beta,\gamma,\delta\}$, expressed via the edge variables.  The transformation, for example, $\sc_{\beta_1,\beta_2}$ acts by
\begin{equation}
\label{eq:sb1b2}
\sc_{\beta_1,\beta_2}:~~~
\begin{array}{c}
\Z_{\beta_1, \alpha_1} \mapsto \Z_{\alpha_1, \beta_2},~~~
\Z_{\beta_2, \alpha_2} \mapsto \dfrac{\Z_{\beta_2, \alpha_2}\Z_{\beta_1, \alpha_1}}{\Z_{\alpha_1, \beta_2}},~~~
\Z_{\beta_3, \alpha_3} \mapsto \Z_{\beta_3, \alpha_3},\\
\Z_{\alpha_1,\beta_2} \mapsto \Z_{\beta_1, \alpha_1},~~~
\Z_{\alpha_2,\beta_3} \mapsto \Z_{\alpha_2, \beta_3},~~~
\Z_{\alpha_3, \beta_1} \mapsto \dfrac{\Z_{\alpha_3, \beta_1}\Z_{\alpha_1, \beta_2}}{\Z_{\beta_1,\alpha_1}}.
\end{array}
\end{equation}
and substituting here $\Z_{\z,\z'}=\z \cdot \z'$ one finds that the action of $\sc_{\beta_1,\beta_2}$ reduces just to permutation of $\beta_1$ and $\beta_2$, the same is true for the other generators $\sc_{\z_1,\z_2}$.

For $q\neq 1$ consider the generators $\mathrm{T}_{\beta_i,\beta_{i+1}}$ \rf{Tbet} which act trivially on $\mathsf{C}$ at all in the autonomous limit. One gets now
\begin{equation}
\label{eq:zigzagsFq}
\mathrm{T}_{\beta_1,\beta_2}:~~~
\begin{array}{c}
\Z_{\beta_1, \alpha_1} \mapsto q^{-1} \Z_{\beta_1, \alpha_1},~~~
\Z_{\beta_2, \alpha_2} \mapsto q \Z_{\beta_2, \alpha_2}\\
\Z_{\alpha_1,\beta_2} \mapsto q \Z_{\alpha_1, \beta_2},~~~
\Z_{\alpha_3, \beta_1} \mapsto q^{-1} \Z_{\alpha_3, \beta_1}
\end{array}
\end{equation}
where $q=\prod_{i,j} \ya_{ij}\yb_{ij}$. Again, after expressing the Casimirs via zig-zags, the action of $\mathrm{T}_{\beta_1,\beta_2}$ is equivalent to $\beta_1 \mapsto q^{-1} \beta_1,~\beta_2 \mapsto q \beta_2$. These formulas suggest that at $q\neq 1$ one can express generators of $\mathsf{C}$ via zig-zags and $q$ by~\footnote{The fractional powers of $q$ in these formulas can be restored using the 'magnetic flux'
interpretation for $q\neq 1$ in non-autonomous case. This interpretation is also consistent with the fact that zig-zags with the different orientations collect fluxes of different signs.}
\begin{equation}
\label{eq:zigzagsZq1}
\begin{array}{ccc}
\Z_{\beta_1, \alpha_1} = q^{\frac{1}{6}}\beta_1\alpha_1,&
\Z_{\beta_2, \alpha_2} = q^{\frac{1}{6}}\beta_2\alpha_2,&
\Z_{\beta_3, \alpha_3} = q^{\frac{1}{6}}\beta_3\alpha_3,\\
\Z_{\alpha_1, \beta_2} = q^{-\frac{1}{6}}\alpha_1\beta_2,&
\Z_{\alpha_2, \beta_3} = q^{-\frac{1}{6}}\alpha_2\beta_3,&
\Z_{\alpha_3, \beta_1} = q^{-\frac{1}{6}}\alpha_3\beta_1
\end{array}
\end{equation}
\begin{equation}
\label{eq:zigzagsZq2}
\begin{array}{cccc}
\Z_{\gamma_1, \delta_1} = q^{-\frac{1}{4}}\gamma_1\delta_1, &
\Z_{\delta_1, \gamma_2} = q^{\frac{1}{4}}\delta_1\gamma_2, &
\Z_{\gamma_2, \delta_2} = q^{-\frac{1}{4}}\gamma_2\delta_2, &
\Z_{\delta_2, \gamma_1} = q^{\frac{1}{4}}\delta_2\gamma_1
\end{array}
\end{equation}
which are consistent with constraints (\ref{eq:constrC}) with $q\neq 1$ if one assumed\footnote{
One can incorporate $\hat{q}\neq 1$ consistently modifying formulas (\ref{eq:casiChor}) and (\ref{eq:casiCver}) by
\begin{equation}
\begin{array}{c}
\Z_{\beta_1, \alpha_1} = \hat{q}^{\frac{1}{5}} \ya_{11}\ya_{12},~~~
\Z_{\beta_2, \alpha_2} = \hat{q}^{\frac{1}{5}}\ya_{21}\ya_{22},~~~
\Z_{\beta_3, \alpha_3} = \hat{q}^{\frac{1}{5}}\ya_{31}\ya_{32},\\
\Z_{\alpha_1, \beta_2} = \hat{q}^{\frac{1}{5}}(\yb_{11}\yb_{12})^{-1},~~~
\Z_{\alpha_2, \beta_3} = \hat{q}^{\frac{1}{5}}(\yb_{21}\yb_{22})^{-1},~~~
\Z_{\alpha_3, \beta_1} = \hat{q}^{\frac{1}{5}}(\yb_{31}\yb_{32})^{-1}
\end{array}
\end{equation}
\begin{equation}
\begin{array}{cc}
\Z_{\gamma_1, \delta_1} = \hat{q}^{\frac{1}{5}}(\ya_{11}\ya_{21}\ya_{31})^{-1},~~~
\Z_{\delta_1, \gamma_2} = \hat{q}^{\frac{1}{5}}\yb_{12}\yb_{22}\yb_{32},\\
\Z_{\gamma_2, \delta_2} = \hat{q}^{\frac{1}{5}}(\ya_{12}\ya_{22}\ya_{32})^{-1},~~~
\Z_{\delta_2, \gamma_1} = \hat{q}^{\frac{1}{5}}\yb_{11}\yb_{21}\yb_{31}
\end{array}
\end{equation}
However, as a meaning of this extension is not clear, we will assume $\hat{q}=1$ in the following.
} $\alpha_{1}\alpha_{2}\alpha_{3}\beta_{1}\beta_{2}\beta_{3}\gamma_1 \gamma_2 \delta_1 \delta_2=\hat{q}=1$. Comparison of  transformation (\ref{eq:sb1b2}) with (\ref{eq:zigzagsZq1}) and (\ref{eq:zigzagsZq2}) leads to the formulas \rf{eq:ZigZagDynNM} for $(N,M)=(3,2)$. The action of remaining generators is defined by
\begin{equation}
\label{eq:ZigZagDynExt}
\begin{array}{lllll}
\Lambda_h: &
\alpha_i \mapsto \alpha_i, & \beta_i \mapsto \beta_i, & \gamma_a \to \gamma_{a+1}, & \delta_a \mapsto \delta_{a+1}, \\
\Lambda_v: &
\alpha_i \mapsto \alpha_{i+1}, & \beta_i \to \beta_{i+1}, & \gamma_a \mapsto \gamma_a, & \delta_a \mapsto \delta_a , \\
\rho: &
\alpha_i \mapsto q^{-\frac{1}{N}}\alpha_{i-1}, & \beta_i \mapsto \beta_i, & \gamma_a \to q^{\frac{1}{M}}\gamma_{a+1}, & \delta_a \mapsto \delta_a.
\end{array}
\end{equation}

\Remark{ Specialities of $N=2$ or $M=2$ case.}{
\ \\
It is well known (see e.g. \cite{Gaiotto}, eq.(3.70)) that spectral curves with a Newton polygon being $2\times N$ rectangle can be mapped to the 'triangle ones' with the catheti $2$ and $2N$ (see Fig.~\ref{fig:polygonRectToTriang}) just by change of variables. Namely, equation
\begin{equation}
S(\lambda,\mu) = P^{+}_N(\mu) \lambda^2 + P_N(\mu)\lambda + P^{-}_N(\mu) = 0
\end{equation}
under $\lambda \mapsto P^{-}_N(\mu)\cdot \lambda^{-1}$ than $S(\lambda,\mu) \mapsto \lambda^{2} P^{-}_N(\mu)^{-1}S(\lambda,\mu) $ turns into
\begin{equation}
S(\lambda,\mu) = \lambda^2 + P_N(\mu)\lambda + P^{+}_N(\mu)P^{-}_N(\mu) = 0.
\end{equation}
\begin{figure}[!h]
\begin{center}
\begin{tikzpicture}

\tikzmath{\Lx=3;\Ly=2;\xs=0;\ys=0;\d=1;}

\draw[dotted,step=\d,shift={(0,0)}] (\xs,\ys) grid (\xs+\d*\Lx, \ys+\d*\Ly);

\begin{scope}[thick]
	\draw (\xs,\ys+\d)--(\xs,\ys);
	\draw (\xs,\ys+2*\d)--(\xs,\ys+\d);	
	
	\draw (\xs+\d,\ys+\d*\Ly)--(\xs,\ys+\d*\Ly);
	\draw (\xs+2*\d,\ys+\d*\Ly)--(\xs+\d,\ys+\d*\Ly);	
	\draw (\xs+2*\d,\ys+\d*\Ly)--(\xs+3*\d,\ys+\d*\Ly);	

	\draw (\xs+\d*\Lx,\ys)--(\xs+\d*\Lx,\ys+\d);
	\draw (\xs+\d*\Lx,\ys+\d)--(\xs+\d*\Lx,\ys+\d*2);
	
	\draw (\xs,\ys)--(\xs+\d,\ys);
	\draw (\xs+\d,\ys)--(\xs+2*\d,\ys);
	\draw (\xs+3*\d,\ys)--(\xs+2*\d,\ys);

\end{scope}

\foreach \x in {0,...,\Lx}
	\foreach \y in {0,...,\Ly}
		\draw[fill] (\xs+\d*\x,\ys+\d*\y) circle[radius=0.05];

\draw[->, thick] (3.5,1)--(4.5,1);

\tikzmath{\Lx=6;\Ly=2;\xs=5;\ys=0;\d=1;}

\draw[dotted,step=\d,shift={(0,0)}] (\xs,\ys) grid (\xs+\d*\Lx, \ys+\d*\Ly);

\begin{scope}[thick]
	\draw (\xs,\ys)--(\xs,\ys+2*\d);	
	\draw (\xs,\ys)--(\xs+6*\d,\ys);	
	\draw (\xs,\ys+2*\d)--(\xs+6*\d,\ys);	

\end{scope}

\foreach \x in {0,...,\Lx}
	\foreach \y in {0,...,\Ly}
		\draw[fill] (\xs+\d*\x,\ys+\d*\y) circle[radius=0.05];

\end{tikzpicture}

\end{center}
\caption{Transformation from rectangle to triangle for $(3,2)$ case.}
\label{fig:polygonRectToTriang}
\end{figure}
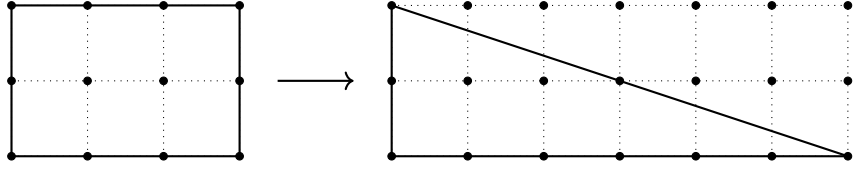
}
For a corresponding cluster integrable system the Poisson quiver from Fig.~\ref{fig:quivNM} can be transformed into the form drawn in Fig.~\ref{fig:quivtriang} -- more common for 'triangular' polygons, studied in detail in \cite{OS:2018}.
\begin{figure}[!h]
\begin{center}
\begin{tikzpicture}

\tikzmath{\Lx=6;\d=1.5;}

\foreach \x in {1,...,\Lx}{
	\draw[quiverVertex] (\d*\x,0) circle;
	\draw[quiverVertex] (\d*\x,\d) circle;
	};

\begin{scope}

\clip(0.4*\d,-0.1*\d) rectangle (\d*\Lx+0.6*\d, 2.1*\d);

\tikzmath{
		for \x in {1,...,\Lx+1}{			
		{
		\draw[styleArrow, thick] (\d*\x,0) -- (\d*\x-\d,0);
		\draw[styleArrow, thick] (\d*\x,\d) -- (\d*\x-\d,\d);
		\draw[styleArrowShifted, thick] (\d*\x-\d,0)--(\d*\x,\d);
		\draw[styleArrowShifted, thick] (\d*\x-\d,\d)--(\d*\x,0);
		};
			};
		};
\end{scope}

\tikzmath{
		for \x in {1,...,int(round(\Lx/2))}{
			\xc = int(round(\Lx/2-\x+1));
			{			
				\node at (2*\d*\x-\d,-0.3*\d) {$\ybbold_{\xc 1}$};
				\node at (2*\d*\x-\d,\d+0.3*\d) {$\ybbold_{\xc 2}$};
				\node at (2*\d*\x,-0.3*\d) {$\yabold_{\xc 2}$};
				\node at (2*\d*\x,\d+0.3*\d) {$\yabold_{\xc 1}$};
			};
		};
};

\end{tikzpicture}

\end{center}
\caption{Quiver for $(3,2)$ case represented in 'triangular' form.}
\label{fig:quivtriang}
\end{figure}
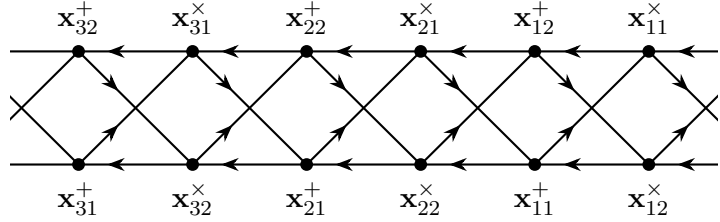
This correspondence results in the 'enhancement' of the symmetry group~\footnote{We are grateful to Y.Yamada for clarification of this point.}: a pair of commuting Weyl groups $A_{N-1}^{(1)} \times A_{N-1}^{(1)}$ is now embedded into larger group $A_{2N-1}^{(1)}$ with the generators
\begin{equation}
\sc_{\alpha_i \beta_{i+1}} = s^{\lb,\lb}_{i1,i2} \ \mu^{\lb}_{i1}\mu^{\lb}_{i2}, \ \ \ \sc_{\beta_{i} \alpha_{i}} = s^{\la,\la}_{i1,i2} \ \mu^{\la}_{i1} \mu^{\la}_{i2},\ \ \ \ i=1,\ldots,N
\end{equation}
Embedding $A_{N-1}^{(1)} \times A_{N-1}^{(1)}\to A_{2N-1}^{(1)}$ is provided by
\begin{equation}
\sc_{\beta_i,\beta_{i+1}} = \sc_{\beta_i \alpha_i} \sc_{\alpha_i \beta_{i+1}} \sc_{\beta_i \alpha_i}, \ \ \
\sc_{\alpha_i,\alpha_{i+1}} = \sc_{\alpha_i \beta_{i+1}} \sc_{\beta_{i+1} \alpha_{i+1}} \sc_{\alpha_i \beta_{i+1}}
\end{equation}
and commutativity of $\sc_{\alpha_i,\alpha_{i+1}}$ and $\sc_{\beta_i,\beta_{i+1}}$ just follows form the relations on 'elementary' generators $\sc_{\beta_i \alpha_i}$, $\sc_{\alpha_i \beta_{i+1}}$. The generators of $A_{2N-1}^{(1)}$ also commute with $\sc_{\delta_i,\delta_{i+1}},\ \sc_{\gamma_i,\gamma_{i+1}}$. The generator $\rho$ is also absorbed. Now it is not a primitive one, but can be presented as a composition
\begin{equation}
\rho = \Lambda_{h}\tilde{\Lambda}_v\prod\limits_{i=1}^{N} \sc_{\alpha_i, \beta_{i+1}}
\end{equation}
where we used 'root' from $\Lambda_v$
\begin{equation}
\tilde{\Lambda}_v:~
\ya_{ia} \mapsto \yb_{i-1,a}, ~~~
\yb_{ia} \mapsto \ya_{i,a},~~~ \mathrm{so} ~~~ \Lambda_v = (\tilde{\Lambda}_v)^2
\end{equation}
so there are no extra 'dimensions' in the lattice of the flows.\\

The action of the enhanced group on Casimirs can be constructed in a way similar to generic case. For example, for the generator $\sc_{\alpha_1,\beta_{2}}$ in $(N,M)=(3,2)$ case from
\begin{equation}
\sc_{\alpha_1,\beta_{2}}:
\begin{array}{llll}
Z_{\beta_1,\alpha_1} \mapsto \dfrac{Z_{\beta_1,\alpha_1}}{Z_{\alpha_1,\beta_2}} &
Z_{\alpha_1,\beta_2} \mapsto \dfrac{1}{Z_{\alpha_1,\beta_2}}, &
Z_{\beta_2,\alpha_2} \mapsto \dfrac{Z_{\beta_2,\alpha_2}}{Z_{\alpha_1,\beta_2}}  \\
Z_{\gamma_1, \delta_1} \mapsto Z_{\alpha_1,\beta_2}Z_{\gamma_1, \delta_1}, &
Z_{\delta_1, \gamma_2} \mapsto Z_{\alpha_1,\beta_2}Z_{\delta_1, \gamma_2}, &
Z_{\gamma_2, \delta_2} \mapsto Z_{\alpha_1,\beta_2}Z_{\gamma_2, \delta_2}, &
Z_{\delta_2, \gamma_1} \mapsto Z_{\alpha_1,\beta_2} Z_{\delta_2, \gamma_1}
\end{array}
\end{equation}
one gets for the zig-zags
\begin{equation}
\sc_{\alpha_1,\beta_{2}}: ~~~
\alpha_1 \mapsto q^{\frac{1}{6}} \beta_2^{-1},\ \ \
\beta_2 \mapsto q^{\frac{1}{6}} \alpha_1^{-1},\ \ \
\gamma_a \delta_a \mapsto q^{-\frac{1}{6}}\alpha_1 \beta_2 \gamma_a \delta_a.
\end{equation}
which contains now 'inversion' of zig-zag, since $\alpha_i$ and $\beta_i$ correspond to the opposite classes in $\H1(\mathbb{T}^2,\mathbb{Z})$. Generally, for the action of $A_{5}^{(1)}$ on zig-zags one gets
\begin{equation}
\begin{array}{rl}
\sc_{\alpha_i \beta_{i+1}}: & \alpha_i \mapsto q^{\frac{1}{6}}\beta_{i+1}^{-1},\ \ \ \beta_{i+1} \mapsto q^{\frac{1}{6}}\alpha_{i}^{-1}, \ \ \  \gamma_a \delta_a \mapsto q^{-\frac{1}{6}}\alpha_i \beta_{i+1} \gamma_a \delta_a \\
\sc_{\beta_{i}\alpha_i}: & \alpha_i \mapsto q^{-\frac{1}{6}}\beta_{i}^{-1},\ \ \ \beta_{i} \mapsto q^{-\frac{1}{6}}\alpha_{i}^{-1}, \ \ \ \gamma_a \delta_a \mapsto q^{\frac{1}{6}}\alpha_i \beta_{i} \gamma_a \delta_a.
\end{array}
\end{equation}

\Remark{Further enhancement for $N=M=2$ 'small square'.}{
\ \\
The group $\Gq$ for this case can be identified with the $q$-Painlev\' e VI symmetry group $W(D_5^{(1)})$ (see e.g. \cite{BGM:2017}). It corresponds naively to the 'double' symmetry enhancement
\begin{equation}
A_{1,\alpha}^{(1)} \times A_{1,\beta}^{(1)}\to A_{3,\alpha,\beta}^{(1)}, \ \ \ A_{1,\gamma}^{(1)} \times A_{1,\delta}^{(1)}\to A_{3,\gamma,\delta}^{(1)}.
\end{equation}
but it turns out moreover that generators of the 'new' extended groups do not commute. For example the generators $\sc_{\alpha_1\beta_2}$ and $\sc_{\delta_1\gamma_2}$ satisfy
\begin{equation}
(\sc_{\alpha_1,\beta_2}\sc_{\gamma_1,\delta_1})^3=1
\end{equation}
and this non-commutativity results in gluing of Dynkin quivers as shown on Fig.~\ref{fig:D5dynkin}.

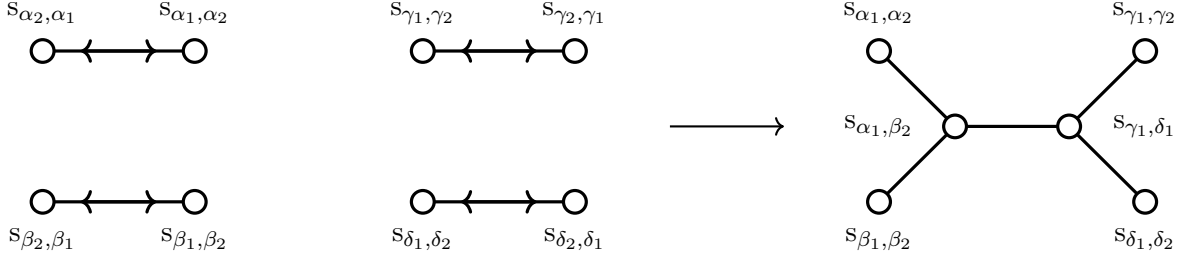
\begin{figure}[h!]
\begin{center}
\begin{tikzpicture}
\tikzmath{\w=1.2;\r=0.15;};

\tikzmath{\xs=-9;\ys=0;};

\draw[line width = \w pt] (\xs-3,\ys+1)--(\xs-1,\ys+1);
\draw[<->, line width = \w pt] (\xs-2.5,\ys+1)--(\xs-1.5,\ys+1);

\draw[line width = \w pt] (\xs-3,\ys-1)--(\xs-1,\ys-1);
\draw[<->, line width = \w pt] (\xs-2.5,\ys-1)--(\xs-1.5,\ys-1);

\draw[line width = \w pt] (\xs+4,\ys+1)--(\xs+2,\ys+1);
\draw[<->, line width = \w pt] (\xs+3.5,\ys+1)--(\xs+2.5,\ys+1);

\draw[line width = \w pt] (\xs+4,\ys-1)--(\xs+2,\ys-1);
\draw[<->, line width = \w pt] (\xs+3.5,\ys-1)--(\xs+2.5,\ys-1);

\draw[fill=white,line width = \w pt,radius=\r] (\xs-1,\ys+1) circle;
\draw[fill=white,line width = \w pt,radius=\r] (\xs-3,\ys+1) circle;

\draw[fill=white,line width = \w pt,radius=\r] (\xs-1,\ys-1) circle;
\draw[fill=white,line width = \w pt,radius=\r] (\xs-3,\ys-1) circle;

\draw[fill=white,line width = \w pt,radius=\r] (\xs+2,\ys+1) circle;
\draw[fill=white,line width = \w pt,radius=\r] (\xs+4,\ys+1) circle;

\draw[fill=white,line width = \w pt,radius=\r] (\xs+2,\ys-1) circle;
\draw[fill=white,line width = \w pt,radius=\r] (\xs+4,\ys-1) circle;

\node[thick] at (\xs-1,\ys+1.5) {$\sc_{\alpha_1,\alpha_2}$};
\node[thick] at (\xs-3,\ys+1.5) {$\sc_{\alpha_2,\alpha_1}$};

\node[thick] at (\xs-1,\ys-1.5) {$\sc_{\beta_1,\beta_2}$};
\node[thick] at (\xs-3,\ys-1.5) {$\sc_{\beta_2,\beta_1}$};

\node[thick] at (\xs+2,\ys+1.5) {$\sc_{\gamma_1,\gamma_2}$};
\node[thick] at (\xs+4,\ys+1.5) {$\sc_{\gamma_2,\gamma_1}$};

\node[thick] at (\xs+2,\ys-1.5) {$\sc_{\delta_1,\delta_2}$};
\node[thick] at (\xs+4,\ys-1.5) {$\sc_{\delta_2,\delta_1}$};

\draw[thick, ->] (-3.75,0)--(-2.25,0);

\tikzmath{\xs=0;\ys=0;};
\draw[line width = \w pt] (\xs,\xs)--(\xs-1,\ys+1);
\draw[line width = \w pt] (\xs,\xs)--(\xs-1,\ys-1);
\draw[line width = \w pt] (\xs,\xs)--(\xs+1.5,\ys);
\draw[line width = \w pt] (\xs+1.5,\xs)--(\xs+2.5,\ys+1);
\draw[line width = \w pt] (\xs+1.5,\xs)--(\xs+2.5,\ys-1);

\draw[fill=white,line width = \w pt,radius=\r] (\xs,\ys) circle;
\draw[fill=white,line width = \w pt,radius=\r] (\xs-1,\ys+1) circle;
\draw[fill=white,line width = \w pt,radius=\r] (\xs-1,\ys-1) circle;

\draw[fill=white,line width = \w pt,radius=\r] (\xs+1.5,\ys) circle;
\draw[fill=white,line width = \w pt,radius=\r] (\xs+2.5,\ys+1) circle;
\draw[fill=white,line width = \w pt,radius=\r] (\xs+2.5,\ys-1) circle;

\node[thick] at (\xs-1,\ys+1.5) {$\sc_{\alpha_1,\alpha_2}$};
\node[thick] at (\xs-1,\ys) {$\sc_{\alpha_1,\beta_2}$};
\node[thick] at (\xs-1,\ys-1.5) {$\sc_{\beta_1,\beta_2}$};
\node[thick] at (\xs+2.5,\ys+1.5) {$\sc_{\gamma_1,\gamma_2}$};
\node[thick] at (\xs+2.5,\ys) {$\sc_{\gamma_1,\delta_1}$};
\node[thick] at (\xs+2.5,\ys-1.5) {$\sc_{\delta_1,\delta_2}$};
\end{tikzpicture}
\end{center}
\caption{Symmetry enhancement from $W\left(A_1^{(1)}\times A_1^{(1)} \times A_1^{(1)} \times A_1^{(1)}\right)$ to $ W(D_5^{(1)})$.}
\label{fig:D5dynkin}
\end{figure}
Another cluster realization of $W\left(D_5^{(1)}\right)$ has been proposed in \cite{BGM:2017}, given by generators
\begin{equation}
\begin{array}{lll}
\sc_0 = s_{11,22}^{\lb,\lb}, &
\sc_1 = s_{12,21}^{\lb,\lb}, &
\sc_2 = s_{11,12}^{\lb,\lb}\mu_{11}^{\lb}\mu_{12}^{\lb} \\
\sc_5 = s_{21,12}^{\la,\la}, &
\sc_4 = s_{11,22}^{\la,\la}, &
\sc_3 = s_{11,21}^{\la,\la}\mu_{11}^{\la}\mu_{21}^{\la}
\end{array}.
\end{equation}
in terms of mutations of the same bipartite graph. In our notation this generators are
\begin{equation}
\begin{array}{lll}
\sc_0 = \sc_{\alpha_1\beta_2}\sc_{\delta_1\gamma_1} \sc_{\gamma_1\gamma_2}\sc_{\delta_1\gamma_1}\sc_{\alpha_1\beta_2}, &
\sc_1 = \sc_{\alpha_1\beta_2}\sc_{\delta_1\gamma_1} \sc_{\delta_1\delta_2}\sc_{\delta_1\gamma_1}\sc_{\alpha_1\beta_2}, &
\sc_2 = \sc_{\alpha_1\beta_2} \\
\sc_5 = \sc_{\gamma_1\delta_1}\sc_{\alpha_1\beta_2}\sc_{\beta_1\beta_2}\sc_{\alpha_1\beta_2}\sc_{\gamma_1\delta_1}, &
\sc_4 = \sc_{\gamma_1\delta_1}\sc_{\alpha_1\beta_2}\sc_{\alpha_1\alpha_2}\sc_{\alpha_1\beta_2}\sc_{\gamma_1\delta_1}, &
\sc_3 = \sc_{\gamma_1\delta_1}.
\end{array}
\end{equation}}
Two presentations can be mapped one to another by conjugation by $\sc_{\alpha_1\beta_2} \sc_{\gamma_1\delta_1} \sc_{\alpha_1\beta_2}$.
\subsection{Towards bilinear equations}
\label{ss:bilinear}

Let us finally turn to the issue of bilinear equations for the cluster tau-functions or
$A$-cluster variables. We postpone rigorous discussion of this issue for a separate publication,
but demonstrate here, how Hirota bilinear equations can arise in the systems, corresponding
to rectangle Newton polygons.

The simplest example of bilinear equations is provided by spider moves, or mutations in a four-valent vertex of the Poisson quiver, see also Fig.~\ref{fig:spiderbip} in Appendix for the transformation of corresponding piece of a bipartite graph. Such transformation induce the only change in $\tau$-variables, which (for all unit coefficients)
\begin{equation}
\tau_0 \mapsto \bar{\tau}_0 = \dfrac{\tau_1 \tau_3 + \tau_2 \tau_4}{\tau_0}~~~ \mathrm{or}~~~ \tau_0 \bar{\tau}_0 = \tau_1 \tau_3 + \tau_2 \tau_4.
\end{equation}
obviously leads to bilinear equation. However, there is no \emph{a priori} reason to get bilinear equations
from generic action by an element of $\Gq$. For example, a single mutation in a six-valent vertex rather
leads to relation, which symbolically has form
\begin{equation}
\tau \bar{\tau} = \tau^3 + \tau^3
\end{equation}
instead of bilinear. Sometimes one can get nevertheless a bilinear relation for a sequence of mutations
without no \emph{a priori} reason for them to hold, see e.g. sect.~2.8 of \cite{BGM:2018}. We are going to
show in this section that the same happens for the transformations, induced by the zig-zag permutations (e.g. $\{\sc_{\beta_i,\beta_{i+1}}\}$ or $\{\sc_{\gamma_a,\gamma_{a+1}}\}$), constructing their explicit action
on tau-variables.

For $A$-cluster algebras\footnote{For the definition of $A$-cluster algebra with coefficients and transition from $\mathcal{X}$ to $A$-cluster algebra see Appendix \ref{ss:clusterAalg}.} the role of Casimir functions is played by 'coefficients' \cite{FZ:2006}, taking values in some tropical semi-field $\mathbb{P}$, see also discussion in \cite{BGM:2018}. For the case of rectangle Newton polygons we label the generators of $\mathbb{P}$ by zig-zags (together with $q$), i.e.
\begin{equation}
\mathbb{P} = \mathrm{Trop}(q,\{\c_{\alpha_i},\c_{\beta_i}\}_{i=1,...,N},\{\c_{\gamma_a},\c_{\delta_a}\}_{i=1,...,M}).
\end{equation}
so that the coefficients are expressed by
\begin{equation}
\cofa_{ia} = q^{\frac{1}{NM}}\dfrac{(\c_{\alpha_{i}}\c_{\beta_{i}})^{\frac{1}{M}}}{(\c_{\gamma_{a}}\c_{\delta_{a}})^{\frac{1}{N}}},~~~
\cofb_{ia} = q^{\frac{1}{NM}}\dfrac{(\c_{\gamma_{a}}\c_{\delta_{a-1}})^{\frac{1}{N}}}{(\c_{\alpha_{i}}\c_{\beta_{i+1}})^{\frac{1}{M}}}.
\end{equation}
The action of transformations $\sc_{\z_i,\z_{i+1}}$ on coefficients in this basis is equivalent to the action on generators of $\mathbb{P}$ like in \rf{eq:ZigZagDynNM} on zig-zags, i.e.
\begin{equation}
\label{eq:ZigZagDynA}
\begin{array}{rll}
\sc_{\alpha_i,\alpha_{i+1}}: &
\c_{\alpha_i} \mapsto q^{\frac{1}{N}}\c_{\alpha_{i+1}}, &
\c_{\alpha_{i+1}} \mapsto q^{-\frac{1}{N}} \c_{\alpha_i},\\
\sc_{\beta_i,\beta_{i+1}}: &
\c_{\beta_i} \mapsto q^{-\frac{1}{N}}\c_{\beta_{i+1}}, &
\c_{\beta_{i+1}} \mapsto q^{\frac{1}{N}}\c_{\beta_i},\\
\sc_{\gamma_a,\gamma_{a+1}}: &
\c_{\gamma_a} \mapsto q^{\frac{1}{M}}\c_{\gamma_{a+1}}, &
\c_{\gamma_{a+1}} \mapsto q^{-\frac{1}{M}}\c_{\gamma_a},\\
\sc_{\delta_a,\delta_{a+1}}: &
\c_{\delta_a} \mapsto q^{-\frac{1}{M}}\c_{\delta_{a+1}}, &
\c_{\delta_{a+1}} \mapsto q^{\frac{1}{M}}\c_{\delta_a}.\\
\end{array}
\end{equation}
Coefficients could be encoded by 'frozen' vertices of quiver. This suggests principle that we assign frozen variables to faces of dual surface, corresponding to zig-zag variables, while mutable variables - to faces of original torus.

Let us now present an example of the action of the generator $\sc_{\beta_1,\beta_2}$ on $\tau$-variables in $(N,M)=(3,2)$ case. An explicit computation gives
\begin{equation}
{\def\arraystretch{2.2}
\left(
\begin{array}{c}
\dfrac{\tbb_{11}}{\tb_{11}}\\
\dfrac{\tba_{11}}{\ta_{11}}\\
\dfrac{\tbb_{12}}{\tb_{12}}\\
\dfrac{\tba_{12}}{\ta_{12}}
\end{array}
\right)
}
=
{\def\arraystretch{2.2}
\left(
\begin{array}{cccc}
\c_{\beta_2}^{\frac{1}{2}} &
q^{\frac{1}{12}}(\c_{\beta_1}\c_{\beta_2})^{\frac{1}{2}}  &
q^{\frac{2}{12}}\c_{\beta_1}^{\frac{1}{2}} &
q^{\frac{3}{12}}\c_{\beta_1} \\
q^{\frac{3}{12}}\c_{\beta_1}^{\frac{1}{2}} &
\c_{\beta_2} &
q^{\frac{1}{12}}\c_{\beta_2}^{\frac{1}{2}}  &
q^{\frac{2}{12}}(\c_{\beta_1}\c_{\beta_2})^{\frac{1}{2}} \\
q^{\frac{2}{12}}\c_{\beta_1}^{\frac{1}{2}} &
q^{\frac{3}{12}}\c_{\beta_1} &
\c_{\beta_2}^{\frac{1}{2}} &
q^{\frac{1}{12}}(\c_{\beta_1}\c_{\beta_2})^{\frac{1}{2}} \\
q^{\frac{1}{12}}\c_{\beta_2}^{\frac{1}{2}} &
q^{\frac{2}{12}}(\c_{\beta_1}\c_{\beta_2})^{\frac{1}{2}} &
q^{\frac{3}{12}}\c_{\beta_1}^{\frac{1}{2}} &
\c_{\beta_2}
\end{array}
\right)
}\cdot C \cdot
{\def\arraystretch{2.2}
\left(
\begin{array}{c}
\dfrac{\tb_{31}\ta_{21}}{\tb_{11}\ta_{11}}\\
\dfrac{\tb_{32}\ta_{21}}{\tb_{12}\ta_{11}}\\
\dfrac{\tb_{32}\ta_{22}}{\tb_{12}\ta_{12}}\\
\dfrac{\tb_{31}\ta_{22}}{\tb_{11}\ta_{12}}
\end{array}
\right)
}
\end{equation}
where $C=\mathrm{diag}\left((\c_{\gamma_1}\c_{\delta})^{\frac{1}{3}},\c_{\alpha_1}^{\frac{1}{2}}(\c_{\delta}/\c_{\delta_1})^{\frac{1}{3}},(\c_{\gamma_2}\c_{\delta})^{\frac{1}{3}},\c_{\alpha_1}^{\frac{1}{2}}(\c_{\delta}/\c_{\delta_2})^{\frac{1}{3}}\right)$, $~\c_{\delta} = \c_{\delta_1} \c_{\delta_2}$. The main point is that the matrix in the r.h.s. is nicely invertible so that these equations can be rewritten in \emph{bilinear form}
\begin{equation}
{\def\arraystretch{2.2}
\left\{
\begin{array}{l}
(\c_{\beta_2}-q^{\frac{1}{3}}\c_{\beta_1})(\c_{\delta}\c_{\gamma_1})^{\frac{1}{3}}\tb_{31}\ta_{21}
=
\c_{\beta_2}^{\frac{1}{2}}\tbb_{11}\ta_{11} - q^{\frac{1}{12}}\c_{\beta_1}^{\frac{1}{2}}\tba_{11}\tb_{11}
\\
(\c_{\beta_2}-q^{\frac{1}{3}}\c_{\beta_1})(\c_{\delta}/\c_{\delta_1})^{\frac{1}{3}}\tb_{32}\ta_{21}
=
\c_{\alpha_1}^{-\frac{1}{2}}\tba_{11}\tb_{12} - q^{\frac{1}{12}}\c_{\alpha_1}^{-\frac{1}{2}}\tbb_{12}\ta_{11}
\\
(\c_{\beta_2}-q^{\frac{1}{3}}\c_{\beta_1})(\c_{\delta}\c_{\gamma_2})^{\frac{1}{3}}\tb_{32}\ta_{22}
=
\c_{\beta_2}^{\frac{1}{2}}\tbb_{12}\ta_{12} - q^{\frac{1}{12}}\c_{\beta_1}^{\frac{1}{2}}\tba_{12}\tb_{12}
\\
(\c_{\beta_2}-q^{\frac{1}{3}}\c_{\beta_1})(\c_{\delta}/\c_{\delta_2})^{\frac{1}{3}}\tb_{31}\ta_{22}
=
\c_{\alpha_1}^{-\frac{1}{2}}\tba_{12}\tb_{11} - q^{\frac{1}{12}}\c_{\alpha_1}^{-\frac{1}{2}}\tbb_{11}\ta_{12}
\end{array}
\right.
}.
\end{equation}
This is actually a generic phenomenon for the zig-zag generators: the same happens, for example, for the generator $\sc_{\delta_1,\delta_2}$ from another component of $\Gq$. One gets explicitly for
the transformation of A-cluster variables
\begin{equation}
t_1	= C_1\cdot C_2 \cdot t_2,
\end{equation}
where
\begin{equation}
{\def\arraystretch{2.5}
\begin{array}{l}
t_1 =\left(
\dfrac{\tbb_{32}}{\tb_{32}} \ \ \
\dfrac{\tba_{32}}{\ta_{32}} \ \ \
\dfrac{\tbb_{22}}{\tb_{22}} \ \ \
\dfrac{\tba_{22}}{\ta_{22}} \ \ \
\dfrac{\tbb_{12}}{\tb_{12}} \ \ \
\dfrac{\tba_{12}}{\ta_{12}}
\right)^{T}
\\
t_2 = \left(
\dfrac{\tb_{31}\ta_{31}}{\tb_{32}\ta_{32}} \ \ \
\dfrac{\tb_{21}\ta_{31}}{\tb_{22}\ta_{32}} \ \ \
\dfrac{\tb_{21}\ta_{21}}{\tb_{22}\ta_{22}} \ \ \
\dfrac{\tb_{11}\ta_{21}}{\tb_{12}\ta_{22}} \ \ \
\dfrac{\tb_{11}\ta_{11}}{\tb_{12}\ta_{12}} \ \ \
\dfrac{\tb_{31}\ta_{11}}{\tb_{32}\ta_{12}}
\right)^{T}
\end{array}
}
\end{equation}
\begin{equation}
C_1=
{\def\arraystretch{2}
\left(
\begin{array}{cccccc}
\c_{\delta_2} &
q^{\frac{1}{12}}\c_{\delta_2}^{\frac{2}{3}} &
q^{\frac{2}{12}}(\c_{\delta_1} \c_{\delta_2}^2)^{\frac{1}{3}} &
q^{\frac{3}{12}}(\c_{\delta_1}\c_{\delta_2})^{\frac{1}{3}} &
q^{\frac{4}{12}}(\c_{\delta_1}^2\c_{\delta_2})^{\frac{1}{3}} &
q^{\frac{5}{12}}\c_{\delta_1}^{\frac{2}{3}}
\\
q^{\frac{5}{12}}\c_{\delta_1} &
\c_{\delta_2}^{\frac{2}{3}} &
q^{\frac{1}{12}}(\c_{\delta_1}\c_{\delta_2}^2)^{\frac{1}{3}} &
q^{\frac{2}{12}}(\c_{\delta_1} \c_{\delta_2})^{\frac{1}{3}} &
q^{\frac{3}{12}}(\c_{\delta_1}^2 \c_{\delta_2})^{\frac{1}{3}} &
q^{\frac{4}{12}}\c_{\delta_1}^{\frac{2}{3}}
\\
q^{\frac{4}{12}}(\c_{\delta_1}^2 \c_{\delta_2})^{\frac{1}{3}} &
q^{\frac{5}{12}}\c_{\delta_1}^{\frac{2}{3}} &
\c_{\delta_2} &
q^{\frac{1}{12}}\c_{\delta_2}^{\frac{2}{3}} &
q^{\frac{2}{12}}(\c_{\delta_1} \c_{\delta_2}^2)^{\frac{1}{3}} &
q^{\frac{3}{12}}(\c_{\delta_1} \c_{\delta_2}^2)^{\frac{1}{3}}
\\
q^{\frac{3}{12}}(\c_{\delta_1}^2 \c_{\delta_2})^{\frac{1}{3}} &
q^{\frac{4}{12}}\c_{\delta_1}^{\frac{2}{3}} &
q^{\frac{5}{12}}\c_{\delta_1} &
\c_{\delta_2}^{\frac{2}{3}} &
q^{\frac{1}{12}}(\c_{\delta_1}\c_{\delta_2}^2)^{\frac{1}{3}} &
q^{\frac{2}{12}}(\c_{\delta_1} \c_{\delta_2})^{\frac{1}{3}}
\\
q^{\frac{2}{12}}(\c_{\delta_1} \c_{\delta_2}^2)^{\frac{1}{3}} &
q^{\frac{3}{12}}(\c_{\delta_1} \c_{\delta_2})^{\frac{1}{3}} &
q^{\frac{4}{12}}(\c_{\delta_1}^2 \c_{\delta_2})^{\frac{1}{3}} &
q^{\frac{5}{12}}\c_{\delta_1}^{\frac{2}{3}} &
\c_{\delta_2} &
q^{\frac{1}{12}}\c_{\delta_2}^{\frac{2}{3}}
\\
q^{\frac{1}{12}}(\c_{\delta_1} \c_{\delta_2}^2)^{\frac{1}{3}} &
q^{\frac{2}{12}}(\c_{\delta_1} \c_{\delta_2})^{\frac{1}{3}} &
q^{\frac{3}{12}}(\c_{\delta_1}^2 \c_{\delta_2})^{\frac{1}{3}} &
q^{\frac{4}{12}}\c_{\delta_1}^{\frac{2}{3}} &
q^{\frac{5}{12}}\c_{\delta_1} &
\c_{\delta_2}^{\frac{2}{3}}
\end{array}
\right)
}
\end{equation}
\begin{equation}
C_2=\mathrm{diag}\left(
(\c_{\alpha}/\c_{\alpha_3})^{\frac{1}{2}}\c_{\gamma_2}^{\frac{1}{3}},
(\c_{\alpha}\c_{\beta_3})^{\frac{1}{2}},
(\c_{\alpha}/\c_{\alpha_2})^{\frac{1}{2}}\c_{\gamma_2}^{\frac{1}{3}},
(\c_{\alpha}\c_{\beta_2})^{\frac{1}{2}},
(\c_{\alpha}/\c_{\alpha_1})^{\frac{1}{2}}\c_{\gamma_2}^{\frac{1}{3}},
(\c_{\alpha}\c_{\beta_1})^{\frac{1}{2}}
\right)
\end{equation}
with $\c_{\alpha} = \c_{\alpha_1} \c_{\alpha_2} \c_{\alpha_3}$. Again, inverting matrix $C_1$ we end up with the set of bilinear equations
\begin{equation}
{\def\arraystretch{2.2}
\left\{
\begin{array}{l}
(\c_{\delta_2}-q^{\frac{1}{2}}\c_{\delta_1})(\c_{\alpha}/\c_{\alpha_3})^{\frac{1}{2}}\tb_{31}\ta_{31}
=
\c_{\gamma_2}^{-\frac{1}{3}}\tbb_{32}\ta_{32} - q^{\frac{1}{12}}\c_{\gamma_2}^{-\frac{1}{3}}\tba_{32}\tb_{32}
\\
(\c_{\delta_2}-q^{\frac{1}{2}}\c_{\delta_1})(\c_{\alpha}\c_{\beta_3})^{\frac{1}{2}}\tb_{21}\ta_{31}
=
\c_{\delta_2}^{\frac{1}{3}}\tba_{32}\tb_{22} - q^{\frac{1}{12}}\c_{\delta_1}^{\frac{1}{3}}\tbb_{22}\ta_{32}
\\
(\c_{\delta_2}-q^{\frac{1}{2}}\c_{\delta_1})(\c_{\alpha}/\c_{\alpha_2})^{\frac{1}{2}}\tb_{21}\ta_{21}
=
\c_{\gamma_2}^{-\frac{1}{3}}\tbb_{22}\ta_{22} - q^{\frac{1}{12}}\c_{\gamma_2}^{-\frac{1}{3}}\tba_{22}\tb_{22}
\\
(\c_{\delta_2}-q^{\frac{1}{2}}\c_{\delta_1})(\c_{\alpha}\c_{\beta_2})^{\frac{1}{2}}\tb_{11}\ta_{21}
=
\c_{\delta_2}^{\frac{1}{3}}\tba_{22}\tb_{12} - q^{\frac{1}{12}}\c_{\delta_1}^{\frac{1}{3}}\tbb_{12}\ta_{22}
\\
(\c_{\delta_2}-q^{\frac{1}{2}}\c_{\delta_1})(\c_{\alpha}/\c_{\alpha_1})^{\frac{1}{2}}\tb_{11}\ta_{11}
=
\c_{\gamma_2}^{-\frac{1}{3}}\tbb_{12}\ta_{12} - q^{\frac{1}{12}}\c_{\gamma_2}^{-\frac{1}{3}}\tba_{12}\tb_{12}
\\
(\c_{\delta_2}-q^{\frac{1}{2}}\c_{\delta_1})(\c_{\alpha}\c_{\beta_1})^{\frac{1}{2}}\tb_{31}\ta_{11}
=
\c_{\delta_2}^{\frac{1}{3}}\tba_{12}\tb_{32} - q^{\frac{1}{12}}\c_{\delta_1}^{\frac{1}{3}}\tbb_{32}\ta_{12}
\end{array}
\right.
}
\end{equation}
It remains yet unclear, how to derive bilinear equations systematically for compositions of elements of $\Gq$. We are going to return to this issue together with discussion of their solutions elsewhere.

\section{Conclusion}

In this paper we have presented extra evidence that cluster integrable systems provide convenient framework for the description of 5d super-symmetric Yang-Mills theory. It has been shown that cluster integrable systems with the Newton polygons $SA(2,\mathbb{Z})$-equivalent to the $N\times M$ rectangles are isomorphic to the $\XXZ$-like spin chains of rank $M$ on $N$ sites (or vice versa) on the 'lowest orbit'. Due to special symmetry of the Kasteleyn operators, defining spectral curves of these systems, it turns to be possible to express the Lax operators of spin chain in terms of the X-cluster variables. Inhomogeneities and twists of the chain can be expressed via (part of) the zig-zag paths on the Goncharov-Kenyon bipartite graphs.

Rectangle Newton polygons generally correspond to linear quiver gauge theories \cite{BPTY:2012} so that
inhomogeneities, 'on-site' Casimirs and twists define the fundamental and bi-fundamental masses together
with the bare couplings on the Yang-Mills side. The proposed cluster description possesses obvious symmetry
between the structure in horizontal and vertical directions so that one gets a natural spectral (or fiber-base or length-rank) duality, interchanging also the rank and length of spin chains. The parameters in dual descriptions are related in the following way:
\begin{equation}
\begin{array}{r|cc}
\mathrm{Length} & N & M \\
\mathrm{Rank} & M & N\\
\mathrm{Spectral~parameters~of~L} & \mu & \lambda \\
\mathrm{Second~spectral~parameter} & \lambda & \mu \\
\mathrm{Twist} & Q=\mathrm{diag}\left( (\delta_1/\gamma_1)^{1/2},...,(\delta_M/\gamma_M)^{1/2}\right) & \tilde{Q}=\mathrm{diag}\left((\alpha_1/\delta_1)^{1/2},...,(\alpha_N/\delta_N)^{1/2}\right)\\
\mathrm{Inhomogeneities} & \left(\beta_1,...,\beta_N\right) & \left(\delta_1^{-1},...,\delta_M^{-1}\right)
\end{array}
\end{equation}
Shear shift of one side of a Newton polygon to the shape of $N\times M$ parallelogram results in the multiplication of the monodromy operator of the spin chain by the cyclic twist matrix.

We have found that the cluster mapping class group $\Gq$ for the 'spin-chain class' always contains a subgroup isomorphic to
\begin{equation}
\Wext\left(A_{N-1,\alpha}^{(1)} \times A_{N-1,\beta}^{(1)} \right) \times \Wext\left(A_{M-1,\gamma}^{(1)}\times A_{M-1,\delta}^{(1)}\right) \rtimes \mathbb{Z}
\end{equation}
whose generators act on zig-zag paths by permutations. Moreover, their action on the $A$-cluster variables gives rise to the $q-$difference bilinear relations. The symmetry enhancement happens in the case $N=2$ (or $M=2$) and results in 'gluing' of two copies of $A_{N-1}^{(1)}$ into $A_{2N-1}^{(1)}$. If both $N=M=2$ the symmetry $\Wext\left(A_{1}^{(1)} \times A_{1}^{(1)} \right) \times \Wext\left(A_{1}^{(1)}\times A_{1}^{(1)}\right) \rtimes \mathbb{Z}$ enhances to the $D_5^{(1)}$ symmetry group of q-PVI equation.

Our first results in this direction actually produce more question than give answers. The following obvious questions (at least!) can be addressed for the further investigations:
\begin{itemize}
\item Trivial rank-$N$ spin chain on a single site once twisted becomes spectrally dual to relativistic Toda chain, see sect.~\ref{ss:twT}. Can we similarly identify the spectral duals of the twisted chains of arbitrary lengths and twists, whose Newton polygons are generic parallelograms -- or even extend this to generic four-gons? This question is also very interesting on the gauge-theory side, where by now only the hyperelliptic case of 'generalized Toda' (four boundary points and all internal points are lying on one line -- pure $SU(N)$ theory with the CS term) was studied in \cite{BGM:2018}.
\item We have derived in sect.~\ref{ss:bilinear} the bilinear relations, coming out of the action of a single 'permutation' generator of $\Gq$ on $A$-cluster variables, acting by transpositions on zig-zags. Is there any systematic principle to derive bilinear equations for compositions of such transformations?
\item In \cite{BS:2016:1}, \cite{BGM:2017}, \cite{Jimbo:2017}, \cite{BGM:2018}, \cite{BS:2018} and \cite{MN:2018} the solutions for q-difference bilinear equations and their degenerations, arising from certain cluster integrable systems, were found in terms of Fourier-transformed Nekrasov functions for the corresponding 5d gauge theories. As partition functions for the 5d linear quiver gauge theories are well known, a natural further step is to show that they solve the bilinear equations found here (and their hypothetical generalizations!).
\end{itemize}

We would like to thank V.Fock, T.Ishibashi, A.Liashyk, A.Shchechkin, I.Vilkoviskiy, Y.Yamada and Y.Zenkevich for the stimulating discussions, and S.Kharchev for providing us with unpublished notes \cite{Kharchev}. We are especially grateful to M.Bershtein and P.Gavrylenko for very important discussions during the whole period of work on the project and their useful remarks after reading the manuscript. The results of Section 4 were obtained under support of the RSF under the grant 19-11-00275, the work of AM was also partially supported by RFBR grant 18-01-00460.
This research was carried out within the HSE University Basic Research Program and funded (jointly) by the Russian Academic Excellence Project '5-100'. 

\appendix	

\section{From quantum to classical spin chains}
\label{ss:limit}
We start with brief overview of the generalities of $R$-matrix formalism for the quantum $\XXZ$-like spin chain, following \cite{Kharchev}, \cite{QGdefs}, and then turn to the details of their classical limit, presenting explicit formulas for the rank 2 and 3 cases.

\subsection{Quantum XXZ spin chain}

Quantum $\mathfrak{gl}_M$ spin chain of $\XXZ$ type can be defined using quantum monodromy matrix $T(u)$, satisfying so-called $RTT$-relations:
\begin{equation}
\label{RTTq}
R(u,v)\cdot (T(u)\otimes \M1) \cdot (\M1 \otimes T(v)) =(\M1 \otimes T(v)) \cdot (T(u)\otimes \M1) \cdot R(u,v).
\end{equation}
Here $T(u)=\sum_{i,j=1}^{M} E_{ij}\otimes T_{ij}(u)$ (two-sided formal series in spectral parameter $u$, as we consider double of $RTT$ algebra) acts in the product of 'auxiliary' space $V=\mathbb{C}^M$ ($E_{ij}\in \mathrm{End}(V)$ -- standard matrix units), and 'quantum' Hilbert space of the chain $\mathcal{H}$, $T_{ij}(u)\in \mathrm{End}(\mathcal{H})$. The trigonometric R-matrix, $R\in \mathrm{End}(V\otimes V)$, is given by:
\begin{equation}
R(u,v) = \sum\limits_{i = 1}^{M} E_{ii}\otimes E_{ii} + \dfrac{\sqrt{u/v}-\sqrt{v/u}}{q\sqrt{u/v}-q^{-1}\sqrt{v/u}} \sum\limits_{i\neq j} E_{ii}\otimes E_{jj} +
\end{equation}
$$+\dfrac{q-q^{-1}}{q\sqrt{u/v}-q^{-1}\sqrt{v/u}} \sum\limits_{i\neq j} (u/v)^{-\frac{1}{2} s_{ij}}E_{ij}\otimes E_{ji}$$
with the sign-factors
\be
\label{sifa}
s_{ij} =
\left\{
\begin{array}{lc}
+1, & i > j \\
-1, & i < j \\
0, & i = j
\end{array}
\right.
\ee
The integrals of motion of the chain come from the coefficients of expansion of the transfer matrix $\mathcal{T}(u) = \mathrm{tr}_V ~T(u) = \sum_{k \in \mathbb{Z}} u^{k} H_k$. Their commutativity immediately follows from the $RTT$-relations \rf{RTTq}:
\begin{equation}
0=[\mathcal{T}(u),\mathcal{T}(v)] = \sum\limits_{m,n=-\infty}^{+\infty} u^{m} v^{n}  [H_m, H_n] ~\Rightarrow ~ [H_m, H_n]=0.
\end{equation}
For the higher-rank case $M>2$ this does not provide the complete set of commuting Hamiltonians, one has to add higher transfer matrices, or take the coefficients of the so-called quantum spectral curve equation
\begin{equation}
S(\lambda,\mu) = \mathrm{det}_q\left(T(\mu)-\lambda\cdot \M1\right)=\sum_{i,j} H_{ij} \lambda^{i} \mu^{j}
\end{equation}
with the quantum determinant is defined by
\begin{equation}
\mathrm{det}_q F(u) = \sum_{{\sigma}}(-1)^{\mathrm{sign}(\sigma)}F_{1,\sigma(1)}(u) F_{2,\sigma(2)}(uq)...F_{M,\sigma(M)}(uq^M)
\end{equation}
The center of the $RTT$ algebra \rf{RTTq} ($[T_{ij}(u),C_k]=0~~ \forall i,j$) is generated by quantum determinant of $T$-operator: $\mathrm{det}_q T(u) = \sum C_k u^k$.

A seminal statement, proven in \cite{QEquivAffine}, claims that the algebra defined by \rf{RTTq} is isomorphic to the quantum affine algebra $\UqhM$. More precisely, there is an isomorphism between the algebra, generated by modes of the currents
\begin{equation}
L^{\pm}(z)=\sum_{k=0}^{+\infty}\sum_{i,j=1}^{M}E_{ij}\otimes L_{i,j}^{\pm}[\pm k]z^{\mp k}.
\end{equation}
satisfying the $RTT$-relations \rf{RTTq}
\begin{equation}
R(u,v)\cdot (L^{\pm}(u)\otimes \M1) \cdot (\M1 \otimes L^{\pm}(v)) =(\M1 \otimes L^{\pm}(v)) \cdot (L^{\pm}(u)\otimes \M1) \cdot R(u,v)
\end{equation}
together with
\be
\label{QAffRel3}
R(u q^{\frac{c}{2}},vq^{-\frac{c}{2}})\cdot (L^{+}(u)\otimes \M1) \cdot (\M1 \otimes L^{-}(v)) =(\M1 \otimes L^{-}(v)) \cdot (L^{+}(u)\otimes \M1) \cdot R(u q^{-\frac{c}{2}},vq^{\frac{c}{2}})
\\
L_{j,i}^{+}[0]=L_{i,j}^{-}[0]=0,\;\;\; L_{k,k}^{+}[0]L_{k,k}^{-}[0]=1,\;\; 1\leq i < j \leq M,\;\; 1 \leq k \leq M
\ee
and quantum affine algebra $\UqhM$ with the central extension $c$. Hence, different integrable systems, constructed from trigonometric $RTT$-relations can be identified with different representations of $\UqhM$.

Among these are spin chains on $N$ sites, exploiting the co-product property that if $T_1(u)$ and $T_2(v)$ both satisfy $RTT$-relations, and act in different quantum spaces, then so does $T=T_1(u)T_2(v)$, where the product is taken over the common auxiliary space. To construct a chain of length $N$, one has to associate an $L$-operator in some representation $\pi_k$ of the $\UqhM$ with each site of the chain, and construct quantum monodromy matrix, taking product in the common auxiliary space
\begin{equation}
T(u) = L^{(N)}(u/u_N)~...~L^{(1)}(u/u_1)Q
\end{equation}
where $u_k\in \mathbb{C}$ are so-called inhomogeneities, $L^{(k)}(u) = \pi_k(L^+)(u)$ and $Q\in \mathrm{End}(V)$ -- a 'twist' matrix, having trivial quantum space. Such approach allows to construct many non-trivial integrable systems by assigning to each site a simple representation of $\UqhM$. Conventional way to do so in case of zero central charge, is to apply first evaluation homomorphism $\mathcal{E}v_z: \UqhM \to \UqM$
\begin{equation}
\begin{array}{llll}
\mathcal{E}v_z(L^{+}_{i,j}[0]) = L^{+}_{i,j}~, &
i\leq j, &
\mathcal{E}v_z(L^{+}_{i,j}[0]) = 0~, &
i\geq j, \\
\mathcal{E}v_z(L^{-}_{i,j}[0]) = 0 ~,&
i\leq j, &
\mathcal{E}v_z(L^{-}_{i,j}[0]) = L^{-}_{i,j}~,&
i\geq j, \\
\mathcal{E}v_z(L^{\pm}_{i,j}[\pm k]) = z^{\pm k} L^{+}_{i,j}~, &
i\leq j,~~k>0&
\mathcal{E}v_z(L^{\pm}_{i,j}[\pm k]) = z^{\pm k} L^{-}_{i,j}~,&
i\geq j,~~k>0,\\
L_{ii}^{+}L_{ii}^{-}=1.
\end{array}
\end{equation}
Positive and negative currents could be collected to:
\begin{equation}
\begin{array}{l}
\mathcal{E}v_z (L^{+}(u)) =  \dfrac{1}{\sqrt{u/z}-\sqrt{z/u}} \left(\sqrt{\dfrac{u}{z}} L^{+} + \sqrt{\dfrac{z}{u}} L^{-}\right) = L_{\mathrm{ev}}(u/z) \\
\mathcal{E}v_z (L^{-}(u)) =  -\dfrac{1}{\sqrt{u/z}-\sqrt{z/u}} \left(\sqrt{\dfrac{u}{z}}  L^{+} + \sqrt{\dfrac{z}{u}} L^{-}\right) = -L_{\mathrm{ev}}(u/z)
\end{array},~~
L^{\pm} = \sum\limits_{i,j=1}^{M} E_{ij}\otimes L^{\pm}_{ij}.
\end{equation}
Homomorphism $\mathcal{E}v_z$ is well defined only for positive or negative sub-algebra at once as geometrical progression for $L^{+}(u)$ converges if $u/v<1$ and for $L^{-}(u)$ if $u/v>1$. But luckily we need only positive part for our purposes. Note also that $L^{+}$ as a matrix is upper triangular, while $L^{-}$ - lower triangular. If we substitute $L_{\mathrm{ev}}$ into $RTT$ relation we can decompose it by degrees of spectral parameters, and get for $L^{\pm}$
\begin{equation}
\label{RTTnonaf1}
R^{+}\cdot (L^{\pm}\otimes \M1) \cdot (\M1 \otimes L^{\pm}) =(\M1 \otimes L^{\pm}) \cdot (L^{\pm}\otimes \M1) \cdot R^{+}
\end{equation}
\begin{equation}
\label{RTTnonaf3}
R^{+}\cdot (L^{+}\otimes \M1) \cdot (\M1 \otimes L^{-}) =(\M1 \otimes L^{-}) \cdot (L^{+}\otimes \M1) \cdot R^{+}
\end{equation}
where we used that $R(u,v)$ can be represented as
\begin{equation}
\left(q\sqrt{u/v}-q^{-1}\sqrt{v/u}\right)\cdot R(u,v) = \sqrt{\dfrac{u}{v}}R^{+}+\sqrt{\dfrac{v}{u}}R^{-}
\end{equation}
with
\begin{equation}
R^{+}= q \sum\limits_{i = 1}^{M} E_{ii}\otimes E_{ii} + \sum\limits_{i\neq j} E_{ii}\otimes E_{jj} + (q-q^{-1})\sum\limits_{i < j} E_{ij}\otimes E_{ji}
\end{equation}
\begin{equation}
R^{-}= q^{-1} \sum\limits_{i = 1}^{M} E_{ii}\otimes E_{ii} + \sum\limits_{i\neq j} E_{ii}\otimes E_{jj} - (q-q^{-1})\sum\limits_{i > j} E_{ij}\otimes E_{ji}
\end{equation}
and relations
\begin{equation}
PR^{\pm}P= \left(R^{\mp}\right)^{-1},~~~ R^{+}-R^{-}=(q-q^{-1})P
\end{equation}
where $P=\sum_{i,j}E_{ij}\otimes E_{ji}$ - permutation matrix. The situation here is similar to the one, which was in the affine context: $RTT$ algebra, now without spectral parameters, generated by matrix elements of $L^{\pm}$ satisfying (\ref{RTTnonaf1})-(\ref{RTTnonaf3}) is isomorphic to the quantum group $\UqM$ in the Drinfeld-Jimbo definitions, if we put
\begin{equation}
L_{i,j}^{+} =
\left\{
\begin{array}{lc}
(q-q^{-1})e_{ji} q^{h_j} & i<j \\
q^{h_i} & i=j \\
0 & i>j
\end{array}
\right. ,~~~
L_{i,j}^{-} =
\left\{
\begin{array}{lc}
0 & i<j \\
q^{-h_i} & i=j \\
(q^{-1}-q)q^{-h_i} e_{ji} & i>j
\end{array}
\right.
\end{equation}
where $e_{i,j}$ - generators of $\UqM$, corresponding to the roots, $h_k$ - to the Cartan sub-algebra \cite{QEquiv, QEquivAffine}. Generators, corresponding to the simple roots $e_i=e_{i,i+1}$, $f_i = e_{i+1,i}$ satisfy relations which are deformation of the usual $\mathfrak{gl}_M$ relations
\begin{equation}
q^{h_a} e_{b} =q^{\delta_{ab}-\delta_{a,b+1}} e_{b}q^{h_a},~~~ q^{h_a} f_{b} =q^{\delta_{a,b+1}-\delta_{ab}} f_{b}q^{h_a}
\end{equation}
\begin{equation}
[e_{a}, f_{b}]=\delta_{ab} \dfrac{q^{h_a} q^{-h_{a+1}} - q^{-h_a} q^{h_{a+1}}}{q-q^{-1}},~~~ q^{h_a}q^{h_b}=q^{h_b}q^{h_a}
\end{equation}
and $q$-deformed Serre relations
\begin{equation}
f_a^2 f_{a-1} - (q+q^{-1}) f_a f_{a-1} f_a + f_{a-1} f_a^2=0,~~~ f_{a-1}^2 f_{a} - (q+q^{-1}) f_{a-1} f_{a} f_{a-1} + f_{a} f_{a-1}^2=0
\end{equation}
\begin{equation}
e_{a}^2 e_{a-1} - (q+q^{-1}) e_{a} e_{a-1} e_{a} + e_{a-1} e_{a}^2=0,~~~e_{a-1}^2 e_{a} - (q+q^{-1}) e_{a-1} e_{a} e_{a-1} + e_{a} e_{a-1}^2=0.
\end{equation}
Non-simple roots could be expressed using recurrence relation
\begin{equation}
e_{a,c} = e_{a,b} e_{b,c} - q e_{b,c} e_{a,b}~,~~~ e_{c,a} = e_{c,b} e_{b,a} - q^{-1} e_{b,a} e_{c,b}~,~~~ a<b<c.
\end{equation}
Algebra $\UqM$ plays here role of deformation of the usual algebra of spin variables. So considering any representation of $\UqM$, we construct $L$-operators satisfying $RTT$ relation, and consequently - integrable spin chain of $\XXZ$ type.

\subsection{Classical XXZ spin chain}

Classical limit of the quantum spin chain appears when we replace algebra of quantum operators in the limit $\hbar\to 0$ by some commutative Poisson algebra of classical dynamical variables. Poisson bracket is coming from the usual prescription
\begin{equation}
\label{eq:commprescr}
\{A,B\} = \lim\limits_{\hbar \to 0}~ \dfrac{\k}{\hbar}[\hat{A},\hat{B}].
\end{equation}
Quantum parameter can be introduced as $q = e^\hbar$. Additional parameter $\k\in \mathbb{C}^{\times}$ provides us with the family of non-isomorphic Poisson algebras. Classical $r$-matrix appears as a first order of the expansion $R(u,v,e^{\hbar})\to \M1\otimes \M1+\hbar\,r(u,v)+O(\hbar^2)$, and looks
\begin{equation}
\label{r1}
r(u,v) = -\dfrac{\sqrt{u/v}+\sqrt{v/u}}{\sqrt{u/v}-\sqrt{v/u}} \sum\limits_{i\neq j} E_{ii}\otimes E_{jj} + \dfrac{2}{\sqrt{u/v}-\sqrt{v/u}} \sum\limits_{i\neq j} (u/v)^{-\frac{1}{2} s_{ij}}E_{ij}\otimes E_{ji}
\end{equation}
Note that we don't assume any dependence of $u$ and $v$ on $\hbar$. This gives for the $\RLL$ relation
\begin{equation}
\label{eq:apprll}
\{L(u)\otimes L(v)\}=\k[L(u)\otimes L(v), r(u/v)],~~~ \{L(u)\otimes L(v)\}=\sum_{ijkl} \{L_{ij}(u)\otimes L_{kl}(v)\}~ E_{ij}\otimes E_{kl}
\end{equation}
with the classical $L$-operator
\begin{equation}
L_{\mathrm{cl+}}(u)=\lim\limits_{\hbar\to 0}~L_{\mathrm{ev}}(u,q=e^{\hbar})=
\end{equation}
$$=\dfrac{1}{u^{\frac{1}{2}}-u^{-\frac{1}{2}}}\left(\sum\limits_{i=1}^{M} \left( u^{\frac{1}{2}}e^{S_i^{0}}+u^{-\frac{1}{2}}e^{-S_i^{0}}\right) E_{ii} + 2u^{\frac{1}{2}} \sum\limits_{i<j} S_{ji} e^{S_j^{0}}E_{ij} -2u^{-\frac{1}{2}} \sum\limits_{i>j} S_{ji} e^{-S_i^{0}} E_{ij}\right)$$
For classical limit of $\UqM$, we assume
\begin{equation}
h_i = S_i^{0}/\hbar ,~~~~~ e_{ij} = S_{ij}/\hbar ~~(e_{i} = S^{+}_{i}/\hbar,~ f_{i} = S^{-}_{i}/\hbar).
\end{equation}
Their Poisson brackets and classical limit of Serre relations are
\begin{equation}
\{S_a^{0}, S_{b}^{\pm}\}= \pm\k(\delta_{ab}-\delta_{a,b+1})S_{b}^{\pm},~~~ \{S_a^+,S_b^-\} = \k\delta_{ab}\sinh(S_a^{0}-S_{a+1}^{0}),~~~ \{S^0_{a}, S^0_{b}\}=0
\end{equation}
\begin{equation}
\{S_a^+ ,\{S_a^+, S_{a-1}^{+}\}\}= \k (S_a^+)^2 S_{a-1}^+,~~~ \{S_{a-1}^+ ,\{S_{a-1}^+, S_{a}^{+}\}\}= \k (S_{a-1}^+)^2 S_{a}^+
\end{equation}
\begin{equation}
\{S_a^- ,\{S_a^-, S_{a-1}^{-}\}\}= \k(S_a^-)^2 S_{a-1}^-,~~~ \{S_{a-1}^- ,\{S_{a-1}^-, S_{a}^{-}\}\}= \k(S_{a-1}^-)^2 S_{a}^-.
\end{equation}
Generators, corresponding to non-simple roots are coming from
\begin{equation}
\k^{-1}\{S_{ab},S_{bc}\}= S_{ac} + S_{ab}S_{bc},~~~ a<b<c
\end{equation}
\begin{equation}
\k^{-1}\{S_{ab},S_{bc}\}= S_{ac} - S_{ab}S_{bc},~~~ a>b>c
\end{equation}
Different Poisson algebra appear, if we put $q=e^{-\hbar}$. $R$-matrix is tending to $R(u,v)\to \M1\otimes \M1-\hbar\,r(u,v)+O(\hbar^2)$ with the same $r$-matrix.  Classical $\RLL$ equation changes sign to
\begin{equation}
\label{r2}
\{L(u)\otimes L(v)\}=\k [r(u/v),L(u)\otimes L(v)].
\end{equation}
$L$-operator, represented through the classical $\UqM$ generators becomes
\begin{equation}
\label{Lspinminus}
L_{\mathrm{cl-}}(u)=\lim\limits_{\hbar\to 0}~L_{\mathrm{ev}}(u,q=e^{-\hbar})=
\end{equation}
$$=\dfrac{1}{u^{\frac{1}{2}}-u^{-\frac{1}{2}}}\left(\sum\limits_{i=1}^{M} \left( u^{\frac{1}{2}}e^{-S_i^{0}}+u^{-\frac{1}{2}}e^{S_i^{0}}\right)E_{ii} - 2u^{\frac{1}{2}} \sum\limits_{i<j} S_{ji} e^{-S_j^{0}}E_{ij} + 2u^{-\frac{1}{2}} \sum\limits_{i>j} S_{ji} e^{S_i^{0}}E_{ij}\right)$$
It is different from the $L_{\mathrm{cl+}}(u)$ only by the change of signs near generators
\begin{equation}
L_{\mathrm{cl+}}(u;S_{ij}, S^0_i) = L_{\mathrm{cl-}}(u;-S_{ij}, -S^0_i).
\end{equation}
Together, this results in that the only relations which change are
\begin{equation}
\k^{-1}\{S_{ab},S_{bc}\}= S_{ac} - S_{ab}S_{bc},~~~ a<b<c
\end{equation}
\begin{equation}
\k^{-1}\{S_{ab},S_{bc}\}= S_{ac} + S_{ab}S_{bc},~~~ a>b>c
\end{equation}
The general recipe to turn expressions from one algebra to another is to invert sign at all the simple generators, and at all the brackets. As a product of $L$-operators in both cases $q=e^{\pm \hbar}$ satisfies classical $RTT$ equation, we can construct classical monodromy matrix
\begin{equation}
T(u) = L^{(N)}(u/u_N)~...~L^{(1)}(u/u_1)Q.
\end{equation}
Commuting Hamiltonians of the classical $\XXZ$ spin chain are coefficients of the classical spectral curve 
\begin{equation}
S(\lambda,\mu)=\det(T(\mu)-\lambda)=\det T(\mu)+...+(-\lambda)^{n-1}\Tr T(\mu)+(-\lambda)^n
\end{equation}
where the Casimir functions are generated by $\det T(\mu)$.

\paragraph{\textbf{Example.}}{ Classical limit of $U_q(\mathfrak{gl}_2)$

Algebra $U_q(\mathfrak{gl}_2)$ has rank $2$ and has $4$ generators - $S_{12}= S^{+},~S_{21}=S^{-},~S_1^{0},~S_2^{0}$. Poisson brackets in the both cases $q=e^{\pm \hbar}$ are similar
\begin{equation}
\{S_1^0,S_2^0\}=0,~~~\{S_1^0,S^{\pm}\}=\pm \k S^{\pm},~~~ \{S_2^0,S^{\pm}\}=\mp \k S^{\pm},~~~ \{S^+,S^-\}= \k\sinh(S_1^0-S_2^0)
\end{equation}
Lax operators are
\begin{equation}
\label{Lax2x2Uq}
L_{\mathrm{cl}\pm}(u) =
\dfrac{1}{u^{\frac{1}{2}}-u^{-\frac{1}{2}}}
\left(
\begin{array}{cc}
u^{\frac{1}{2}}e^{\pm S_1^{0}}+u^{-\frac{1}{2}}e^{\mp S_1^{0}} &
 \pm 2u^{\frac{1}{2}} S^{-} e^{\pm S_2^{0}}\\
\mp 2u^{-\frac{1}{2}}S^{+} e^{\mp S_2^{0}} &
u^{\frac{1}{2}}e^{\pm S_2^{0}}+u^{-\frac{1}{2}}e^{\mp S_2^{0}}
\end{array}
\right).
\end{equation}
Casimirs are generated by
\begin{equation}
\det L_{\mathrm{cl\pm}}(u) = \dfrac{1}{(u^{\frac{1}{2}}-u^{-\frac{1}{2}})^2}\left(u e^{\pm S^0}+u^{-1} e^{\mp S^0}+2(\cosh(S^0_1-S^0_2)+2S^{+}S^{-})\right)
\end{equation}
So there are two independent Casimirs - total projection of the spin $S^0=S_1^{0}+S_2^{0}$ and 'square' of the spin (or quadratic Casimir) $C_2=\cosh(S^0_1-S^0_2)+2S^{+}S^{-}$. If their values are fixed, the resulting symplectic leaf is $2$-dimensional. Coordinates $S_1^{0}-S_2^0$ and $S^{+}/S^{-}$ could be chosen on it, for example. Form usual for $\mathfrak{sl}_2$ chain appears if we transform Lax operator by
\begin{equation}
\label{eq:2x2transfroms}
u \mapsto u \,e^{\mp(S_1^0+S_2^0)}, ~~~\mathrm{then} ~~~
L(u) \mapsto
(u^{\frac{1}{2}}-u^{-\frac{1}{2}})
\left(
\begin{array}{cc}
u^{-1/2} & 0\\
0 & 1
\end{array}
\right)
\cdot L(u) \cdot
\left(
\begin{array}{cc}
u^{1/2} & 0\\
0 & 1
\end{array}
\right)
\end{equation}
Introducing variables
\begin{equation}
\tilde{S}^{z}= \frac{1}{2}\left(S_1^{0}-S_2^{0}\right),~~~\tilde{S}^{+} = S^{+} e^{\pm \tilde{S}^z},~~~
\tilde{S}^{-} = S^{-} e^{\mp \tilde{S}^z}
\end{equation}
Lax operators becomes
\begin{equation}
\label{Lax2x2lit}
L_{\mathrm{cl}\pm}(u) =
\left(
\begin{array}{cc}
u^{\frac{1}{2}}e^{\pm \tilde{S}^{z}}+u^{-\frac{1}{2}}e^{\mp \tilde{S}^z} &
\pm 2 \tilde{S}^{-}\\
\mp 2\tilde{S}^{+}&
u^{\frac{1}{2}}e^{\mp \tilde{S}^{z}}+u^{-\frac{1}{2}}e^{\pm \tilde{S}^{z}}
\end{array}
\right).
\end{equation}
New variables satisfy almost the same relations
\begin{equation}
\label{eq:gl2spinlit}
\{\tilde{S}^z,\tilde{S}^{\pm}\}=\pm \k \tilde{S}^{z},~~~ \{\tilde{S}^+,\tilde{S}^-\}= \k\sinh(2 \tilde{S}^z).
\end{equation}
Note that only in $2\times 2$ case it is possible to eliminate spectral parameter from the off-diagonal elements of matrix.
}

\paragraph{\textbf{Example.}}{ Classical limit of $U_q(\mathfrak{gl}_3)$\\

Algebra $U_q(\mathfrak{gl}_3)$ has rank $3$ and $9$ generators. In the classical limit, generators corresponding to the simple positive roots are $S_{12}=S_1^{+},~ S_{23}=S_2^{+}$. If $q=e^{\hbar}$, only non-simple positive root $S_{13}$ can be defined using relation
\begin{equation}
\k^{-1} \{S_1^{+},S_2^{+}\} = S_1^{+} S_2^{+} + S_{13}.
\end{equation}
Substituting this into Serre relations
\begin{equation}
\{S_1^{+},\{S_1^{+},S_2^{+}\}\} = \k^2 (S_1^{+})^2 S_2^{+}, ~~~ \{S_2^{+},\{S_2^{+},S_1^{+}\}\} = \k^2 (S_2^{+})^2 S_1^{+}
\end{equation}
we can get two remaining brackets
\begin{equation}
\{S_1^{+}, S_{13}\} = -\k S_1^{+} S_{13}, ~~~ \{S_2^{+}, S_{13}\} = \k S_2^{+} S_{13}
\end{equation}
Analogously, $S_{21}=S_1^{-},~ S_{32}=S_2^{-},~ S_{31}$ are negative generators, with the brackets
\begin{equation}
\k^{-1}\{S_1^{-},S_2^{-}\} = S_1^{-} S_2^{-} - S_{31}
\end{equation}
\begin{equation}
\{S_1^{-}, S_{31}\} = -\k\, S_1^{-} S_{31}, ~~~ \{S_2^{-}, S_{31}\} = \k\, S_2^{-} S_{31}
\end{equation}
Cartan part has got three commuting generators $S_1^{0},~ S_2^{0},~S_3^{0}$. Their Poisson brackets with other generators are
\begin{equation}
\begin{array}{lll}
\{S_1^{0},S_1^\pm\}= \pm \k\, S_1^\pm, & \{S_2^{0},S_1^\pm\}= \mp \k\, S_1^\pm, & \{S_3^{0},S_1^\pm\}= 0 \\
\{S_1^{0},S_2^\pm\}= 0, & \{S_2^{0},S_2^\pm\}= \pm \k\, S_2^\pm, & \{S_3^{0},S_2^\pm\}= \mp \k\, S_2^\pm \\
\{S_1^{0},S_3^\pm \}= \mp \k\, S_3^\pm, & \{S_2^{0},S_3^\pm\}= 0, & \{S_3^{0},S_3^\pm\}= \pm \k\, S_3^\pm
\end{array}
\end{equation}
Or generally
\begin{equation}
\{S^0_k,S_{ij}\}= \k\,(\delta_{ik}-\delta_{jk})S_{ij}
\end{equation}
For the Poisson brackets between positive and negative simple roots we have got
\begin{equation}
\{S_1^{+},S_1^-\}=\k\, \sinh \left(S_1^{0}-S_2^{0}\right),~~~ \{S_2^{+},S_2^-\}=\k\, \sinh \left(S_2^{0}-S_3^{0}\right), ~~~ \{S_3^{+},S_3^-\}=\k\, \sinh \left(S_3^{0}-S_1^{0}\right)
\end{equation}
\begin{equation}
\{S_1^{\pm},S_2^\mp \}=0,~~~\{S_2^{\pm},S_3^\mp \}=0,~~~ \{S_3^{\pm},S_1^\mp \}=0
\end{equation}
Finally, for non-simple roots using Jacobi identity
\begin{equation}
\{S_{13},S_1^- \}=-\k\, S_2^{+}e^{S_1^0 - S_2^0},~~~\{S_{13},S_2^- \}=\k\, S_1^{+}e^{S_3^0 - S_2^0}
\end{equation}
\begin{equation}
\{S_{31},S_1^+ \}=-\k\, S_2^{-}e^{S_1^0 - S_2^0},~~~\{S_{31},S_2^+ \}=\k\, S_1^{-}e^{S_3^0 - S_2^0}
\end{equation}
\begin{equation}
\{S_{13},S_{31} \}=\k\, \sinh(S_1^{0}-S_3^{0})
\end{equation}
In agreement with the general prescription, the relations, which are being changing, if we choose $q=e^{-\hbar}$, are
\begin{equation}
\k^{-1}\,\{S_1^{+},S_2^{+}\} = -S_1^{+} S_2^{+} + S_{13}, ~~~ \{S_1^{+}, S_{13}\} = \k\, S_1^{+} S_{13}, ~~~ \{S_2^{+}, S_{13}\} = -\k\,S_2^{+} S_{13}
\end{equation}
\begin{equation}
\k^{-1}\, \{S_1^{-},S_2^{-}\} = -S_1^{-} S_2^{-} - S_{31}, ~~~ \{S_1^{-}, S_{31}\} = \k\, S_1^{-} S_{31}, ~~~ \{S_2^{-}, S_{31}\} = -\k\, S_2^{-} S_{31}
\end{equation}
\begin{equation}
\{S_{13},S_1^- \}=-\k\, S_2^{+}e^{-S_1^0 + S_2^0},~~~\{S_{13},S_2^- \}=\k\, S_1^{+}e^{-S_3^0 + S_2^0}
\end{equation}
\begin{equation}
\{S_{31},S_1^+ \}=-\k\, S_2^{-}e^{-S_1^0 + S_2^0},~~~\{S_{31},S_2^+ \}=\k\, S_1^{-}e^{-S_3^0 + S_2^0}
\end{equation}
The Lax operator is
\begin{equation}
L_{\mathrm{cl}+}(u) =
\dfrac{1}{u^{\frac{1}{2}}-u^{-\frac{1}{2}}}
\left(
\begin{array}{ccc}
u^{\frac{1}{2}}e^{S_1^{0}}+u^{-\frac{1}{2}}e^{-S_1^{0}} & 2u^{\frac{1}{2}} S^{-}_1 e^{S_2^{0}} & 2u^{\frac{1}{2}} S^{-}_{31} e^{S_3^{0}} \\
-2u^{-\frac{1}{2}}S^{+}_1 e^{-S_2^{0}} & u^{\frac{1}{2}}e^{S_2^{0}}+u^{-\frac{1}{2}}e^{-S_2^{0}} & 2u^{\frac{1}{2}} S^{-}_2 e^{S_3^{0}} \\
-2u^{-\frac{1}{2}}S^{+}_{13} e^{-S_3^{0}} & -2u^{-\frac{1}{2}}S^{+}_2 e^{-S_3^{0}} & u^{\frac{1}{2}}e^{S_3^{0}}+u^{-\frac{1}{2}}e^{-S_3^{0}}
\end{array}
\right)
\end{equation}
which gives generating function of the Casimirs:
$$C(u)=\det L_{\mathrm{cl+}}(u) = \dfrac{1}{(u^{\frac{1}{2}}-u^{-\frac{1}{2}})^3}\left(u^{3/2}e^{-S^0} + u^{1/2} e^{-S^0} C_3^{+} + u^{-1/2} e^{S^0} C_3^- + u^{-3/2} e^{S^0}\right)$$
where
\begin{equation}
S^0 =S^0_1+S^0_2+S^0_3
\end{equation}
\begin{equation}
C_3^{+}= e^{2S^0_1}+e^{2S^0_2}+e^{2S^0_3}+4 e^{S^0_1+S^0_2} S_{12} S_{21}+4 e^{S^0_1+S^0_3} S_{13} S_{31}+4 e^{S^0_2+S^0_3} S_{23} S_{32}+8 e^{S^0_1+S^0_3} S_{32} S_{21} S_{13}
\end{equation}
$$C_3^-=e^{-2S^0_3}+e^{-2S^0_2}+e^{-2S^0_1}+4 e^{-S^0_1-S^0_2} S_{12} S_{21}+4 e^{-S^0_1-S^0_3} S_{13} S_{31}+4 e^{-S^0_2-S^0_3} S_{23} S_{32}-8 e^{-S^0_1-S^0_3} S_{12} S_{23} S_{31}$$
note that if we pass from $L_{\mathrm{cl+}}$ to $L_{\mathrm{cl-}}$, Casimirs would change $S^0 \to -S^{0}$, $C_3^- \to C_3^+$, $C_3^+ \to C_3^-$.
}

\section{Cluster integrable system}
\label{ss:cluster}

Here we remind some basics of the cluster classical integrable systems, which have two equivalent constructions:
\begin{itemize}
	\item A combinatorial way \cite{GK:2011} assigns to a convex Newton polygon a bipartite graph $\Gamma$ on torus $\mathbb{T}^2$. The cluster variables $\{x_i\}$ are then just monodromies of $\mathbb{C}^\times$-valued connection around the faces of $\Gamma$. The spectral curve of integrable system is given by dimer's partition function on $\Gamma\subset\mathbb{T}^2$.
	\item A group theory construction exploits the Poisson submanifolds or double Bruhat cells, parameterized by cyclically irreducible words in $(W\times W)^\sharp$ (the co-extended double affine Weyl group of $\widehat{PGL}(N)$) \cite{FM:2014}. The structure of cluster Poisson variety is coming from restriction of standard trigonometric $r$-matrix bracket \cite{FG:2005}, while the integrals of motion are given by $\mathrm{Ad}$-invariant functions on the Poisson submanifold.
\end{itemize}

\subsection{$\mathcal{X}$-cluster variety}

$\mathcal{X}$-cluster variety is defined by the set of split toric charts $(\mathbb{C}^{\times})^d$ assigned with $d\times d$ integer-valued and skew-symmetric \textit{exchange matrix} $\varepsilon$. Such a pair is called \textit{seed} or \textit{cluster seed}. Coordinate functions $x_i\in \mathbb{C}^{\times}$ on these charts are Poisson variables with the logarithmically constant Poisson bracket
\begin{equation}
\label{clustbr}
\{x_i,x_j\} = \varepsilon_{ij} x_i x_j.
\end{equation}
The matrix $\varepsilon$ can be encoded by \textit{quiver} $\mathcal{Q}$ -- an oriented graph, whose vertices are labeled by cluster variables, and number of arrows from vertex $i$ to $j$ is equal\footnote{The arrows from any vertex to itself are forbidden and any two opposite arrows should be annihilated.} to $\varepsilon_{ij}$.
Generally the Poisson bracket \rf{clustbr} has Casimir functions $Z(x)$, $\{Z,x_i\}=0$ for any $x_i$, the number of independent Casimir functions coincides with the dimension of kernel of $\varepsilon$.

\begin{figure}[!h]
\centering
	\begin{tikzpicture}
	\node at (-2,1) {$\varepsilon=$};	
	\matrix at (0,1) [matrix of nodes,row sep=0, left delimiter={(}, right delimiter={)}]
	{
 		0 &  2 & 0 & -2 \\
 		-2 &  0 & 2 & 0 \\
 		0 &  -2 & 0 & 2 \\
 		2 &  0 & -2 & 0 \\
	};

	\tikzmath{\xs=3;\bend=15;};
	\draw[quiverVertex] (\xs+0,0) circle;
	\node at (\xs-0.3,2) {$1$};
	\draw[quiverVertex] (\xs+0,2) circle;
	\node at (\xs+2+0.3,2) {$2$};
	\draw[quiverVertex] (\xs+2,0) circle;
	\node at (\xs+2+0.3,0) {$3$};
	\draw[quiverVertex] (\xs+2,2) circle;
	\node at (\xs-0.3,0	) {$4$};

	\path (\xs,0) edge[styleQuiverEdge, bend right=\bend] (\xs,2);
	\path (\xs,0) edge[styleQuiverEdge, bend left=\bend] (\xs,2);	
	
	\path (\xs,2) edge[styleQuiverEdge, bend right=\bend] (\xs+2,2);
	\path (\xs,2) edge[styleQuiverEdge, bend left=\bend] (\xs+2,2);	
	
	\path (\xs+2,2) edge[styleQuiverEdge, bend right=\bend] (\xs+2,0);
	\path (\xs+2,2) edge[styleQuiverEdge, bend left=\bend] (\xs+2,0);
	
	\path (\xs+2,0) edge[styleQuiverEdge, bend right=\bend] (\xs,0);
	\path (\xs+2,0) edge[styleQuiverEdge, bend left=\bend] (\xs,0);
	
\end{tikzpicture}
\caption{An example of skew-symmetric matrix $\varepsilon$ (two-particle Toda chain) and corresponding quiver. Poisson bracket has two Casimir functions $\Z=x_1 x_3$ and $q=x_1 x_2 x_3 x_4$.}
\label{fig:todaex}
\end{figure}

The cluster seeds are glued together by special coordinate birational transformations -- \textit{mutations} $\mu_k: (\{x_i\},\varepsilon) \to (\{x'_i\},\varepsilon')$, assigned to each vertex of the quiver $\mathcal{Q}$ or variable $x_k$:
\begin{equation}
\label{eq:mutations}
x_i\mapsto x_i'=
\left\{
\begin{array}{lcl}
x_i^{-1} & , & i=k\\
x_i \left(1+x_k^{\mathrm{sgn}\, \varepsilon_{ik}}\right)^{\varepsilon_{ik}} & , & i\neq k
\end{array}
\right.
\end{equation}
\begin{equation}
\label{eq:mutationsquiv}
\varepsilon_{ij} \mapsto \varepsilon_{ij}' =
\left\{
\begin{array}{lcl}
 - \varepsilon_{ij} & , & i=k ~\mathrm{or}~j = k,\\
\varepsilon_{ij}+\dfrac{\varepsilon_{ik}|\varepsilon_{kj}|+\varepsilon_{kj}|\varepsilon_{ik}|}{2} & , &  \mathrm{otherwise}
\end{array}
\right. .
\end{equation}
The transformation of exchange matrix can be easily reformulated as transformation of corresponding quiver. Mutations are the Poisson maps, i.e.
\begin{equation}
\{x'_i,x'_j\}=\varepsilon'_{ij} x'_i x'_j,
\end{equation}
Collection of seeds glued by mutations is called \textit{$\mathcal{X}$-cluster variety}.

Denote by $\mathcal{G}_\mathcal{Q}$ the stabilizer of the quiver $\mathcal{Q}$ -- the group consisting of composition of mutations and permutations of the vertices, which maps quiver $\mathcal{Q}$ to itself: such transformations
nevertheless generate non-trivial maps of the cluster variables $\{x_i\}$. This group is called the \emph{mapping class group} of $\mathcal{X}$-cluster variety.

\subsection{From bipartite graph to cluster integrable system}

\paragraph{$\mathcal{X}$-cluster variety from bipartite graph.} For a bipartite graph $\Gamma \hookrightarrow \T2$ embedded in torus (without self-intersections) the vertices are divided into black and white subsets $B,W\subset C_0(\Gamma)$ so that the black vertices are connected only with the white ones and visa versa. We chose orientation of edges from black to white, and assume that graph is connected and all 2-valent vertices are contracted.
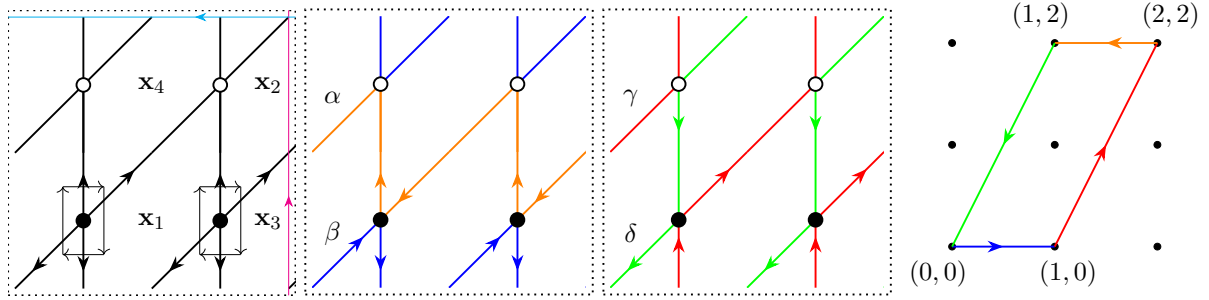
\begin{figure}[h!]
\begin{center}
\scalebox{0.9}{
\begin{tikzpicture}
\tikzmath{
	int \Lx,\Ly;
	\Lx=2;\Ly=1;
	\scale=1;
	\dofset=0.1;
	\cofset=0.1;
}

\begin{scope}[scale=\scale]

\clip(-\cofset,-\cofset) rectangle (2*\Lx + \cofset, 4*\Ly + \cofset);

\tikzmath{
		for \i in {0,...,\Lx}{
			{			
				\draw  (\i*2,0) pic[scale=\scale] {toda_one_text};
			};
		};
};

\node at (2, 3) {$\mathbf{x}_4$};
\node at (3.7, 3) {$\mathbf{x}_2$};
\node at (3.7, 1) {$\mathbf{x}_3$};
\node at (2, 1) {$\mathbf{x}_1$};

\draw[->] (0.7,1.5)--(1.3,1.5);
\draw[->] (1.3,1.5)--(1.3,0.5);
\draw[->] (1.3,0.5)--(0.7,0.5);
\draw[->] (0.7,0.5)--(0.7,1.5);

\draw[->] (2.7,1.5)--(3.3,1.5);
\draw[->] (3.3,1.5)--(3.3,0.5);
\draw[->] (3.3,0.5)--(2.7,0.5);
\draw[->] (2.7,0.5)--(2.7,1.5);

\draw[dotted, thick] (-\dofset, -\dofset) -- (2*\Lx+\dofset,-\dofset) -- (2*\Lx+\dofset,4*\Ly+\dofset) -- (-\dofset,4*\Ly+\dofset) -- (-\dofset,-\dofset);

\draw[thin, color = \colB, opacity=1, styleArrowShifted] (0.1-\cofset+2*\Lx,-\dofset) -- (0.1-\cofset+2*\Lx,4*\Ly+\dofset);
\draw[thin, color = \colA, opacity=1, styleArrowShifted] (\dofset+2*\Lx, 0.1-\cofset+4*\Ly)--(-\dofset,0.1-\cofset+4*\Ly);

\end{scope}
\end{tikzpicture}
\begin{tikzpicture}
\tikzmath{
	int \Lx,\Ly;
	\Lx=2;\Ly=1;
	\scale=1;
	\dofset=0.1;
	\cofset=0;
}

\begin{scope}[scale=\scale]

\clip(-\cofset,-\cofset) rectangle (2*\Lx + \cofset, 4*\Ly + \cofset);

\tikzmath{
		for \i in {0,...,\Lx}{
			{			
				\draw  (\i*2,0) pic[scale=\scale] {toda_casi_hor};
			};
		};
};

\end{scope}

\begin{scope}[scale=\scale]

\node at (0.3, 2.8) {$\alpha$};
\node at (0.3, 0.8) {$\beta$};

\draw[dotted, thick] (-\dofset, -\dofset) -- (2*\Lx+\dofset,-\dofset) -- (2*\Lx+\dofset,4*\Ly+\dofset) -- (-\dofset,4*\Ly+\dofset) -- (-\dofset,-\dofset);

\end{scope}
\end{tikzpicture}
\begin{tikzpicture}
\tikzmath{
	int \Lx,\Ly;
	\Lx=2;\Ly=1;
	\scale=1;
	\dofset=0.1;
	\cofset=0;
}

\begin{scope}[scale=\scale]

\clip(-\cofset,-\cofset) rectangle (2*\Lx + \cofset, 4*\Ly + \cofset);

\tikzmath{
		for \i in {0,...,\Lx}{
			{			
				\draw  (\i*2,0) pic[scale=\scale] {toda_casi_hor_twist};
			};
		};
};

\end{scope}

\begin{scope}[scale=\scale]

\node at (0.3, 2.8) {$\gamma$};
\node at (0.3, 0.8) {$\delta$};

\draw[dotted, thick] (-\dofset, -\dofset) -- (2*\Lx+\dofset,-\dofset) -- (2*\Lx+\dofset,4*\Ly+\dofset) -- (-\dofset,4*\Ly+\dofset) -- (-\dofset,-\dofset);

\end{scope}
\end{tikzpicture}
\begin{tikzpicture}

\tikzmath{\Lx=2;\Ly=2;\xs=0;\ys=0;\d=1.5;}


\foreach \x in {0,...,\Lx}
	\foreach \y in {0,...,\Ly}
		\draw[fill] (\xs+\d*\x,\ys+\d*\y) circle[radius=0.05];

\draw[blue, thick, styleArrowShort] (\xs,\ys)--(\xs+\d,\ys);
\draw[red, thick, styleArrowShort] (\xs+\d,\ys)--(\xs+\d*2,\ys+\d*2);
\draw[orange, thick, styleArrowShort] (\xs+2*\d,\ys+\d*2)--(\xs+\d,\ys+\d*2);
\draw[green, thick, styleArrowShort] (\xs+\d,\ys+2*\d)--(\xs,\ys);

\node at (-0.2+\d,2*\d+0.4) {$(1,2)$};
\node at (2*\d+0.2,2*\d+0.4) {$(2,2)$};
\node at (-0.2,-0.4) {$(0,0)$};
\node at (\d+0.2,-0.4) {$(1,0)$};

\draw[white] (-0.5,0.3)--(-0.5,0.6);

\end{tikzpicture}

}
\caption{Left: Bipartite graph for the two-particle Toda chain, small arrows give the Poisson structure
from Fig.~\ref{fig:todaex}. Center: zig-zag paths $\alpha,\beta,\gamma,\delta$. Right: Newton polygon obtained from zig-zags as elements $\H1(\T2,\mathbb{Z})$. Numbers are labeling degrees of $(\lambda,\mu)$ in spectral curve (\ref{eq:spectodaapp}).}
\label{fig:todaexbip}
\end{center}
\end{figure}

The coordinates $X=\{x_\gamma \in \mathbb{C}^{\times}\,|\,\gamma \in \H1(\Gamma,\mathbb{Z})\}$ of GK integrable system are multiplicative functions on $\H1(\Gamma,\mathbb{Z})$, considered as an Abelian group, i.e. $x_{\gamma_1}x_{\gamma_2}=x_{\gamma_1 + \gamma_2}$. Any element $\gamma \in \H1(\Gamma,\mathbb{Z})$ can be decomposed as
\begin{equation}
\label{torusHdecomp}
\gamma = n_A\gamma_A + n_B\gamma_B + \sum_i k_i f_i,~~ n_A,n_B,k_i\in \mathbb{Z}
\end{equation}
where $\gamma_A, \gamma_B$ form a basis in $\H1(\T2,\mathbb{Z})$, while $\mathsf{F}=\{f_i\}=\{\partial B_i\}$ is the set of faces or boundaries of the disks $\T2 \setminus \Gamma = \sqcup_i B_i $ with the orientation induced from surface, generating $\H1(\Gamma,\mathbb{Z})/\H1(\T2,\mathbb{Z})$ modulo single relation $\sum_k f_k=0$. Therefore, there is an exact sequence
\begin{equation}
0 \to \mathbb{Z} \to \mathsf{F} \to \H1(\Gamma,\mathbb{Z}) \to \H1(\T2,\mathbb{Z}) \to 0
\end{equation}
The set of face variables $\{x_{f}| f \in \mathsf{F}\}$ are coordinates on the toric chart on $\XG$, these are the $X$-cluster variables, transforming rationally under the cluster mutations or 'spider moves' of bipartite graph. Relation $\sum_k f_k=0$ can be relaxed, this results in \emph{deautonomization} $q=\prod_i x_{f_i} \neq 1$ of
cluster integrable system and leads to non-trivial $q$-dynamics.

Exchange matrix of the cluster seed is given by intersection form on
the dual surface $\Td2$, obtained from $\Gamma$
by gluing disks, which become faces of $\Td2$, to zig-zag paths $\mathsf{Z}$\,\footnote{Zig-zags could be easily
  found, as they are presented by paths on $\Gamma$, which turn
  maximally right at each black vertex, and turn maximally left at
  each white one. In the central and right pictures from
  Fig.~\ref{fig:todaexbip} the zig-zag paths for Toda chain on two
  sites are drawn.}, and forgetting structure of the torus. Embedding $\hat{\pi}:\Gamma \hookrightarrow \Td2$
allows to consider any cycle $\gamma \in \H1(\Gamma,\mathbb{Z})$ as an
element of $\H1(\Td2,\mathbb{Z})$, which is equipped with
non-degenerate skew-symmetric intersection form
$\langle \cdot\, ,\cdot \rangle: \H1(\Td2,\mathbb{Z})\times \H1(\Td2,\mathbb{Z}) \to
\mathbb{Z}$, which defines the Poisson bracket on $X$ by \be
\label{PBgamma}
\{x_{\gamma},x_{\gamma'}\} = \langle \hat{\pi} \gamma,\hat{\pi} \gamma'\rangle\, x_{\gamma}x_{\gamma'}.
\ee
Intersection form computed on faces $\mathsf{F}$ give exchange matrix of cluster seed $\varepsilon_{ij} = \langle f_i,f_j\rangle$. Effective way of writing this matrix is the following: for each black vertex, draw arrows in the clockwise direction between each pair of consecutive faces, which have this vertex as a corner. Then matrix element $\varepsilon_{ij}$ is equal to the alternated number of arrows from $f_i$ to $f_j$, see Fig.~\ref{fig:todaexbip}.

Classes of zig zag paths in $\H1(\T2,\mathbb{Z})$ are in one to one correspondence with the boundary intervals of Newton polygon $\overline{\Delta}$, and this correspondence is the simple way to build $\Delta$ by bipartite graph.
They are trivial in $\H1(\Td2,\mathbb{Z})$ so that the variables $x_\z$, corresponding to the zig-zag paths $\z \in \H1(\Gamma,\mathbb{Z})$ are Casimir functions of the bracket \rf{PBgamma}, i.e.
\be
\{x_\z,x_\gamma\} = 0,\ \ \ \ \z \in \mathsf{Z},\ \forall \gamma\in H_1(\Gamma,\mathbb{Z}).
\ee
 Zig-zag variables $x_{z_i}$ always present non-trivial elements from $\H1(\T2,\mathbb{Z})$, so single zig-zag itself cannot be expressed via the cluster variables. On the central and right pictures from Fig.~\ref{fig:todaexbip} zig-zag paths for the Toda chain on two sites are drawn. Casimir $\mathsf{Z}$ from the example from previous sub-section in terms of it is given by $\Z=x_{\alpha}x_{\beta}$.

\paragraph{Spectral curve.} Now, we are ready to construct Hamiltonians of integrable system, which is given by dimer partition function on it.\\

Perfect matching on bipartite graph $\Gamma$ is such configuration of edges $D\subset C_1(\Gamma)$ that each vertex has one adjacent edge from $D$. Such configurations has specific property that $\partial D=W-B$. Fixing any $D_0\subset  C_1(\Gamma)$ we can put an element $D-D_0\in C_1(\Gamma)$, which is closed, into correspondence to any perfect matching. Any $D-D_0$ under decomposition (\ref{torusHdecomp}) can be presented as
\begin{equation}
D-D_0 = n_A(D-D_0)\gamma_A + n_B(D-D_0)\gamma_B + \sum_i k_i(D-D_0) f_i
\end{equation}
Denoting variables $x_{i} = x_{f_i}, ~ \lambda=x_{\gamma_A},~ \mu=x_{\gamma_B}$ \textit{dimer partition function} of bipartite graph $\Gamma$ could be defined as
\begin{equation}
\mathcal{Z}_{\Gamma,D_0}(\lambda,\mu) = \sum_D \lambda^{n_A(D-D_0)}\mu^{n_B(D-D_0)} \prod\limits_{i} x_i^{k_i(D-D_0)}
\end{equation}
Equation $Z_\Gamma(\lambda,\mu)=0$ with $\lambda,\mu \in \mathbb{C}^{\times}$ defines curve $C\subset \mathbb{C}^{\times}\times \mathbb{C}^{\times}$. The curve $C$ is spectral curve of integrable system. Collecting terms corresponding to the same degrees of spectral parameters
\begin{equation}
\mathcal{Z}_{\Gamma,D_0}(\lambda,\mu) = \sum\limits_{(i,j)\in N} \lambda^{i}\mu^{j} \mathcal{H}_{ij},~~~ N \subset \mathbb{Z}^{2}
\end{equation}
we get Hamiltonians $\mathcal{H}_{ij}$ of the Goncharov-Kenyon integrable system with the Newton polygon $\Delta$. Change of the base configuration $D_0$ just multiplies partition function by monomial
\begin{equation}
\mathcal{Z}_{D_0'} = x_{D_0-D_0'} \mathcal{Z}_{D_0},~~~ x_{D_0-D_0'}=\lambda^{n_A(D_0-D_0')}\mu^{n_B(D_0-D_0')} \prod\limits_{i} x_i^{k_i(D_0-D_0')}.
\end{equation}

In \cite{GK:2011} authors proved for the special choose of $D_0$ that the model is integrable - i.e. that $\{\mathcal{H}_{ij},\mathcal{H}_{kl}\}=0$, and the number of independent Hamiltonians is half of the dimension of phase space. Hamiltonians which correspond to the boundary integral points of $N$ are Casimirs of Poisson bracket, and have to be fixed, to get symplectic leaf with non-degenerate Poisson bracket.
Boundary intervals of $\partial N$ are in one to one correspondence with zig-zag paths. Vector presenting boundary interval coincides with the class of corresponding zig-zag in $\H1(\T2,\mathbb{Z})$. Choice of $D_0$ proposed in \cite{GK:2011} are so that $\mathcal{H}_{ij}=1$ for one corner of Newton polygon.

Important detail which remained out of the scope yet is that we have to choose concrete representative in $\H1(\Gamma,\mathbb{Z})$ for cycles $\gamma_A$ and $\gamma_B$. Usually, spectral parameters are expected to commute with all dynamical variables of the system, so $\gamma_A$ and $\gamma_B$ have to be chosen as an integral combinations of zig-zag paths. However, it is not always possible -- even for the simplest example of bipartite graph from Fig.~\ref{fig:todaexbip}, zig-zags are $(1,0), (1,2), (-1,0), (-1,-2)$-cycles, and subgroup generated by them in $\H1(\T2,\mathbb{Z})$ has index two. Generally, this order is $d=|\H1(\T2,\mathbb{Z})/\mathsf{Z}|$, so by choosing spectral parameters expressed via zig-zags, we get Hamiltonians depending on fractional powers of cluster variables $x_i^{1/d}$. Convenient choosing of spectral parameters normalization is so that three Hamiltonians in three corners of Newton polygon become equal to unit.

On right panel of Fig.~\ref{fig:todaexperf} perfect matchings for bipartite graph from Fig.~\ref{fig:todaexbip} are drawn. Selecting third matching in the first row as a $D_0$, and spectral parameters by $\lambda=x_{\gamma_A},~\mu=x_{\gamma_B}$ with $\gamma_A=\beta,~\gamma_B=-\frac{1}{2}(\beta+\delta)$, one gets spectral curve
\begin{equation}
\label{eq:spectodaapp}
\mathcal{Z} = 1+\lambda+\lambda \mu^2 + \lambda^2 \mu^2 \Z^{-1} + \lambda\mu \left(\sqrt{x_1 x_4}+\Z^{-1}\sqrt{\dfrac{x_1}{x_4}} + \sqrt{\dfrac{x_4}{x_1}}+\dfrac{1}{\sqrt{x_1 x_4}} \right).
\end{equation}
Coefficient at $\lambda\mu$ is precisely Hamiltonian of closed relativistic Toda chain on two sites. Newton polygon of this curve coincides with the one obtained from zig-zags and drawn on the right panel of Fig.~\ref{fig:todaexbip}.
\begin{figure}[!h]
\begin{center}
\scalebox{0.7}{
\begin{tikzpicture}
\tikzmath{
	\scale=1;
	\dofset=0.01;
	\cofset=0.05;
}

\begin{scope}[scale=\scale]
\tikzmath{\xs=-5;\ys=-2.5;}

\clip(\xs-\cofset,\ys-\cofset) rectangle (\xs+4 + \cofset,\ys+ 4 + \cofset);

\draw (\xs+0,\ys+0) pic[scale=\scale] {toda_one_text};
\draw (\xs+2,\ys+0) pic[scale=\scale] {toda_one_text};

\node[font=\bfseries,scale=1.3] at (\xs+0.7,\ys+1.9) {a};
\node[font=\bfseries,scale=1.3] at (\xs+1.5,\ys+2.0) {b};
\node[font=\bfseries,scale=1.3] at (\xs+1.3,\ys+0.6) {c};
\node[font=\bfseries,scale=1.3] at (\xs+0.4,\ys+0.9) {d};

\node[font=\bfseries,scale=1.3] at (\xs+2+0.7,\ys+1.9) {e};
\node[font=\bfseries,scale=1.3] at (\xs+2+1.5,\ys+2.0) {f};
\node[font=\bfseries,scale=1.3] at (\xs+2+1.3,\ys+0.6) {g};
\node[font=\bfseries,scale=1.3] at (\xs+2+0.4,\ys+0.9) {h};

\node[font=\bfseries,scale=1] at (\xs+1.5,\ys+1) {$1$};
\node[font=\bfseries,scale=1] at (\xs+2+1.5,\ys+1) {$2$};
\node[font=\bfseries,scale=1] at (\xs+2+1.5,\ys+2+1) {$2$};
\node[font=\bfseries,scale=1] at (\xs+1.5,\ys+2+1) {$1$};

\draw[dotted, thick] (\xs-\dofset, \ys-\dofset) -- (\xs+4+\dofset,\ys-\dofset) -- (\xs+4+\dofset,\ys+4+\dofset) -- (\xs-\dofset,\ys+4+\dofset) -- (\xs-\dofset,\ys-\dofset);
\end{scope}

\tikzmath{\xs=0;\ys=0;}

\begin{scope}[scale=\scale]

\clip(\xs-\cofset,\ys-\cofset) rectangle (\xs+4 + \cofset,\ys+ 4 + \cofset);

\draw (\xs+0,\ys+0) pic[scale=\scale] {toda_one_text_noarrow};
\draw (\xs+2,\ys+0) pic[scale=\scale] {toda_one_text_noarrow};

\draw[blue, ultra thick, styleArrowShort] (\xs+1,\ys+1) -- (\xs+3,\ys+3);
\draw[blue, ultra thick, styleArrowShort] (\xs+2-0.1,\ys+4)--(\xs+1-0.1,\ys+3);
\draw[blue, ultra thick, styleArrowShort] (\xs+3-0.1,\ys+1)--(\xs+2-0.1,\ys+0);

\draw[red, ultra thick, styleArrowShort] (\xs+1,\ys+3)--(\xs+2,\ys+4);
\draw[red, ultra thick, styleArrowShort] (\xs+2,\ys+0)--(\xs+3,\ys+1);

\draw[red, ultra thick, styleArrowShort] (\xs,\ys)--(\xs+1,\ys+1);
\draw[red, ultra thick, styleArrowShort] (\xs+3,\ys+3)--(\xs+4,\ys+4);

\draw[dotted, thick] (\xs-\dofset, \ys-\dofset) -- (\xs+4+\dofset,\ys-\dofset) -- (\xs+4+\dofset,\ys+4+\dofset) -- (\xs-\dofset,\ys+4+\dofset) -- (\xs-\dofset,\ys-\dofset);
\end{scope}

\node[font=\bfseries,scale=1.3] at (\xs+2,\ys-0.5) {$\mu^{-1},~ bh$};

\tikzmath{\xs=5;\ys=0;}
\begin{scope}[scale=\scale]

\clip(\xs-\cofset,\ys-\cofset) rectangle (\xs+4 + \cofset,\ys+ 4 + \cofset);

\draw (\xs+0,\ys+0) pic[scale=\scale] {toda_one_text_noarrow};
\draw (\xs+2,\ys+0) pic[scale=\scale] {toda_one_text_noarrow};

\draw[blue, ultra thick, styleArrowShort] (\xs+1,\ys+1) -- (\xs+3,\ys+3);
\draw[blue, ultra thick, styleArrowShort] (\xs,\ys+2) -- (\xs+1,\ys+3);
\draw[blue, ultra thick, styleArrowShort] (\xs+3,\ys+1) -- (\xs+4,\ys+2);

\draw[red, ultra thick, styleArrowShort] (\xs+1,\ys+3)--(\xs+2,\ys+4);
\draw[red, ultra thick, styleArrowShort] (\xs+2,\ys+0)--(\xs+3,\ys+1);

\draw[red, ultra thick, styleArrowShort] (\xs,\ys)--(\xs+1,\ys+1);
\draw[red, ultra thick, styleArrowShort] (\xs+3,\ys+3)--(\xs+4,\ys+4);

\draw[dotted, thick] (\xs-\dofset, \ys-\dofset) -- (\xs+4+\dofset,\ys-\dofset) -- (\xs+4+\dofset,\ys+4+\dofset) -- (\xs-\dofset,\ys+4+\dofset) -- (\xs-\dofset,\ys-\dofset);
\end{scope}

\node[font=\bfseries,scale=1.3] at (\xs+2,\ys-0.5) {$\lambda,~ bf$};

\tikzmath{\xs=10;\ys=0;}
\begin{scope}[scale=\scale]

\clip(\xs-\cofset,\ys-\cofset) rectangle (\xs+4 + \cofset,\ys+ 4 + \cofset);

\draw (\xs+0,\ys+0) pic[scale=\scale] {toda_one_text_noarrow};
\draw (\xs+2,\ys+0) pic[scale=\scale] {toda_one_text_noarrow};

\draw[blue, ultra thick, styleArrowShort] (\xs-0.1+2,\ys+4)--(\xs-0.1+1,\ys+3);
\draw[blue, ultra thick, styleArrowShort] (\xs-0.1+3,\ys+1)--(\xs-0.1+2,\ys+0);

\draw[blue, ultra thick, styleArrowShort] (\xs+1-0.1,\ys+1)--(\xs-0.1,\ys);
\draw[blue, ultra thick, styleArrowShort] (\xs+4-0.1,\ys+4)--(\xs+3-0.1,\ys+3);

\draw[red, ultra thick, styleArrowShort] (\xs+1,\ys+3)--(\xs+2,\ys+4);
\draw[red, ultra thick, styleArrowShort] (\xs+2,\ys+0)--(\xs+3,\ys+1);

\draw[red, ultra thick, styleArrowShort] (\xs,\ys)--(\xs+1,\ys+1);
\draw[red, ultra thick, styleArrowShort] (\xs+3,\ys+3)--(\xs+4,\ys+4);

\draw[dotted, thick] (\xs-\dofset, \ys-\dofset) -- (\xs+4+\dofset,\ys-\dofset) -- (\xs+4+\dofset,\ys+4+\dofset) -- (\xs-\dofset,\ys+4+\dofset) -- (\xs-\dofset,\ys-\dofset);
\end{scope}

\node[font=\bfseries,scale=1.3] at (\xs+2,\ys-0.5) {$\lambda^{-1} \mu^{-2},~ hd$};

\tikzmath{\xs=15;\ys=0;}
\begin{scope}[scale=\scale]

\clip(\xs-\cofset,\ys-\cofset) rectangle (\xs+4 + \cofset,\ys+ 4 + \cofset);

\draw (\xs+0,\ys+0) pic[scale=\scale] {toda_one_text_noarrow};
\draw (\xs+2,\ys+0) pic[scale=\scale] {toda_one_text_noarrow};

\draw[blue, ultra thick, styleArrowShort] (\xs,\ys+2) -- (\xs+1,\ys+3);
\draw[blue, ultra thick, styleArrowShort] (\xs+3,\ys+1) -- (\xs+4,\ys+2);

\draw[blue, ultra thick, styleArrowShort] (\xs+1-0.1,\ys+1)--(\xs-0.1,\ys);
\draw[blue, ultra thick, styleArrowShort] (\xs+4-0.1,\ys+4)--(\xs+3-0.1,\ys+3);

\draw[red, ultra thick, styleArrowShort] (\xs+1,\ys+3)--(\xs+2,\ys+4);
\draw[red, ultra thick, styleArrowShort] (\xs+2,\ys+0)--(\xs+3,\ys+1);

\draw[red, ultra thick, styleArrowShort] (\xs,\ys)--(\xs+1,\ys+1);
\draw[red, ultra thick, styleArrowShort] (\xs+3,\ys+3)--(\xs+4,\ys+4);

\draw[dotted, thick] (\xs-\dofset, \ys-\dofset) -- (\xs+4+\dofset,\ys-\dofset) -- (\xs+4+\dofset,\ys+4+\dofset) -- (\xs-\dofset,\ys+4+\dofset) -- (\xs-\dofset,\ys-\dofset);
\end{scope}

\node[font=\bfseries,scale=1.3] at (\xs+2,\ys-0.5) {$\mu^{-1},~ df$};

\tikzmath{\xs=0;\ys=-5;}
\begin{scope}[scale=\scale]

\clip(\xs-\cofset,\ys-\cofset) rectangle (\xs+4 + \cofset,\ys+ 4 + \cofset);

\draw (\xs+0,\ys+0) pic[scale=\scale] {toda_one_text_noarrow};
\draw (\xs+2,\ys+0) pic[scale=\scale] {toda_one_text_noarrow};

\draw[blue, ultra thick, styleArrowShort] (\xs+1,\ys+1) -- (\xs+1,\ys+3);
\draw[blue, ultra thick, styleArrowShort] (\xs+3,\ys+1) -- (\xs+3,\ys+3);

\draw[red, ultra thick, styleArrowShort] (\xs+1,\ys+3)--(\xs+2,\ys+4);
\draw[red, ultra thick, styleArrowShort] (\xs+2,\ys+0)--(\xs+3,\ys+1);

\draw[red, ultra thick, styleArrowShort] (\xs,\ys)--(\xs+1,\ys+1);
\draw[red, ultra thick, styleArrowShort] (\xs+3,\ys+3)--(\xs+4,\ys+4);

\draw[dotted, thick] (\xs-\dofset, \ys-\dofset) -- (\xs+4+\dofset,\ys-\dofset) -- (\xs+4+\dofset,\ys+4+\dofset) -- (\xs-\dofset,\ys+4+\dofset) -- (\xs-\dofset,\ys-\dofset);
\end{scope}

\node[font=\bfseries,scale=1.3] at (\xs+2,\ys-0.5) {$ae$};

\tikzmath{\xs=5;\ys=-5;}
\begin{scope}[scale=\scale]

\clip(\xs-\cofset,\ys-\cofset) rectangle (\xs+4 + \cofset,\ys+ 4 + \cofset);

\draw (\xs+0,\ys+0) pic[scale=\scale] {toda_one_text_noarrow};
\draw (\xs+2,\ys+0) pic[scale=\scale] {toda_one_text_noarrow};

\draw[blue, ultra thick, styleArrow] (\xs+1,\ys+1)--(\xs+1,\ys);
\draw[blue, ultra thick, styleArrow] (\xs+1,\ys+4)--(\xs+1,\ys+3);
\draw[blue, ultra thick, styleArrowShort] (\xs+3,\ys+1) -- (\xs+3,\ys+3);

\draw[red, ultra thick, styleArrowShort] (\xs+1,\ys+3)--(\xs+2,\ys+4);
\draw[red, ultra thick, styleArrowShort] (\xs+2,\ys+0)--(\xs+3,\ys+1);

\draw[red, ultra thick, styleArrowShort] (\xs,\ys)--(\xs+1,\ys+1);
\draw[red, ultra thick, styleArrowShort] (\xs+3,\ys+3)--(\xs+4,\ys+4);

\draw[dotted, thick] (\xs-\dofset, \ys-\dofset) -- (\xs+4+\dofset,\ys-\dofset) -- (\xs+4+\dofset,\ys+4+\dofset) -- (\xs-\dofset,\ys+4+\dofset) -- (\xs-\dofset,\ys-\dofset);
\end{scope}

\node[font=\bfseries,scale=1.3] at (\xs+2,\ys-0.5) {$\mu^{-1},~ ce$};

\tikzmath{\xs=10;\ys=-5;}
\begin{scope}[scale=\scale]

\clip(\xs-\cofset,\ys-\cofset) rectangle (\xs+4 + \cofset,\ys+ 4 + \cofset);

\draw (\xs+0,\ys+0) pic[scale=\scale] {toda_one_text_noarrow};
\draw (\xs+2,\ys+0) pic[scale=\scale] {toda_one_text_noarrow};

\draw[blue, ultra thick, styleArrow] (\xs+3,\ys+1)--(\xs+3,\ys);
\draw[blue, ultra thick, styleArrow] (\xs+3,\ys+4)--(\xs+3,\ys+3);
\draw[blue, ultra thick, styleArrowShort] (\xs+1,\ys+1) -- (\xs+1,\ys+3);

\draw[red, ultra thick, styleArrowShort] (\xs+1,\ys+3)--(\xs+2,\ys+4);
\draw[red, ultra thick, styleArrowShort] (\xs+2,\ys+0)--(\xs+3,\ys+1);

\draw[red, ultra thick, styleArrowShort] (\xs,\ys)--(\xs+1,\ys+1);
\draw[red, ultra thick, styleArrowShort] (\xs+3,\ys+3)--(\xs+4,\ys+4);

\draw[dotted, thick] (\xs-\dofset, \ys-\dofset) -- (\xs+4+\dofset,\ys-\dofset) -- (\xs+4+\dofset,\ys+4+\dofset) -- (\xs-\dofset,\ys+4+\dofset) -- (\xs-\dofset,\ys-\dofset);
\end{scope}

\node[font=\bfseries,scale=1.3] at (\xs+2,\ys-0.5) {$\mu^{-1},~ ga$};

\tikzmath{\xs=15;\ys=-5;}
\begin{scope}[scale=\scale]

\clip(\xs-\cofset,\ys-\cofset) rectangle (\xs+4 + \cofset,\ys+ 4 + \cofset);

\draw (\xs+0,\ys+0) pic[scale=\scale] {toda_one_text_noarrow};
\draw (\xs+2,\ys+0) pic[scale=\scale] {toda_one_text_noarrow};

\draw[blue, ultra thick, styleArrow] (\xs+1,\ys+1)--(\xs+1,\ys);
\draw[blue, ultra thick, styleArrow] (\xs+1,\ys+4)--(\xs+1,\ys+3);
\draw[blue, ultra thick, styleArrow] (\xs+3,\ys+1)--(\xs+3,\ys);
\draw[blue, ultra thick, styleArrow] (\xs+3,\ys+4)--(\xs+3,\ys+3);

\draw[red, ultra thick, styleArrowShort] (\xs+1,\ys+3)--(\xs+2,\ys+4);
\draw[red, ultra thick, styleArrowShort] (\xs+2,\ys+0)--(\xs+3,\ys+1);

\draw[red, ultra thick, styleArrowShort] (\xs,\ys)--(\xs+1,\ys+1);
\draw[red, ultra thick, styleArrowShort] (\xs+3,\ys+3)--(\xs+4,\ys+4);

\draw[dotted, thick] (\xs-\dofset, \ys-\dofset) -- (\xs+4+\dofset,\ys-\dofset) -- (\xs+4+\dofset,\ys+4+\dofset) -- (\xs-\dofset,\ys+4+\dofset) -- (\xs-\dofset,\ys-\dofset);
\end{scope}

\node[font=\bfseries,scale=1.3] at (\xs+2,\ys-0.5) {$\mu^{-2},~ cg$};

\end{tikzpicture}
}
\caption{Left: bipartite graph with the edge weights. Small integers are to enumerate black and white vertices. Right: Perfect matchings for bipartite graph from Fig.~\ref{fig:todaexbip} are in blue. Red color indicates reference matching $D_0$. Weights of the corresponding contributions to determinant of Kasteleyn operator are written below.}
\label{fig:todaexperf}
\end{center}
\end{figure}
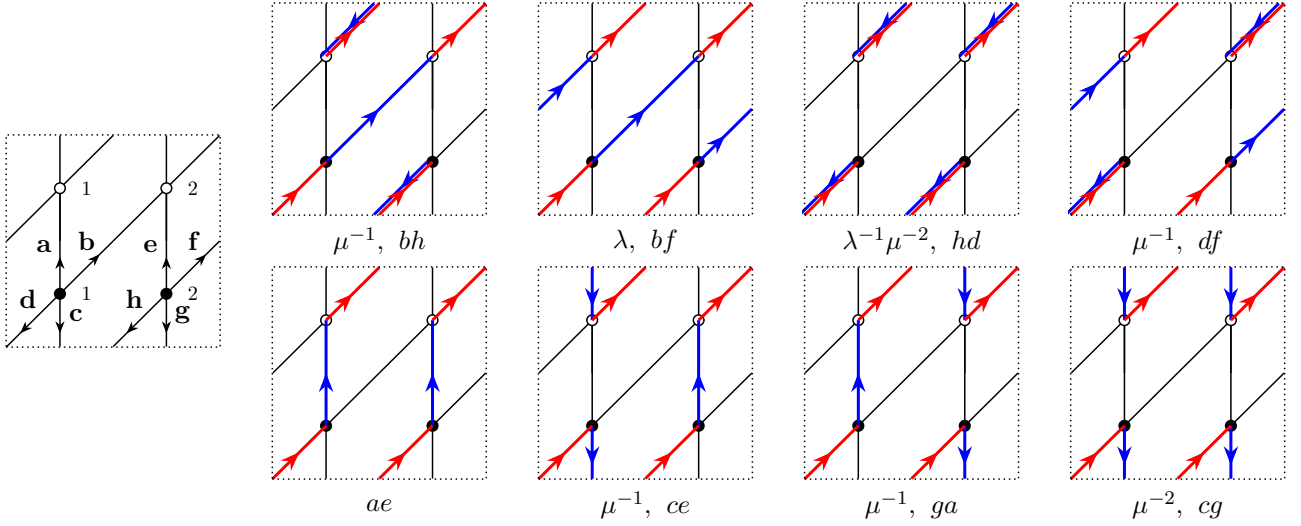

\paragraph{Kasteleyn operator.} Dimer partition function can be computed using the \textit{Kasteleyn operator}. To define it, first, consider discrete linear bundle with connection $a$ on bipartite graph $\Gamma$. In trivialization this means that we associate 1-d vector space $\mathbb{C}$ with each vertex of $\Gamma$, and discrete monodromy $a_e\in \mathbb{C}^{\times}$ with each edge $e$ oriented from black to white. For the edge with the opposite orientation set $a_{-e}=a_e^{-1}$. This definition can be extended to any $\gamma \in C_1(\Gamma,\mathbb{Z})$ by $a_{\gamma_1+\gamma_2} = a_{\gamma_1}a_{\gamma_2}$. Dynamical variables $x_{\gamma}$ used above, could be naturally associated with monodromies taken over cycles, i.e. $x_\gamma = a_\gamma$ if $\partial \gamma=0$. It is problematic to introduce Poisson structure for variables $a_e$, as they are not 'gauge invariant'. We can perform gauge transformation $e^{i\phi_k}$ at each vertex $k$, which results in change $a_{e_{ij}} \to a_{e_{ij}}e^{i(\phi_i-\phi_j)}$ of edge variables\footnote{Actually, Poisson structure could be introduced even for non-closed loops. Interested reader can find one in Appendix of \cite{GK:2011}.}. However monodromies $x_{\gamma}$ are well defined, so their bracket is given by (\ref{PBgamma}).

Second ingredient in this construction is discrete spin structure - multiplicative map $K_e: C_1(\Gamma)\to \{\pm 1\}$, which assigns $\pm 1$ to each edge $e$ in such a way that for any face $B_i$
\begin{equation}
K_{\partial B_i} = (-1)^{l(B_i)/2+1}
\end{equation}
where $l(B_i)$ - number of edges, adjacent to $B_i$.

Finally, the third ingredient is choosing of two oriented cycles $h_A$ and $h_B$ on $\T2$, which cross edges of $\Gamma$ transversally, and as elements of $\H1(\T2,\mathbb{Z})$ they present $[h_A]=\gamma_B$ and $[h_B]=\gamma_A$ (indices $A$ and $B$ are indeed interchanged). We denote by $\langle e,h_{A,B}\rangle$ intersection index of edge $e$ with cycle $h_{A,B}$. It is $+1$ if edge cross cycle from the left to the right, if you look along cycle. Bringing all ingredients together, Kasteleyn operator $\Kast: \mathbb{C}^{|B|} \to \mathbb{C}^{|W|}$ of graph $\Gamma$ is a $|B|$ by $|W|$ (which are equal) matrix
\begin{equation}
\Kast = \sum\limits_{i=1}^{|B|}\sum\limits_{j=1}^{|W|} \Kast_{ij} E_{ij},~~~ \Kast_{ij}(\lambda,\mu)= a_{e_{ij}}K_{e_{ij}} \lambda^{\langle e_{ij},h_A \rangle} \mu^{\langle e_{ij},h_B \rangle}
\end{equation}
where we assume that $a_{e_{ij}}$ is zero, if there are no edges between vertices $i$ and $j$. Note that as an operator acting $\mathbb{C}^{|B|} \to \mathbb{C}^{|W|}$, it acts from the right on row vectors. It could be shown that
\begin{equation}
\det \Kast(\tilde{\lambda},\tilde{\mu}) = \sum_D (-1)^{s([D])}  \tilde{\lambda}^{\langle D,h_A \rangle} \tilde{\mu}^{\langle D,h_B \rangle} a_D
\end{equation}
where summation goes over all perfect matchings, sign $(-1)^{s}$ will depend only on the resulting class in homology of perfect matching after normalization. Parameters $\tilde{\lambda}$ and $\tilde{\mu}$ are different from $\lambda$ and $\mu$ used above. They do not indicate belonging of contribution to any particular homology class, as $D$ are not closed. To make it so, we have to subtract some 'reference configuration' $D_0$, and choose pair of elements $\zeta_A, \zeta_B\in \H1(\T2,\mathbb{Z})$, presenting $A$ and $B$ cycles on $\T2$. Precise relation between determinant of Kasteleyn operator and dimer partition function is
\begin{equation}
\mathcal{Z}'_{\Gamma,D_0}(\lambda,\mu) = \sum\limits_{(i,j)\in N} (-1)^{s(i,j)} \lambda^{i}\mu^{j} \mathcal{H}_{ij}=\left.\dfrac{\det \Kast(\tilde{\lambda},\tilde{\mu})}{a_{D_0}\tilde{\lambda}^{\langle D_0,h_A \rangle}\tilde{\mu}^{\langle D_0,h_B \rangle}}\right|_{\tilde{\lambda}\to \lambda/a_{\zeta_A},\,\tilde{\mu}\to \mu/a_{\zeta_B}}
\end{equation}
For our example, cycles $h_A$ and $h_B$ are shown in Fig.~\ref{fig:todaexbip}, left. Weights of perfect matchings are written under the pictures in Fig.~\ref{fig:todaexperf}. It can be easily seen that sum over them could be computed by
\begin{equation}
\det \Kast(\lambda,\mu) =
\det
\left(
\begin{array}{cc}
a + \mu^{-1} \, c & -b - \lambda^{-1}\mu^{-1}\,d \\
\lambda \, f + \mu^{-1} \, h & e + \mu^{-1} \, g
\end{array}
\right)
=
\end{equation}
$$
=\mu^{-1}\, b h + \lambda\, b f + \lambda^{-1} \mu^{-2}\, h d + \mu^{-1}\, d f + a e + \mu^{-1}\, c e + \mu^{-1}\, a g +\mu^{-2}\, c g
$$
Dividing it by $a_{D_0}=\lambda^{-1}\mu^{-2}hd$ and rescaling $\lambda \to \lambda/x_{\beta},~\mu \to \mu/x_{-\frac{1}{2}(\beta+\delta)}$ with $x_\beta = \dfrac{cg}{hd},~x_\delta = \dfrac{dh}{ae}$, one immediately gets spectral curve (\ref{eq:spectodaapp}).

\paragraph{Spider moves.} There is a special class of mutations for the quivers constructed from bipartite graph called {\it spider moves} \cite{GK:2011}. If mutation is performed at four-valent vertex corresponding to four-gonal face of bipartite graph, one can change bipartite graph as shown in Fig.~\ref{fig:spiderbip}, left, and redefine weights on the edges in such a way that dimer partition function remains unchanged. Cluster variables expressed by edges are changing as they should under corresponding mutations (see Fig.~\ref{fig:spiderbip}, right, for the change of quiver).
\begin{figure}[!h]
\begin{center}
\begin{tikzpicture}

\begin{scope}[scale = 0.7]

\draw[thick] (0.75,0.75)--(1.5,3);
\draw[thick] (1.5,3)--(3.75,3.75);
\draw[thick] (3,1.5)--(3.75,3.75);
\draw[thick] (3,1.5)--(0.75,0.75);

\draw[thick] (0.75,0.75)--(0.5,0.5);
\draw[thick] (0.5,4)--(1.5,3);
\draw[thick] (3,1.5)--(4,0.5);
\draw[thick] (3.75,3.75)--(4,4);

\draw[blackCircle] (0.75,3.75) circle;
\draw[blackCircle] (3.75,0.75) circle;
\draw[blackCircle] (0.75,0.75) circle;
\draw[blackCircle] (3.75,3.75) circle;

\draw[whiteCircle] (1.5,3) circle;
\draw[whiteCircle] (3,1.5) circle;

\draw[thick, ->] (4.75,2.25)--(5.75,2.25);

\tikzmath{\xs=6;};

\draw[thick] (\xs+1.5,1.5)--(\xs+0.75,3.75);
\draw[thick] (\xs+0.75,3.75)--(\xs+3,3);
\draw[thick] (\xs+3,3)--(\xs+3.75,0.75);
\draw[thick] (\xs+3.75,0.75)--(\xs+1.5,1.5);

\draw[thick] (\xs+1.5,1.5)--(\xs+0.5,0.5);
\draw[thick] (\xs+0.5,4)--(\xs+0.75,3.75);
\draw[thick] (\xs+3.75,0.75)--(\xs+4,0.5);
\draw[thick] (\xs+3,3)--(\xs+4,4);

\draw[blackCircle] (\xs+0.75,3.75) circle;
\draw[blackCircle] (\xs+3.75,0.75) circle;
\draw[blackCircle] (\xs+0.75,0.75) circle;
\draw[blackCircle] (\xs+3.75,3.75) circle;

\draw[whiteCircle] (\xs+1.5,1.5) circle;
\draw[whiteCircle] (\xs+3,3) circle;

\end{scope}

\tikzmath{\xs=9;\ys=0.5;}

\draw[styleQuiverEdge, gray] (\xs+1,\ys+2) -- (\xs+0,\ys+1);
\draw[styleQuiverEdge, gray] (\xs+1,\ys+0) -- (\xs+2,\ys+1);

\draw[styleQuiverEdge] (\xs+0,\ys+1) -- (\xs+1,\ys+1);
\draw[styleQuiverEdge] (\xs+1,\ys+1) -- (\xs+1,\ys+0);
\draw[styleQuiverEdge] (\xs+1,\ys+1) -- (\xs+1,\ys+2);
\draw[styleQuiverEdge] (\xs+2,\ys+1) -- (\xs+1,\ys+1);

\draw[quiverVertex] (\xs+0,\ys+1) circle;
\draw[quiverVertex] (\xs+1,\ys+0) circle;
\draw[quiverVertex] (\xs+1,\ys+2) circle;
\draw[quiverVertex] (\xs+2,\ys+1) circle;
\draw[quiverVertex] (\xs+1,\ys+1) circle;

\node at (\xs+1.3,\ys+1.3) {$0$};
\node at (\xs+1.5,\ys+2) {$1$};
\node at (\xs+2,\ys+0.5) {$2$};
\node at (\xs+0.5,\ys+0) {$3$};
\node at (\xs+0,\ys+1.5) {$4$};

\draw[->, thick] (\xs+3,\ys+1) -- (\xs+4,\ys+1);

\draw[styleQuiverEdge] (\xs+6,\ys+1) -- (\xs+5,\ys+1);
\draw[styleQuiverEdge] (\xs+6,\ys+0) -- (\xs+6,\ys+1);
\draw[styleQuiverEdge] (\xs+6,\ys+2) -- (\xs+6,\ys+1);
\draw[styleQuiverEdge] (\xs+6,\ys+1) -- (\xs+7,\ys+1);

\draw[styleQuiverEdge, gray] (\xs+7,\ys+1) -- (\xs+6,\ys+2);
\draw[styleQuiverEdge, gray] (\xs+5,\ys+1) -- (\xs+6,\ys+0);

\draw[quiverVertex] (\xs+5,\ys+1) circle;
\draw[quiverVertex] (\xs+6,\ys+0) circle;
\draw[quiverVertex] (\xs+6,\ys+2) circle;
\draw[quiverVertex] (\xs+7,\ys+1) circle;
\draw[quiverVertex] (\xs+6,\ys+1) circle;

\node at (\xs+6.3,\ys+1.3) {$0$};
\node at (\xs+6.5,\ys+2) {$1$};
\node at (\xs+7,\ys+0.5) {$2$};
\node at (\xs+5.5,\ys+0) {$3$};
\node at (\xs+5,\ys+1.5) {$4$};

\end{tikzpicture}

\end{center}
\caption{Left: Transformation of bipartite graph under spider move. Right: mutation of quiver under spider-move. We draw only the edges, connecting $1,2,3,4$ with $0$, affected by the mutation.}
\label{fig:spiderbip}
\end{figure}
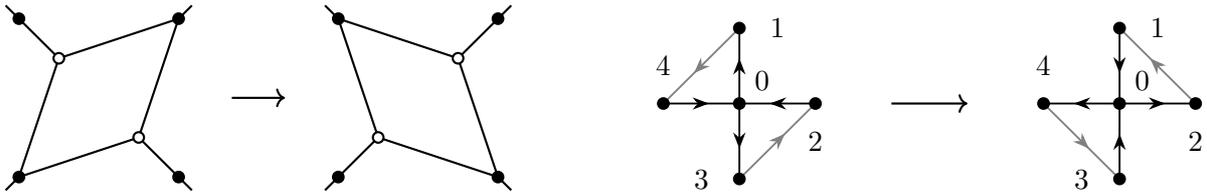

\subsection{Cluster algebras}
\label{ss:clusterAalg}

Dual description to $\mathcal{X}$-cluster language is $A$-cluster language \cite{FZ:2006}. To define $A$-cluster algebra, starting from the quiver $Q$, we associate with each vertex $i\in \mathcal{Q}$ of quiver a pair of variables $(\tau_i,\cof_i)$, $\cof$-variables are often called the {\it coefficients}, and $\tau$-variables usually refereed as {\it cluster variables}~\footnote{In contrast to original papers, see e.g. \cite{FZ:2006}, we denote them as
$\tau$-variables, since they satisfy some bilinear relations, as shown in Sect.~\ref{ss:bilinear}.}. The coefficients take values in {\it tropical semi-field} $\mathbb{P} = \mathrm{Trop}(\c_1,...,\c_r)=\langle \c_1^{n_1}...\c_r^{n_r}\,|\, n_{\alpha}\in \mathbb{Q}\,\rangle$ -- a set equipped with the pair of operations:
\begin{equation}
\begin{array}{l}
\c_1^{n_1}...\,\c_r^{n_r}\odot \c_1^{m_1}...\,\c_r^{m_r} = \c_1^{n_1+m_1}...\,\c_r^{n_r+m_r}\\
\c_1^{n_1}...\,\c_r^{n_r}\oplus \c_1^{m_1}...\,\c_r^{m_r} = \c_1^{\mathrm{min}(n_1,m_1)}...\,\c_r^{\mathrm{min}(n_r,m_r)}
\end{array}
\end{equation}
It can be easily seen that with respect to $\odot$ element $1=\c_1^{0}...\,\c_r^{0}$ is unit, and each element $\cof = \c_1^{n_1}...\, \c_r^{n_r}$ has inverse $\c^{-1} = \c_1^{-n_1}...\, \c_r^{-n_r}$. Both operations are commutative, we also have distributivity $a\odot(b\oplus c)=a\odot b \oplus a\odot c$. It is convenient for our purposes to allow fractional powers of $\c$, so that elements $\c_1,...,\c_r$ generate whole $\mathbb{P}$. Field $\mathcal{F}$ of rational functions in $\{\tau_i\}_{i\in \mathcal{Q}}$ with coefficients in $\mathbb{P}$ is called {\it ambient field}. {\it Cluster seed} is a set $(\mathcal{Q},\{\tau_i,\cof_i \}_{i\in \mathcal{Q}})$. Mutation $\mu_k$ transforms seed into some other seed $(\mathcal{Q'},\{\tau'_i,\cof '_i \}_{i\in \mathcal{Q'}})$ with $\mathcal{Q}'$ related to $\mathcal{Q}$ by the rules (\ref{eq:mutationsquiv}), while new cluster variables $\tau'_i \in \mathcal{F}$ and coefficients $\cof'_i \in \mathbb{P}$ are defined by
\begin{equation}
\cof_i'=
\left\{
\begin{array}{lcl}
\cof_i^{-1} & , & i=k\\
\cof_i \left(1\oplus \cof_k^{\mathrm{sgn}\, \varepsilon_{ik}}\right)^{\varepsilon_{ik}} & , & i\neq k
\end{array}
\right.
\end{equation}
\begin{equation}
\tau_k' = \dfrac{\cof_k\prod_{i=1}^{|\mathcal{Q}|} \tau_i^{[\varepsilon_{ik}]_+}+\prod_{i=1}^{|\mathcal{Q}|} \tau_i^{[-\varepsilon_{ik}]_+}}{(1\oplus \cof_k)\tau_k},\ \ \ \tau_i' = \tau_i\ \mathrm{if}\ i\neq k
\end{equation}
where $[a]_+=\mathrm{max}(0,a)$. Alternative point of view on coefficients is to consider generators of $\mathbb{P}$ as {\it frozen variables}, placed in additional vertices of quiver, where mutations are forbidden. If coefficients are expressed through the generators $\{\c_\alpha\}$ by $\cof_i = \c_1^{n_{1,i}}...\,\c_r^{n_{r,i}}$ for some fixed seed, we introduce another matrix $b$ which contains exchange matrix $\varepsilon$ as a block
\begin{equation}
b =
\left(
\begin{array}{c}
\boxed{\varepsilon} \\
\boxed{N}
\end{array}
\right)
,~\mathrm{where}~
N=\left(
\begin{array}{cccc}
n_{1,1} & n_{1,2} & ... & n_{1,|\mathcal{Q}|} \\
n_{2,1} & n_{2,2} & ... & n_{2,|\mathcal{Q}|} \\
... & ... & ... & ... \\
n_{r,1} & n_{r,2} & ... & n_{r,|\mathcal{Q}|}
\end{array}
\right).
\end{equation}
This can be viewed as an addition of $r$ vertices with the variables $\tau_{|\mathcal{Q}|+1}=\c_1,...\, ,\tau_{|\mathcal{Q}|+r}=\c_r$  to the quiver, and connection of each vertex containing $\tau_{|\mathcal{Q}|+\alpha}$ with the vertices containing $\tau_i,\, i<|\mathcal{Q}|$ by $n_{\alpha,i}$ arrows. We will denote extended quiver by $\hat{\mathcal{Q}}$ so that $|\hat{\mathcal{Q}}|=|\mathcal{Q}|+r$. Mutation rules for $\tau$ variables get unified form
\begin{equation}
\label{eq:mutAfroz}
\tau_k' = \dfrac{\prod_{i=1}^{|\hat{\mathcal{Q}}|} \tau_i^{[b_{ik}]_+}+\prod_{i=1}^{|\hat{\mathcal{Q}}|} \tau_i^{[-b_{ik}]_+}}{\tau_k},\ \ \ \tau_i' = \tau_i\ \mathrm{if}\ i\neq k,
\end{equation}
while mutation rules for coefficients are no longer needed - they transformations are taken into account by transformations of extended quiver with frozen variables. The map from $A$-cluster variables to $X$-cluster variables is given by
\begin{equation}
\label{eq:AtoX}
x_i = \prod\limits_{k=1}^{|\hat{\mathcal{Q}}|} \tau_k^{b_{ki}},
~~~~
1\leq i \leq |\mathcal{Q}|
.
\end{equation}
Under this map the frozen variables (i.e. coefficients) parameterize the Casimir functions of Poisson algebra $\{\Z,\,\cdot\,\}=0$, which are monomials $\Z=\prod_i x_i^{c_i}$ in $X$-variables, defined by the property that $\sum_i \varepsilon_{ij} c_j = 0$. If one takes all unit coefficients, the Casimirs
\begin{equation}
\Z = \prod\limits_{i=1}^{|\mathcal{Q}|} x_i^{c_i} = \prod\limits_{i=1}^{|\mathcal{Q}|} \prod\limits_{k=1}^{|\mathcal{Q}|} \tau_k^{\varepsilon_{ki}c_i} = \prod\limits_{k=1}^{|\mathcal{Q}|} \tau_k^{\sum_i \varepsilon_{ki}c_i}=1,
\end{equation}
become trivial. Mutation rules (\ref{eq:mutAfroz}) and (\ref{eq:mutations}) are consistent with (\ref{eq:AtoX}).

In the example of relativistic affine Toda chain with two particles (or on two sites) one gets two Casimir functions $Z$ and $q$. Extended exchange matrix, chosen following \cite{BGM:2017}, is drawn at Fig.~\ref{fig:todaexA}.
\begin{figure}[!h]
\centering
	\begin{tikzpicture}
	\node at (-2,1) {$b=$};	
	\matrix at (0,1) [matrix of nodes,row sep=0, left delimiter={(}, right delimiter={)}]
	{
 		0 &  2 & 0 & -2 \\
 		-2 &  0 & 2 & 0 \\
 		0 &  -2 & 0 & 2 \\
 		2 &  0 & -2 & 0 \\
 		2 &  0 & 2 & 0 \\
 		2 & -2 & 2 & -2 \\
	};

	\tikzmath{\xs=5;\ys=-0.5;\bend=15;};
	\draw[quiverVertex] (\xs+0,\ys+0) circle;
	\node at (\xs-0.3,\ys+2) {$1$};
	\draw[quiverVertex] (\xs+0,\ys+2) circle;
	\node at (\xs+2+0.3,\ys+2) {$2$};
	\draw[quiverVertex] (\xs+2,\ys+0) circle;
	\node at (\xs+2+0.3,\ys+0) {$3$};
	\draw[quiverVertex] (\xs+2,\ys+2) circle;
	\node at (\xs-0.3,\ys+0) {$4$};
	\draw[blue, quiverVertexBlue] (\xs+3,\ys+3) circle;
	\node at (\xs+3+0.3,\ys+3) {$5$};
	\draw[blue, quiverVertexBlue] (\xs+1,\ys+1) circle;
	\node at (\xs+1-0.3,\ys+1) {$6$};

	\path (\xs,\ys+0) edge[styleQuiverEdge, bend right=\bend] (\xs,\ys+2);
	\path (\xs,\ys+0) edge[styleQuiverEdge, bend left=\bend] (\xs,\ys+2);	
	
	\path (\xs,\ys+2) edge[styleQuiverEdge, bend right=\bend] (\xs+2,\ys+2);
	\path (\xs,\ys+2) edge[styleQuiverEdge, bend left=\bend] (\xs+2,\ys+2);	
	
	\path (\xs+2,\ys+2) edge[styleQuiverEdge, bend right=\bend] (\xs+2,\ys+0);
	\path (\xs+2,\ys+2) edge[styleQuiverEdge, bend left=\bend] (\xs+2,\ys+0);
	
	\path (\xs+2,\ys+0) edge[styleQuiverEdge, bend right=\bend] (\xs,\ys+0);
	\path (\xs+2,\ys+0) edge[styleQuiverEdge, bend left=\bend] (\xs,\ys+0);
	
	
	\path (\xs+2,\ys+2) edge[blue, styleQuiverEdge, bend right=\bend] (\xs+1,\ys+1);
	\path (\xs+2,\ys+2) edge[blue, styleQuiverEdge, bend left=\bend] (\xs+1,\ys+1);
	\path (\xs+1,\ys+1) edge[blue, styleQuiverEdge, bend right=\bend] (\xs+2,\ys+0);
	\path (\xs+1,\ys+1) edge[blue, styleQuiverEdge, bend left=\bend] (\xs+2,\ys+0);
	\path (\xs,\ys+0) edge[blue, styleQuiverEdge, bend right=\bend] (\xs+1,\ys+1);
	\path (\xs,\ys+0) edge[blue, styleQuiverEdge, bend left=\bend] (\xs+1,\ys+1);		
	\path (\xs+1,\ys+1) edge[blue, styleQuiverEdge, bend right=\bend] (\xs,\ys+2);
	\path (\xs+1,\ys+1) edge[blue, styleQuiverEdge, bend left=\bend] (\xs,\ys+2);	
	
	\path (\xs+3,\ys+3) edge[blue, styleQuiverEdge, bend right=20] (\xs,\ys+2);
	\path (\xs+3,\ys+3) edge[blue, styleQuiverEdge, bend left=-5] (\xs,\ys+2);	
	
	\path (\xs+3,\ys+3) edge[blue, styleQuiverEdge, bend right=-5] (\xs+2,\ys+0);	
	\path (\xs+3,\ys+3) edge[blue, styleQuiverEdge, bend left=20] (\xs+2,\ys+0);
	
\end{tikzpicture}
\caption{Left: extended exchange matrix $b$ for $\widehat{SL_2}$ Toda chain. Right: extended quiver with frozen vertices shown by blue.}
\label{fig:todaexA}
\end{figure}
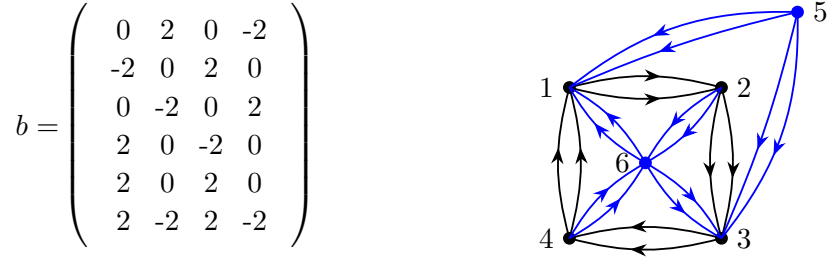

The coefficients can be read from the matrix $b$ (two lowest rows)
\begin{equation}
\cof_1 = (\tau_5 \tau_6)^{2},\ \ \ \cof_2 = \tau_6^{-2}, \ \ \ \cof_3 = (\tau_5 \tau_6)^{2},\ \ \ \cof_4 = \tau_6^{-2}
\end{equation}
Introducing (cf. with Fig.~\ref{fig:todaex}) $\tau_5 = q^{\frac{1}{4}}$, $\tau_6 = \Z^{\frac{1}{4}}$, one can write dynamics of cluster variables under $q$-Painlev\'e flow $T=s_{1,2}s_{3,4}\mu_1 \mu_3$, $T\tau_i = \overline{\tau}_i = \tau_i(qZ)$, as
\begin{equation}
\overline{(\tau_1,\tau_2,\tau_3,\tau_4)}=\left(\tau_2,\dfrac{\tau_2^2 + q^{\frac{1}{2}}\Z^{\frac{1}{2}}\tau_4^2}{\tau_1},\tau_4,\dfrac{\tau_4^2 + q^{\frac{1}{2}}\Z^{\frac{1}{2}}\tau_2^2}{\tau_3}\right).
\end{equation}
Eliminating $\tau_2$ and $\tau_4$, one turns it into bilinear form of $q$-Painlev\'e $A_7^{(1)'}$ equation
\begin{equation}
\overline{\tau}_1 \underline{\tau}_1 = \tau_1^2 + \Z^{\frac{1}{2}}\tau_3^2, \ \ \
\overline{\tau}_3 \underline{\tau}_3 = \tau_3^2 + \Z^{\frac{1}{2}}\tau_1^2.
\end{equation}

\section{Proof of the $\RLL$ relation for cluster L-matrices}
\label{ss:RLL}

Here some details of proof of (\ref{RTTClusterMain}) are collected.
Recall the definitions \rf{Lclust} (here and below $i,j=1,\ldots,M$)
\be
\label{LclustC}
L_{ij}(\mu) =
\dfrac{1}{\mu^{\frac{1}{2}}-\mu^{-\frac{1}{2}}}
\left\{
\begin{array}{ll}
i=j, & \mu^{\frac{1}{2}}z_i^{-2}+\mu^{-\frac{1}{2}}z_i^2 \\
i\neq j, & \mu^{-\frac{s_{ij}}{2}} (z_j^2+z_j^{-2})\dfrac{\tau_j}{\tau_i}
\end{array}
\right. ,~~~
\tau_i=w_i\prod\limits_{k=1}^{M}z_k^{s_{ki}}~.
\ee
where the variables $z_i,w_i$ have Poisson brackets
\begin{equation}
\{z_i,w_j\}=\frac{1}{4}\delta_{ij}z_i w_j,~~~ \{z_i,z_j\}=\{w_i,w_j\}=0.
\end{equation}
It is useful to note that
\begin{equation}
\{z_i, \tau_j\}=\frac{1}{4}\delta_{ij} z_i \tau_j,~~~ \{\tau_i, \tau_j\}=-\frac{1}{2}s_{ij}\tau_i \tau_j.
\end{equation}
In addition to the sign-factors \rf{sifa}
we also introduce~\footnote{Notation $k\in (ij)$ means that we consider $i,j,k$ on the circle $\mathbb{Z}/M\mathbb{Z}$, with $k$ in the oriented interval from $i$ to $j$.}
\begin{equation}
s_{ij}^{k} =
\left\{
\begin{array}{lc}
+1, & k\in (ij) \\
-1, & k\in (ji) \\
0, & k = i,j
\end{array}
\right.
\end{equation}
which satisfies
\begin{equation}
\label{rel2}
s^k_{ij} = -s^k_{ji},\;\;\; s^k_{ij} = s^i_{jk},\;\;\; s^k_{ij} = s_{ij}+s_{jk}+s_{ki}.
\end{equation}
From definitions \rf{LclustC}
\be
\label{rel4}
z_k^2 = -\dfrac{L_{kk}(\lambda)\sqrt{\mu}-L_{kk}(\mu)\sqrt{\lambda}}{\sqrt{\lambda/\mu}-\sqrt{\mu/\lambda}},\;\;\;
z_k^{-2} = \dfrac{L_{kk}(\lambda)/\sqrt{\mu}-L_{kk}(\mu)/\sqrt{\lambda}}{\sqrt{\lambda/\mu}-\sqrt{\mu/\lambda}}
\\
L_{ij}(\lambda)L_{kl}(\mu)=\lambda^{-\frac{1}{2}s_{ij}+\frac{1}{2}s_{kl}} \mu^{\frac{1}{2}s_{ij}-\frac{1}{2}s_{kl}} L_{ij}(\mu)L_{kl}(\lambda),~~~ i\neq j,~~~ k\neq l.
\ee
We take an anzatz
\begin{equation}
\label{rProof}
\tilde{r}(a)=\sum\limits_{k=1}^{M} f_k(a) E_{kk}\otimes E_{kk} + \sum\limits_{m\neq n} g_{mn}(a) E_{mn}\otimes E_{nm}
\end{equation}
and show that one can choose $f_{k}$ and $g_{mn}$ such that equation
\begin{equation}
\{L(\lambda)\otimes L(\mu)\} = [\tilde{r}(\lambda/\mu),L(\lambda)\otimes L(\mu)]
\end{equation}
holds. By direct computation it can be shown that $(a \neq i \neq j \neq k \neq l)$:
\begin{equation}
\label{tabbr}
\begin{array}{l|ll}
 & a.\; \{L(\lambda)\otimes L(\mu)\} & b.\; [\tilde{r}(\lambda/\mu),L(\lambda)\otimes L(\mu)] \\
\hline
1.\; E_{ii}\otimes E_{jj} & 0 & 0\\
2.\; E_{aa}\otimes E_{ij} & 0 & g_{ai} L_{ia}(\lambda)L_{aj}(\mu)-g_{ja} L_{aj}(\lambda)L_{ia}(\mu)\\
3.\; E_{aa}\otimes E_{aj} & A L_{aa}(\lambda)L_{aj}(\mu) - B_{aj} L_{aj}(\lambda)L_{aa}(\mu) & f_{a} L_{aa}(\lambda)L_{aj}(\mu) - g_{ja} L_{aj}(\lambda)L_{aa}(\mu)\\
4.\; E_{aa}\otimes E_{ia} & -A L_{aa}(\lambda)L_{ia}(\mu) + B_{ia} L_{ia}(\lambda)L_{aa}(\mu) & -f_{a} L_{aa}(\lambda)L_{ia}(\mu) + g_{ai} L_{ia}(\lambda)L_{aa}(\mu)\\
5.\; E_{ij}\otimes E_{ji} & B_{ji} (L_{jj}(\lambda)L_{ii}(\mu)-L_{ii}(\lambda)L_{jj}(\mu)) & g_{ij} L_{jj}(\lambda)L_{ii}(\mu)-g_{ij} L_{ii}(\lambda)L_{jj}(\mu)\\
6.\; E_{ij}\otimes E_{kl} & \frac{1}{2}(s_{ij}^k+s_{ji}^l)L_{ij}(\lambda)L_{kl}(\mu) & g_{ik} L_{kj}(\lambda)L_{il}(\mu)-g_{lj} L_{il}(\lambda)L_{kj}(\mu)\\
7.\; E_{ij}\otimes E_{ia} & - \frac{1}{2}s^a_{ij} L_{ij}(\lambda)L_{ia}(\mu) & f_{i} L_{ij}(\lambda)L_{ia}(\mu) - g_{aj} L_{ia}(\lambda)L_{ij}(\mu)\\
8.\; E_{ij}\otimes E_{aj} & \frac{1}{2}s^a_{ij} L_{ij}(\lambda)L_{aj}(\mu) & -f_{j} L_{ij}(\lambda)L_{aj}(\mu) + g_{ia} L_{aj}(\lambda)L_{ij}(\mu)\\
9.\; E_{ij}\otimes E_{ja} &  B_{ji} L_{jj}(\lambda)L_{ia}(\mu) - B_{ja} L_{ia}(\lambda)L_{jj}(\mu) & g_{ij} L_{jj}(\lambda)L_{ia}(\mu)-g_{aj} L_{ia}(\lambda)L_{jj}(\mu)\\
10.\; E_{ij}\otimes E_{ai} & - B_{ji} L_{ii}(\lambda)L_{aj}(\mu) + B_{ai} L_{aj}(\lambda)L_{ii}(\mu) & -g_{ij} L_{ii}(\lambda)L_{aj}(\mu) + g_{ia} L_{aj}(\lambda)L_{ii}(\mu)\\
\end{array}
\end{equation}
with
\begin{equation}
A=A(\sqrt{\lambda/\mu}) = \frac{1}{2}\dfrac{\sqrt{\lambda/\mu}+\sqrt{\mu/\lambda}}{\sqrt{\lambda/\mu}-\sqrt{\mu/\lambda}},\;\;\; B_{ij} = B_{ij}(\sqrt{\lambda/\mu}) = \dfrac{(\lambda/\mu)^{\frac{1}{2}s_{ij}}}{\sqrt{\lambda/\mu}-\sqrt{\mu/\lambda}}.
\end{equation}
Computations in 1,2,7,8.a) are straightforward. In 3, 4, 5.a) relation (\ref{rel4}) has to be used. 9,10.a) can be obtained by application of (\ref{rel4}) and (\ref{rel2}):
\be
\{L_{ij}(\lambda),L_{ja}(\mu)\} = -\frac{1}{2}\lambda^{-\frac{1}{2}s_{ij}} \mu^{-\frac{1}{2}s_{ja}} \dfrac{\tau_a}{\tau_i}(s_{ij}^{a}(z_j^2 + z_j^{-2})(z_a^2+z_a^{-2}) + (z_j^2 - z_j^{-2})(z_a^2+z_a^{-2}))=
\\
=-\frac{1}{2}\lambda^{-\frac{1}{2}s_{ij}} \mu^{-\frac{1}{2}s_{ja}} (z_a^2+z_a^{-2})\dfrac{\tau_a}{\tau_i}((s_{ij}^{a} + 1)z_j^2 +(s_{ij}^{a} - 1) z_j^{-2})) =
\\
=
-s_{ij}^{a} \lambda^{-\frac{1}{2}s_{ij}} \mu^{-\frac{1}{2}s_{ja}} (z_a^2+z_a^{-2})\dfrac{\tau_a}{\tau_i}z_j^{2s_{ij}^{a}}=
\\
=\dfrac{\lambda^{-\frac{1}{2}s_{ij}+\frac{1}{2}s_{ia}} \mu^{-\frac{1}{2}s_{ja}}}{\sqrt{\lambda/\mu}-\sqrt{\mu/\lambda}} L_{ia}(\lambda)\left[L_{jj}(\lambda)\mu^{\frac{1}{2}s_{ij}^a}-L_{jj}(\mu)\lambda^{\frac{1}{2}s_{ij}^a}\right] =
\\
=
\dfrac{(\lambda/\mu)^{\frac{1}{2}s_{ji}} L_{jj}(\lambda) L_{ia}(\mu) - (\lambda/\mu)^{\frac{1}{2}s_{ja}}  L_{ia}(\lambda) L_{jj}(\mu)}{\sqrt{\lambda/\mu}-\sqrt{\mu/\lambda}}
\ee
Looking at the table (\ref{tabbr}) we can suggest that the last two columns are equal, if we put
\begin{equation}
f_i = A(\sqrt{\lambda/\mu}),\;\; g_{ij} = B_{ji}(\sqrt{\lambda/\mu})
\end{equation}
For 1-5 and 9-10 it is obvious. For 6, 7, 8 it is easier to move from the right to the left. For 6, using (\ref{rel2}):
\be
g_{ik}L_{kj}(\lambda)L_{il}(\mu)-g_{lj}L_{il}(\lambda)L_{kj}(\mu) =
\\
=
\dfrac{\lambda^{-\frac{1}{2}s_{ik}-\frac{1}{2}s_{kj}}\mu^{-\frac{1}{2}s_{ki}-\frac{1}{2}s_{il}}-\lambda^{-\frac{1}{2}s_{lj}-\frac{1}{2}s_{il}}\mu^{-\frac{1}{2}s_{jl}-\frac{1}{2}s_{kj}}}{\sqrt{\lambda/\mu}-\sqrt{\mu/\lambda}} \dfrac{\tau_j}{\tau_k}\dfrac{\tau_l}{\tau_i}(z_j^2+z_j^{-2})(z_l^2+z_l^{-2})=
\\
=\dfrac{\lambda^{-\frac{1}{2}s_{ik}^j}\mu^{-\frac{1}{2}s_{ki}^l}-\lambda^{-\frac{1}{2}s_{lj}^i}\mu^{-\frac{1}{2}s_{jl}^k}}{\sqrt{\lambda/\mu}-\sqrt{\mu/\lambda}} L_{ij}(\lambda)L_{kl}(\mu)
\ee
All possible relative positions of the indices $i,j,k,l$ can be encoded in the table
\begin{equation}
\begin{array}{cccc|c}
s_{ik}^j & s_{ki}^l & s_{lj}^i & s_{jl}^k & s_{ij}^k+s_{ji}^l\\
\hline
+1 & +1 & +1 & +1 & 0\\
+1 & -1 & +1 & -1 & 0\\
+1 & -1 & -1 & +1 & -2\\
-1 & +1 & -1 & +1 & 0\\
-1 & +1 & +1 & -1 & +2\\
-1 & -1 & -1 & -1 & 0
\end{array}
\end{equation}
which shows that 6.a) and 6.b) from (\ref{tabbr}) are equal. For 7.b):
\begin{equation}
f_{i}L_{ij}(\lambda)L_{ia}(\mu)-g_{aj}L_{ia}(\lambda)L_{ij}(\mu) =
\end{equation}
$$=\dfrac{1}{2}\dfrac{(\sqrt{\lambda/\mu}+\sqrt{\mu/\lambda})\lambda^{-\frac{1}{2}s_{ij}}\mu^{-\frac{1}{2}s_{ia}}-2\lambda^{-\frac{1}{2}s_{aj}-\frac{1}{2}s_{ia}}\mu^{-\frac{1}{2}s_{ja}-\frac{1}{2}s_{ij}}}{\sqrt{\lambda/\mu}-\sqrt{\mu/\lambda}}\dfrac{\tau_j}{\tau_i}\dfrac{\tau_a}{\tau_i}(z_j^2+z_j^{-2})(z_a^2+z_a^{-2})=$$
$$=\dfrac{1}{2}\dfrac{\sqrt{\lambda/\mu}+\sqrt{\mu/\lambda}-2(\lambda/\mu)^{-\frac{1}{2}s_{ia}^j}}{\sqrt{\lambda/\mu}-\sqrt{\mu/\lambda}}L_{ij}(\lambda)L_{ia}(\mu)=-\dfrac{1}{2}s_{ij}^a L_{ij}(\lambda)L_{ia}(\mu)$$
which is equal to 7.a). Similarly for 8 a) and b). To show that (\ref{rProof}) is equal to (\ref{r1}) multiplied by $\frac{1}{2}$, we have to note that
\begin{equation}
\sum_{k=1}^{M} E_{kk}\otimes E_{kk} = \M1\otimes\M1 - \sum_{i\neq j} E_{ii}\otimes E_{jj}
\end{equation}
and $\M1\otimes\M1$ is commuting with anything, so can be always added to the $r$-matrix with the arbitrary coefficient, without any change of the relations.


\begin{thebibliography}{99}

\bibitem[BGM17]{BGM:2017}
M.Bershtein, P.Gavrylenko, A.Marshakov,
{\it Cluster integrable systems, q-Painlev\'e equations and their quantization};
JHEP (2018) 077;
[\href{http://arxiv.org/abs/1711.02063}{{\tt arXiv:1711.02063}}].

\bibitem[BGM18]{BGM:2018}
M.Bershtein, P.Gavrylenko, A.Marshakov,
{\it Cluster Toda chains and Nekrasov functions};
Theor Math Phys (2019) 198:157
[\href{http://arxiv.org/abs/1804.10145}{{\tt arXiv:1804.10145}}].

\bibitem[BGT]{BGT:2017}
G.Bonelli, A.Grassi, A.Tanzini,
{\it Quantum curves and q-deformed Painlevé equations};
A. Lett Math Phys (2019)
[\href{https://arxiv.org/abs/1710.11603}{{\tt arXiv:1710.11603}}].

\bibitem[BPTY]{BPTY:2012} L.Bao, E.Pomoni, M.Taki, F.Yagi
\emph{M5-branes, toric diagrams and gauge theory duality,}
JHEP, \textbf{105} (2012); [\href{https://arxiv.org/abs/1112.5228v2}{{\tt arXiv:1112.5228}}].

\bibitem[BS]{BS}
V.~Bazhanov, S.~Sergeev,
{\it Zamolodchikov's tetrahedron equation and hidden structure of quantum groups}, Journal of Physics A 39 (2006) 13;
[\href{https://arxiv.org/abs/hep-th/0509181}{{\tt arXiv:hep-th/0509181}}].

\bibitem[BS16]{BS:2016:1}M. Bershtein and A. Shchechkin, \textit{$q$-deformed Painlev\'e tau function and $q$-deformed conformal blocks}, J. Phys. A: Math. and Theor.,  \textbf{50 8} (2017) 085202;  [\href{http://arxiv.org/abs/1608.02566}{{\tt arXiv:1608.02566}}].

\bibitem[BS18]{BS:2018}M. Bershtein and A. Shchechkin, \textit{Painlev\'e equations from Nakajima-Yoshioka blow-up relations} [\href{https://arxiv.org/abs/1811.04050}{{\tt arXiv:1811.04050}}].

\bibitem[DF]{QEquivAffine}
J.T.~Ding, I.B.~Frenkel,
{\it Isomorphism of two realizations of quantum affine algebra $U_q(\mathfrak{gl}(n))$};
Comm.\ Math.\ Phys.\ {\bf 156} 2 (1993), 277--300;\
[\href{https://projecteuclid.org:443/euclid.cmp/1104253628}{{\tt https://projecteuclid.org:443/euclid.cmp/1104253628}}].

\bibitem[EFS]{EFS}
R.~Eager, S.~Franco, K.~Schaeffer,
{\it Dimer Models and Integrable Systems};
JHEP {\bf 2012} 106 (2012)
[\href{https://arxiv.org/abs/1107.1244v2}{{\tt arXiv:1107.1244}}].

\bibitem[F]{Fock} V.V.Fock
\emph{Inverse spectral problem for GK integrable system,}
[\href{https://arxiv.org/abs/1503.00289}{{\tt arXiv:1503.00289}}].

\bibitem[FG05]{FG:2005} V. V. Fock, A. B. Goncharov, Cluster X-varieties, amalgamation and Poisson-Lie groups. In	 Algebraic Geometry Theory and Number Theory, pp. 27–68, Progr. Math., 253, Birkhäuser	 Boston, Boston, MA, 2006; [\href{http://arxiv.org/abs/math.RT/0508408}{{\tt arXiv:math.RT/0508408}}].
	
\bibitem[FM97]{FM:1997}
V.~V.~Fock and A.~Marshakov, {\em A Note on Quantum Groups and Relativistic Toda Theory},
Nucl.Phys. {\bf 56B} (Proc. Suppl.) (1997) 208-214.	

\bibitem[FM14]{FM:2014} V. V. Fock and A.~Marshakov, \emph{Loop groups, Clusters, Dimers and Integrable systems}, in Geometry and Quantization of Moduli Spaces 1--65;
[\href{http://arxiv.org/abs/1401.1606}{{\tt arXiv:1401.1606}}].

\bibitem[FZ]{FZ:2006} S.Fomin, A.Zelevinsky
\emph{Cluster algebras IV: Coefficients,}
Compositio Mathematica, \textbf{143}(1) (2007), 112-164; [\href{https://arxiv.org/abs/math/0602259}{{\tt arXiv:0602259}}].

\bibitem[FHM]{FHM}
S.~Franco, Y.~Hatsuda, M.~Marino,
{\it Exact quantization conditions for cluster integrable systems};
Journal of Statistical Mechanics: Theory and Experiment, {\bf 6} 6 (2016)
[\href{https://arxiv.org/abs/1512.03061}{{\tt arXiv:1512.03061}}].

\bibitem[FRT]{QEquiv}
L.D. Faddeev, N.Yu.Reshetikhin, L.A.Takhtajan,
{\it Quantization of Lie groups and Lie algebras};\
Algebra\ and\ Analysis\ (Russian) 1.1 (1989), 118-206

	

\bibitem[GK]{GK:2011}A. B. Goncharov and R. Kenyon, \emph{Dimers and cluster integrable systems}, Ann.  Sci. Ec. Norm. Sup ́(2013) \textbf{46, 5}, 747--813; [\href{http://arxiv.org/abs/1107.5588}{{\tt arXiv:1107.5588}}].

\bibitem[GIL]{GIL:1207}
O. Gamayun, N. Iorgov, O. Lisovyy, \textit{Conformal field theory of Painlev\'e VI}, JHEP \textbf{1210}, (2012), 38;[\href{http://arxiv.org/abs/1207.0787}{{\tt arXiv:1207.0787}}].

\bibitem[GMN]{Gaiotto} D. Gaiotto, G. Moore, A. Neitzke \emph{Wall-crossing, Hitchin Systems, and the WKB Approximation}, Adv in Math
\textbf{234}, (2013), 239--403
[\href{http://arxiv.org/abs/0907.3987}{{\tt arXiv:0907.3987}}].

\bibitem[SZ]{SZ} M. Semenyakin, Y. Zenkevich. \emph{To appear}.

\bibitem[GKMMM]{GKMMM}
A.~Gorsky, I.~Krichever, A.~Marshakov, A.~Mironov and A.~Morozov,
{\it Integrability and Seiberg-Witten exact solution};
Phys.\ Lett.\ B {\bf 355}, 466 (1995)
[\href{https://arxiv.org/abs/hep-th/9505035}{{\tt arXiv:hep-th/9505035}}].

\bibitem[HI]{Hone:2014}A.~Hone, R.~Inoue \emph{Discrete Painlev\'e equations from $Y$-systems}, J. Phys. A: Math. and Theor., \textbf{47 (47)} 2014;
[\href{http://arxiv.org/abs/1405.5379}{{\tt arXiv:1405.5379}}].

\bibitem[IIO]{IIO} R.~Inoue, T.~Ishibashi, H.~Oya \emph{Cluster realizations of Weyl groups and higher Teichm\"uller theory},
[\href{http://arxiv.org/abs/1902.02716}{{\tt arXiv:1902.02716}}].

\bibitem[JNS]{Jimbo:2017} M.~Jimbo, H.~Nagoya, and H.~Sakai, \emph{CFT approach to the $q$-Painlev\'e VI equation}; J. Int. Syst. (2017) \textbf{2}, 1–27 [\href{http://arxiv.org/abs/1706.01940}{{\tt arXiv:1706.01940}}].

\bibitem[K]{Kharchev} S.~Kharchev, \emph{Unpublished.}

\bibitem[M]{M:2012}
A.~Marshakov,
{\it Loop groups, Clusters, Dimers and Integrable systems};\
Journal of Geometry and Physics, {\bf 67}, p. 16-36.;
[\href{https://arxiv.org/abs/1207.1869}{{\tt arXiv:1207.1869}}].

\bibitem[MN]{MN:2018}
Y.~Matsuhira, H.~Nagoya,
{\it Combinatorial expressions for the tau functions of q-Painlevé V and III equations};\
[\href{https://arxiv.org/abs/1811.03285}{{\tt arXiv:1811.03285}}].

\bibitem[MM]{MM:1997}
A.~Marshakov, A.~Mironov,
{\it 5d and 6d Supersymmetric Gauge Theories: Prepotentials from Integrable Systems};
Nucl.\ Phys.\ B {\bf 518}, 1 (1998) 59-91
[\href{https://arxiv.org/abs/hep-th/9711156}{{\tt arXiv:hep-th/9711156}}].

\bibitem[MMRZZ]{MMRZZ}
A.~Mironov, A.~Morozov, B.~Runov, Y.~Zenkevich, A.~Zotov,
{\it Spectral dualities in XXZ spin chains and five dimensional gauge theories};
JHEP {\bf 1312} (2013) 034
[\href{https://arxiv.org/abs/1307.1502}{{\tt arXiv:hep-th/1307.1502}}].


\bibitem[N]{N:1996}
N.~Nekrasov,
{\it Five Dimensional Gauge Theories and Relativistic Integrable Systems};
Nuclear Physics B, {\bf 531}-1, 323-344.\
[\href{https://arxiv.org/abs/hep-th/9609219}{{\tt arXiv:hep-th/9609219}}].

\bibitem[O15]{Okubo:2015} N.~Okubo, \emph{Bilinear equations and q-discrete Painlev\'e equations satisfied by variables and coefficients in cluster algebras}, J. Phys. A: Math. Theor. \textbf{48} 355201;
[\href{http://arxiv.org/abs/1505.03067}{{\tt arXiv:1505.03067}}].

\bibitem[O17]{Okubo:2017} N.~Okubo, \emph{Co-primeness preserving higher dimensional extension of $q$-discrete Painlev\'e I, II equations};
[\href{http://arxiv.org/abs/1704.05403}{{\tt arXiv:1704.05403}}].

\bibitem[OS]{OS:2018}N.Okubo, T.Suzuki \emph{Generalized $q$-Painlev\'e VI systems of type $(A_{2n+1}+A_1+A_1)^{(1)}$ arising from cluster algebra}
[\href{http://arxiv.org/abs/1810.03252}{{\tt arXiv:1810.03252}}].

\bibitem[OPS]{QGdefs}
A.~Oskin, S.~Pakuliak, A.~Silantyev,
{\it On the universal weight function for the quantum affine algebra $U_q(\hat{\mathfrak{gl}}_N)$};\
Lett.\ Math.\ Phys.\  {\bf 91} (2010) 167;
[\href{https://arxiv.org/abs/0711.2821v2}{{\tt arXiv:0711.2821}}].

\bibitem[ORV]{ORV} A.~Okounkov, N.~Reshetikhin, C.~Vafa, \emph{Quantum Calabi-Yau and Classical Crystals}, [\href{https://arxiv.org/abs/hep-th/0309208}{{\tt hep-th/0309208}}].

\bibitem[R]{R:1990} S.N.M. Ruijsenaars, \emph{”Relativistic Toda systems”},
Commun.Math. Phys., {\bf 133}:217 (1990), 753-760.\
[\href{https://projecteuclid.org/euclid.cmp/1104201396}{{\tt euclid.cmp/1104201396}}].

\bibitem[S]{S:1996}
N.~Seiberg,
{\it Five Dimensional SUSY Field Theories, Non-trivial Fixed Points and String Dynamics};
Phys.Lett.B, {\bf 388} (1996), 753-760.\
[\href{https://arxiv.org/abs/hep-th/9608111v2}{{\tt arXiv:hep-th/9608111}}].	

\bibitem[SW]{SW:1994}
N.~Seiberg, E.~Witten,
{\it Monopole Condensation, And Confinement In N=2 Supersymmetric Yang-Mills Theory};
Nuclear Physics B, {\bf 426} 1 (1994), 19-52.\
[\href{https://arxiv.org/abs/hep-th/9407087v1}{{\tt arXiv:hep-th/9407087}}].	

\end{thebibliography}
\end{document}